\documentclass[aps,pre,amsfonts,amssymb,amsmath,floatfix, onecolumn, superscriptaddress]{revtex4-2}

\usepackage{color}
\usepackage{graphicx}
\usepackage{afterpage} 
\usepackage{multirow}
\usepackage[english]{babel}

\graphicspath{{.}{Topology_of_PDLs_arxiv/}}

\newcommand{\phiact}{\ensuremath{\phi^{\rm act}}}
\newcommand{\phiarr}{\ensuremath{\phi^{\rm LAT}}}

\newcommand{\LL}{\mathcal{L}}
\newcommand{\TT}{\mathcal{T}}

\newcommand{\UU}{\mathcal{U}}
\newcommand{\EE}{\mathcal{E}}

\newcommand{\FF}{\mathcal{F}}
\newcommand{\BB}{\mathcal{B}}

\newcommand{\ZZ}{\mathcal{Z}}
\newcommand{\hh}{\mathcal{h}}
\newcommand{\ttt}{\mathcal{t}}

\newcommand{\Vstar}{V_*}
\newcommand{\Rstar}{R_*}

\newcommand{\uu}{\mathbf{u}}
\newcommand{\dd}{\partial}

\newcommand{\KULAK}{Department of Mathematics, KU Leuven Campus Kortrijk, Belgium}
\newcommand{\ISI}{iSi Health - KU Leuven Institute of Physics-based Modeling for In Silico Health}
\newcommand{\LUMC}{Leiden University Medical Center, the Netherlands}

\makeatletter\AtBeginDocument{\let\@elt\relax}\makeatother 

\begin{document}

\begin{abstract}
Several excitable systems, such as the heart, self-organize into complex spatio-temporal patterns that involve wave collisions, wave breaks, and rotating vortices, of which the dynamics are incompletely understood. Recently, conduction block lines in two-dimensional media were recognized as phase defects, on which quasi-particles can be defined. These particles also form bound states, one of which corresponds to the classical phase singularity. Here, we relate the quasi-particles to the structure of the dynamical attractor in state space and extend the framework to three spatial dimensions. We reveal that 3D excitable media are governed by phase defect surfaces, i.e. branes, and three flavors of topologically preserved curves, i.e. strings: heads, tails, and pivot curves. We identify previously coined twistons as points of co-dimension three at the crossing of a head curve and a pivot curve. Our framework predicts splitting and branching phase defect surfaces that can connect multiple classical filaments, thereby proposing a new mechanism for the origin, perpetuation, and control of complex excitation patterns, including cardiac fibrillation.
\end{abstract}

\title{Strings, branes and twistons: topological analysis of phase defects in excitable media such as the heart}

\author{Louise Arno}
\affiliation{\KULAK}
\affiliation{\ISI} 
\author{Desmond Kabus}
\affiliation{\KULAK}
\affiliation{\ISI} 
\affiliation{\LUMC}
\author{Hans Dierckx}
\affiliation{\KULAK}
\affiliation{\ISI} 
\date{\today}

\maketitle

\section{Introduction}

\subsection{Motivation}

Excitable media are characterized by the property that a small stimulus can evoke a large response, after which the system gradually returns to its initial state. Some of the most interesting natural systems gain their complex dynamics from the interaction between coupled excitable elements, including electrical pulses in the heart \cite{Zipes:1995}, brain \cite{Bressloff:2014}, and intestines \cite{Sanders2019}, and pandemic spread \cite{noble_geographic_1974}. In these systems, the precise spatio-temporal pattern emerges from self-organization and tends to determine the global state of the system. E.g. in pandemics, the activation tends to come in waves, and in the cardiac context, the sequence of propagating pulses determines whether the heart is functioning normally or exhibits a rhythm disorder. The formalism presented in this paper was designed with cardiac applications in mind but is expected to be also useful in other excitable systems. 

Over the past decades, it was found that excitable systems can be better understood by looking at the self-emerging phenomena that arise in them. These include traveling wave fronts, where elements become excited, and wave backs, where elements return to the resting state. When a wave front hits a wave back, e.g. due to heterogeneity of the medium, a wave block occurs, which is known as a conduction block in a cardiac context \cite{Janse:1980}. Its end points, called pivots, are of interest for ablation strategies during atrial fibrillation \cite{seitz_af_2017}. Such an event of conduction block may initiate the creation of rotating vortices, called spiral waves in two spatial dimensions (2D) and scroll waves in three spatial dimensions (3D). In the cardiac context, these emergent patterns are also called `rotors', see the numerical examples in Fig. \ref{fig:spirals}. 

Since the local excitation of cardiomyocytes invokes local mechanical contraction, abnormal patterns such as rotors cause life-threatening arrhythmias. Cardiac arrhythmias are among the leading cause of death worldwide \cite{WHO:2020}, partly since the basic mechanisms that underlie arrhythmias such as atrial or ventricular fibrillation are incompletely understood \cite{cronin_2019,mujovic_catheter_2017}. A main concern with the current theory of rotors and arrhythmias \cite{PanfilovDierckx:2017} is that previous efforts on spiral and scroll wave dynamics have focused on long-lived rotating patterns, such that their time-averaged drift properties and stability could be derived from first principles \cite{Keener:1988,Biktashev:1994}. In both clinical reality and optical voltage mapping experiments \cite{Salama:1987}, however, the excited regions during fibrillation and complex tachycardias are aperiodic and fragmented, and rotors persisting for more than one rotation are seldom observed \cite{tabereaux_mechanisms_2009,Annonni:2018,Arno:2021}. 

Nonetheless, in the thicker regions of the cardiac wall, it was found useful to consider the instantaneous rotation point of 2D spiral waves and rotation axes of 3D scroll waves, called phase singularities (PSs) and filaments, respectively \cite{Winfree:1984,Clayton:2005}. In the regime of long-lived rotors, observed in numerical simulations, filaments are detected as PSs, also termed spiral wave tip, on both the inner endocardial and outer epicardial boundary of the myocardial wall. Therefore, one expects that a simultaneous recording of the endo- and epicardial activation yields similar patterns of PSs, but this is not found clinically. In this clinical context a mismatch between endo- and epicardial patterns is observed, both in the cardiac atria \cite{kharbanda_simultaneous_2020} and ventricles \cite{tung_simultaneous_2020}. 

The aim of this paper is to present a consistent theoretical description of the emerging excitation structures that can arise in the bulk of 3D excitable media, unifying the concepts of rotors, both short- and long-lived, and conduction blocks. To this purpose, we build further on the 2D phase defect theory involving quasi-particles and Feynman-like diagrams that we recently proposed \cite{Arno:2023feynman}. 

\subsection{Outline} 

In the remainder of this introductory section, we first review the classical interpretation of linear-core meander and filament dynamics. Then, we discuss a recent interpretation of linear spiral wave cores as phase discontinuities \cite{Tomii:2021,Arno:2021} and how they give rise to quasi-particles \cite{Arno:2023feynman}. In Sec. \ref{sec:methods}, we describe how numerical examples were computed. The main results are in Sec. \ref{sec:statespace}, where the quasi-particles in excitation patterns are introduced by the state space, and in Sec. \ref{sec:3D}, where a theoretical framework is outlined that refines the scroll wave filament concept and reconciles it with linear rotor cores. First, we formulate the theory, and then we predict the existence of a few new states, which are found in numerical experiments. In Sec. \ref{sec:discussion} the results are discussed within and beyond a cardiac electrophysiology context, where links to other fields of mathematics and physics are given. Concluding remarks follow in Sec. \ref{sec:conclusions}.

 \subsection{Brief review of spirals, scrolls and filaments}

\begin{figure*}
\begin{tabular}{ccc}
   (a)  & (b) & (c)  \\
   \includegraphics[height=4cm]{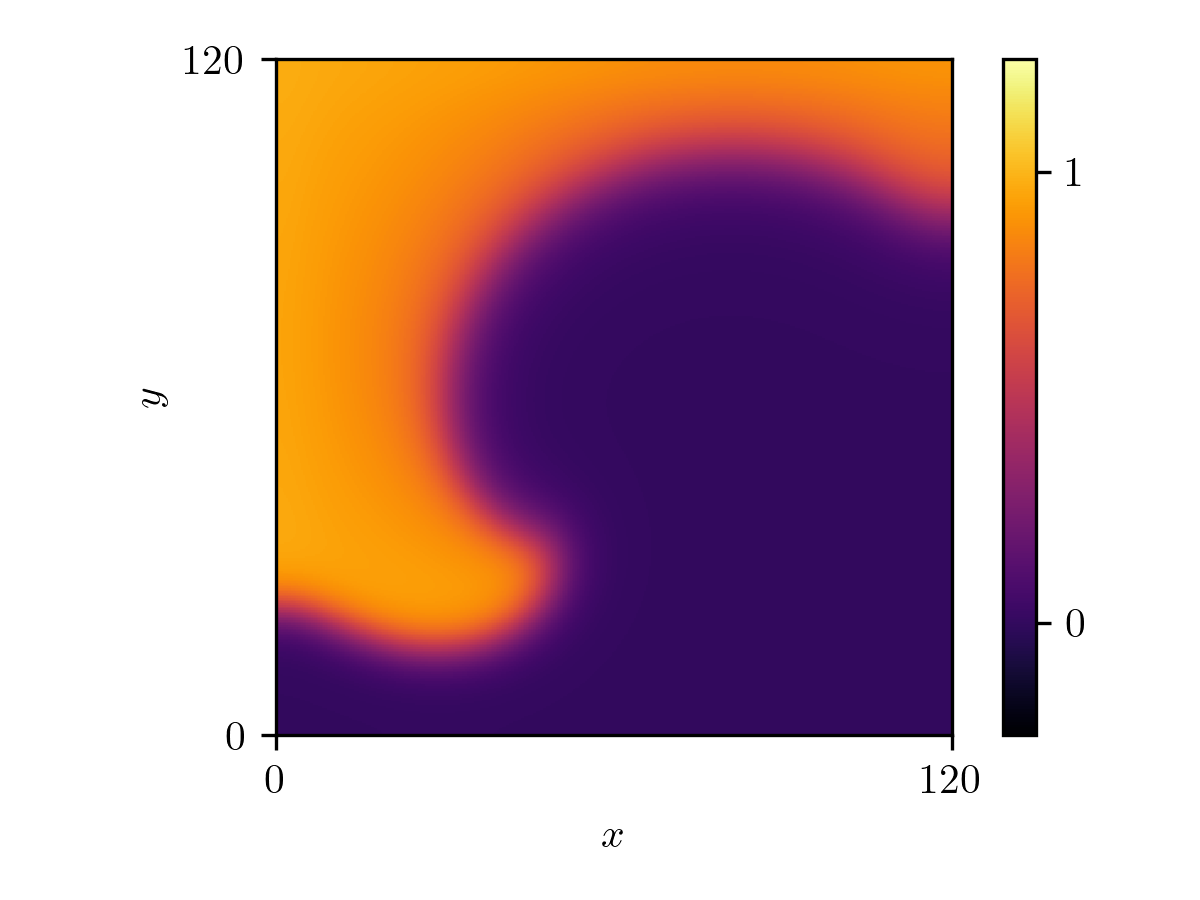}
  & \includegraphics[height=4cm]{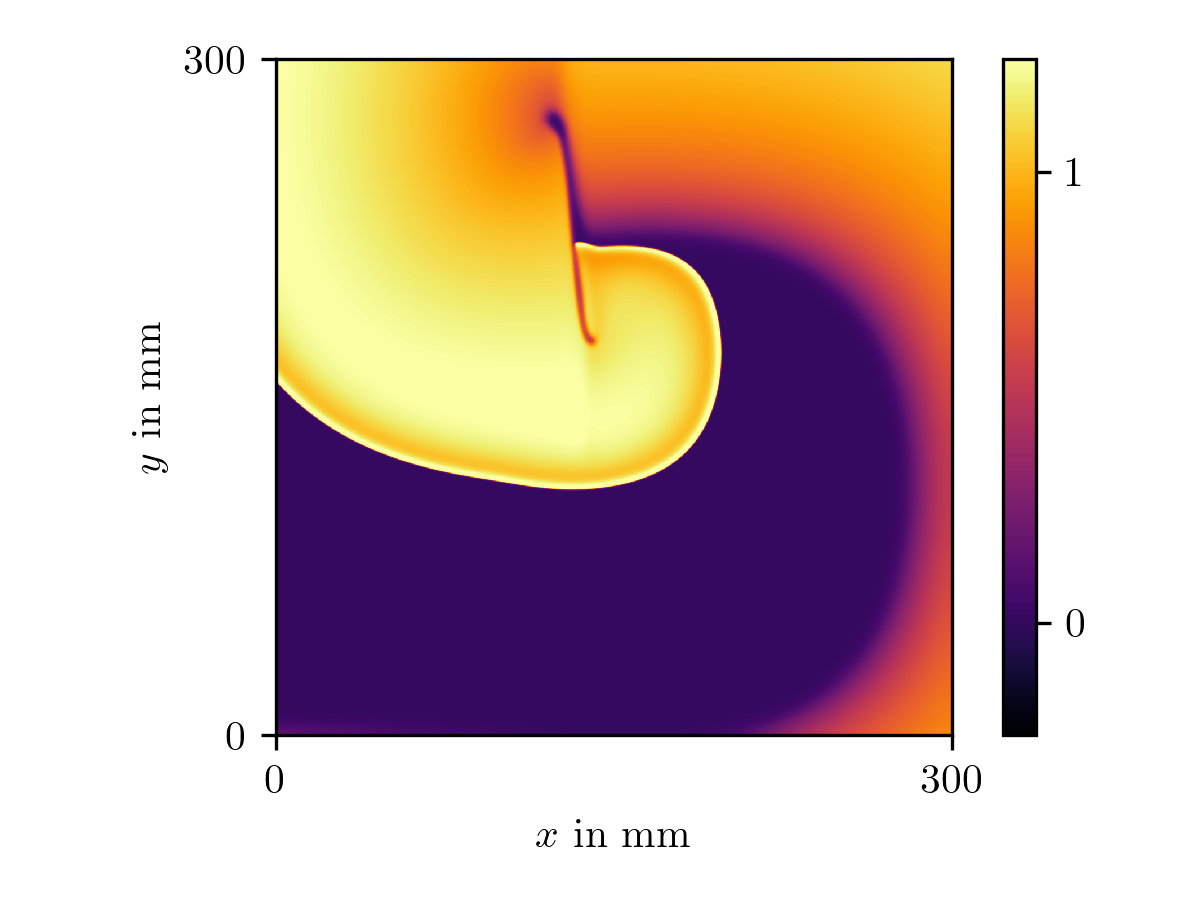}
&  \includegraphics[height=4cm]{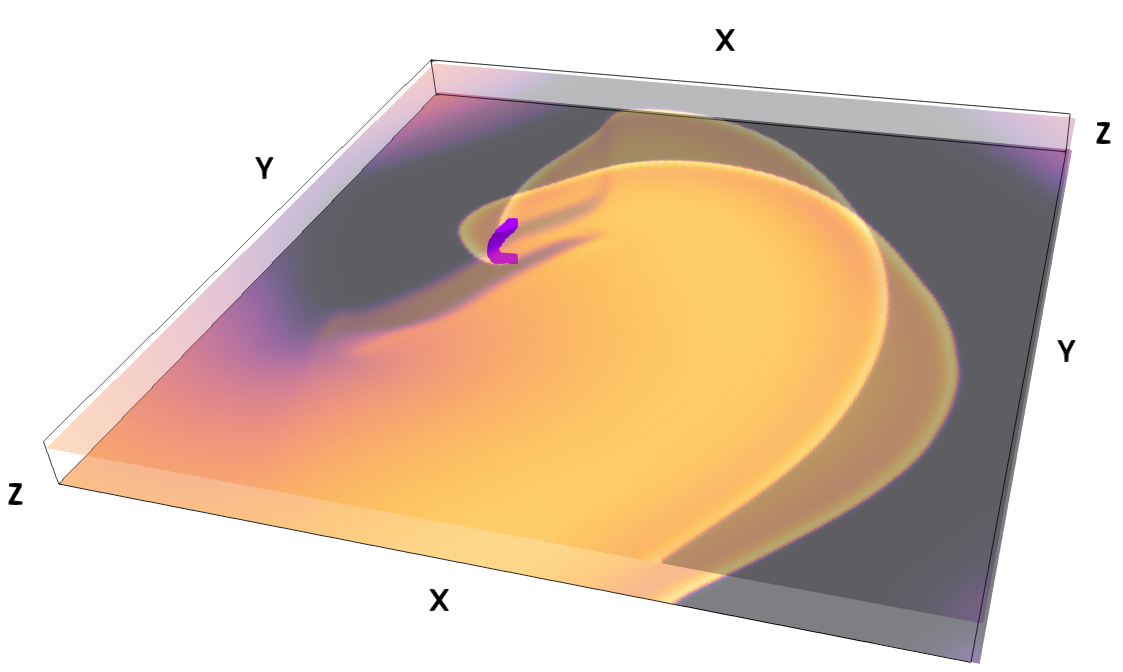}
\end{tabular}
\caption{Key excitable structures in this paper: from simple to complex excitation patterns. Simulated patterns are (a) circular-core spiral wave, using the Aliev-Panfilov (AP) cardiac excitation model \cite{Aliev:1996} in an isotropic medium, (b) biperiodically rotating (meandering) spiral wave with linear core in the Bueno-Orovio-Cherry-Fenton (BOCF) model\cite{BuenoOrovio:2008}, (c) 3D rotating scroll wave in the BOCF model \cite{BuenoOrovio:2008} for the anisotropic case with $D_1= 1$ and $D_2=1/5$ for which the fibre angle varies from $-60 ^{\circ}$ (bottom) to $60 ^{\circ}$ (top). \label{fig:spirals}
}
\end{figure*}

Fig. \ref{fig:spirals}a and b show two types of spiral waves, produced by time-stepping of the reaction-diffusion equations, also known as the cardiac monodomain model 
\begin{align}
    \dd_t \uu =  \Delta \mathbf{P} \uu +  \mathbf{F}(\uu), \label{RDE}
\end{align}
with $\Delta$ the Laplacian operator. This partial differential equation states how a column matrix $\uu(\vec{r},t)$ of $m$ state variables evolves in time. The constant matrix $\mathbf{P} = \rm{diag}(P_{11},0,...0)$ makes sure that only the transmembrane potential $V$ diffuses. In panel a., the local non-linear kinetics $\mathbf{F}(\uu)$ were chosen according to Aliev and Panfilov \cite{Aliev:1996}, which after an initial transient process produces a rigidly rotating spiral structure. In the range of simple activation-recovery modes, also a biperiodic meandering motion of the phase singularity was found, originating from a Hopf bifurcation in the medium parameters \cite{Barkley:1990b}. Another type of meander is shown in Fig. \ref{fig:spirals}b: If the excitation has a marked plateau phase, as occurs in the heart, the spiral wave tip needs to travel around the tissue it previously excited, producing a star-shaped or linear tip trajectory \cite{Fast:1990,Fenton:1998}. The different self-dynamics of the rotor core have been associated with different cardiac arrhythmias \cite{zipes_chapter_2004}.

During experiments and observations, spiral waves have a linear-shaped core \cite{Kabus:2022} and are often short-lived \cite{Arno:2021}. Therefore, research has been focusing on the description of single linear-core rotors \cite{Marcotte:2016,Dierckx:2017,Dierckx:2017PRL}. However, this 2D theory, which represents a spiral wave by its rotation point, is generally only applicable to long-lived rotors. Additionally, the framework does not include conduction blocks, observed in various cardiac data that possibly create spiral waves.
Therefore, there is a need for a framework that links rotors to conduction blocks and includes short-lived rotating waves. Our recent work addresses this question using quasi-particles in two spatial dimensions \cite{Arno:2023feynman}. 

In three spatial dimensions, the spiral wave becomes a scroll wave, rotating around a central core region. In the case of rigidly rotating spirals, one can define a clear local rotation axis, the rotor filament, around which the scroll wave turns \cite{Winfree:1984,Clayton:2005}. Filament dynamics in this regime have been studied theoretically with great success \cite{Keener:1986,Keener:1998,Keener:1991b,Keener:1992,Keener:1995b,Keener:1995a,Wellner:2002,Verschelde:2007,Dierckx:2012,Dierckx:2019} and its self-induced motion was shown to crucially depend on the filament curvature and twist of the scroll wave around the filament. Fig. \ref{fig:spirals}c shows an example of a 3D scroll wave together with its filament. Fenton and Karma found in numerical simulations that when anisotropy of myofibers is present, the filament exhibits accumulated regions of local twist, which they termed `twiston', and is discussed in relation to cardiac arrhythmias \cite{Fenton:1998}. 

However, the classical filament theory has several shortcomings: As in the two-dimensional case, it generally does not cover linear cores and it assumes long-lived scrolls. Additionally, the 3D theory does not explain the mismatch between endo- and epicardial patterns observed by clinicians \cite{kharbanda_simultaneous_2020,tung_simultaneous_2020}. The key result of this paper is to relieve the most important shortcomings and present a topologically consistent view on the possible intramural structures.

To reach this goal, we first revisit the classical filament definition via the activation phase and how a reparameterization of this phase led to an alternative description of the core of a cardiac rotor. 

\subsection{Overview of phase and phase defects in excitable media\label{sec:overview}}

\begin{figure*}
    \centering
    \includegraphics{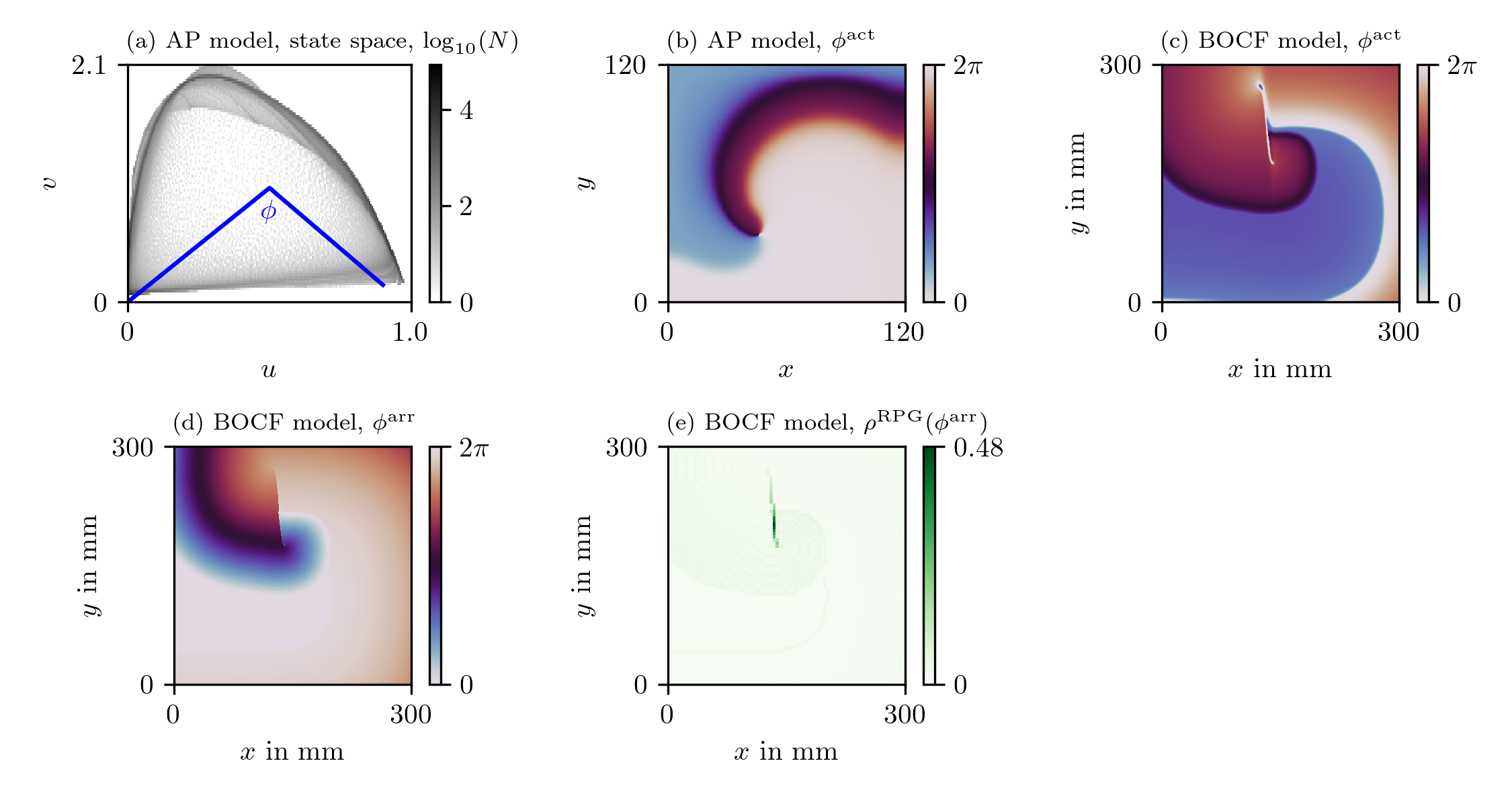}
    \caption{Overview of classical phase analysis and recently introduced phase defects. (a) Plotting two state variables with respect to each other allows us to define a phase angle in this plane. (b) The circular-core spiral wave from Fig. \ref{fig:spirals}a colored according to the phase, showing a point where all phases meet, the phase singularity. (c) The linear-core spiral wave from Fig. \ref{fig:spirals}b colored using the activation phase $\phi^{act}$. (d) Same spiral wave, using a phase $\phiarr$ based on local activation time, such that the wave front is no longer emphasized \cite{Kabus:2022}. Instead, a phase defect line or conduction block line is observed at the rotor core. (e) The phase defect density $\rho = ||\vec{\nabla} \phi ||$ allows to localize the phase defect \cite{Kabus:2022}. The data were down-sampled to make the otherwise few pixels thin line more visible.}
    \label{fig:phasesummary}
\end{figure*}

Excitable systems are closely related to oscillatory media, in which every element activates periodically, rather than returning to a resting state. Examples of oscillatory systems include chemical clock reactions \cite{Zhabotinsky:1973}, coupled oscillators in studies of synchronization phenomena \cite{strogatz2003sync} and the heart's natural pacemaker cells in the sinoatrial node. Due to the approximate periodic activity, oscillatory media have been described by a spatial phase $\phi \in [0, 2\pi)$ that tracks the progression of an element along the cycle, which can be seen in Fig. \ref{fig:phasesummary}a. Early on, it was recognized that many biological processes are naturally captured by a cyclic variable, or phase \cite{Winfree:1974}.

Classical phase analysis starts from two linear independent observables in the system, say $V$ and $R$ \cite{Gray:1995b,Bray:2003,Clayton:2005}. Here, $V(\vec{r},t)$ may be the transmembrane potential, and $R(\vec{r},t)$ either a recovery variable (in simulations) or a time-delayed or Hilbert-transformed $V$. Then, the activation phase is introduced as the polar coordinate around a point $\Vstar$, $\Rstar$ chosen inside the cyclic loop of the cell by
\begin{align}
    \phi^{\rm act}= \mathrm{atan2}(R - \Rstar,V - \Vstar) \label{eq:phase}
\end{align}
as shown in Fig. \ref{fig:phasesummary}a. 

This observation gave rise to phase analysis, in which the phase tracks the position on the typical excitation path in state space \cite{Gray:1998, Clayton:2005}. In emerging patterns such as rotating spiral waves, it was found that the phase changes over $\pm 2\pi$ if one circumscribes the rotation point of the spiral wave, from which it was concluded that a phase singularity (PS) is located near this site. An example of such PS, for which all different phases meet in this point, can be seen in Fig. \ref{fig:phasesummary}b. In three spatial dimensions (3D), this PS forms the filament curve discussed above. Filaments persist over time and can only end on the medium boundaries, in other words, they are `topologically protected' \cite{Pertsov:2000}. 

Recently, it was proposed that at the center of meandering rotors with linear core lies an extended phase discontinuity \cite{Tomii:2021} or phase defect line (PDL) \cite{Arno:2021}. This nature of linear cores was overlooked for a long time, since applying filtering or diffusion to the signals from experiments or simulation indeed creates a single point that is the phase singularity. Fig. \ref{fig:phasesummary}c shows a phase map of the linear-core spiral of Fig. \ref{fig:spirals}b. In addition to the linear core, there are also strong phase gradients near the wave front and back. The latter variations are natural in the system and result purely from the phase definition. If one alters the phase definition to 
\begin{align}
  \phiarr = 2\pi \tanh \left( \frac{t - T(\vec{r})}{\tau}\right) 
\end{align}
with $T$ the latest local activation time of a point, the phase will only vary strongly over the defect, see panel d. \cite{Arno:2021}. Quantifying the magnitude of the phase gradient allows us to detect the linear core or phase defect line explicitly, see Fig. \ref{fig:phasesummary}e or alternative methods in \cite{Tomii:2021,Kabus:2022}. 

In our previous publication \cite{Arno:2023feynman} we observed three different curves in an excitable surface pattern: wave fronts, wave backs, and PDLs. The end points of these three lines were called `heads', `tails', and `pivots', respectively, forming the building blocks of cardiac patterns. Here, we introduce these points from a state-space viewpoint and extend their notion to three dimensions. 

\section{Numerical methods \label{sec:methods}}

\subsection{Numerical simulations}
The results presented in this paper are valid for a wide range of excitable systems, which do not necessarily need to obey the reaction-diffusion eq. \eqref{RDE}. For definiteness, we produced the numerical examples in this paper by time-stepping of eq. \eqref{RDE} using the forward Euler method. See Tab. \ref{tab:sims} for simulation parameters. The reaction-kinetics function $\mathbf{F}(\uu)$ depends on the model, for which the ones used in this paper are indicated in the corresponding figures and Tab. \ref{tab:sims}. To include rotational anisotropy, the Laplacian $\Delta$ in eq. \eqref{RDE} was replaced by $\nabla \mathbf{D} \nabla$ with $\nabla$ the gradient and  $\mathbf{D}$ a $3\times 3$- matrix with elements $D^{ij} = D_1 \delta^{ij} + (D_1-D_2) e_f^i e_f^j$, $i,j \in \{ 1,2,3 \}$. Here, $\delta^{ij}$ represents the kronecker delta. The values of the diffusivities $D_1$ and $ D_2$ and the orientation of the fibre direction $\vec{e}_f$ depend on the different examples, indicated in the corresponding figures. 

\begin{table}[]
    \centering
    \caption{Overview of simulation used throughout the paper. $dx$ and $dt$ are the space and time step. Kinetics refer to different cardiac monodomain models: Aliev-Panfilov (AP) \cite{Aliev:1996}, Bueno-Orovio-Cherry-Fenton (BOCF) in epicardial settings (EPI) \cite{BuenoOrovio:2008}, Fenton-Karma (FK) in MLR-I settings \cite{Fenton:1998}. }
    \begin{tabular}{|c c c c c|}
    \hline
       Fig.  & dx & dt & domain size & kinetics \\
       \hline \hline
       1a, 2a-b  & $1$ & $0.01$  & $400 \times 400$& AP \\
       \hline
       1b, 2c-e   & $0.25$ mm &$0.009$ ms & $1200 \times 1200$ & BOCF (EPI)\\
       \hline
       1c, 6d  & $0.31$ mm & $0.1$ ms &  $375 \times 375 \times 20$ &  BOCF (EPI) \\
       \hline
       3b-c  & $1$ & $0.009$ & $100 \times 100$ & AP \\
       3d-e  & $0.262$ mm &   $0.163$ ms & $450 \times 450$ & FK (MLR-I) \\
       \hline
       6c  & $0.262$ mm &   $0.163$ ms & $450 \times 450 \times 20$ & FK (MLR-I) \\
       \hline
       9b, 12c & $0.31$ mm & $0.1$ ms &  $600 \times 600 \times 20$ & BOCF (EPI)  \\
       \hline
       10  & 1 & 0.007 & $100 \times 100 \times 20$& AP \\
       11  & $0.25$ mm &$0.009$ ms & $1200 \times 1200 \times 20$ & BOCF (EPI) \\
       \hline
       13  & $0.5$ mm& $0.1$ mm & $168\times 208 \times 231$ & BOCF (EPI) \\
       \hline
    \end{tabular}
    \label{tab:sims}
\end{table}

\subsection{Post-processing}
Calculation of the phase was following Kabus \textit{et al.} 2022 \citep{Kabus:2022}. Wave fronts and backs were selected as the isochrones which indicate the transition of going from an excited region E, defined as $V> \Vstar$ to an unexcited region, defined as $V \leq \Vstar$. The distinction between front and back was done via a second threshold $\Rstar$. Both values of $\Vstar$ and $\Rstar$ depend on the chosen dynamics. Phase defects were tracked as the points for which the phase coherence \citep{Kabus:2022} is above a threshold, depending on the chosen model. The positions of heads and tails were calculated as the points where the transition of going from a front and back to a phase defect, respectively. The pivots were calculated as points with extreme curvature. The compound particles were drawn manually on the 2D simulation results. 

\section{Definition of quasi-particles from a state-space viewpoint \label{sec:theory}} 
 
\subsection{Topological structures\label{sec:statespace}}

\begin{table}[]
  \caption{Overview of designated regions in state space and the physical domain, listing the colors used throughout this paper. If space has $n$ dimension and a structure is $d$-dimensional, its co-dimension $c = n-d$. Thus heads, tails and edges are a point in 2D and a curve in 3D.}
    \label{tab:structures}
    \centering
    \begin{tabular}{|c|c|c|c|}
    \hline
    symbol & name &co-dimension & color \\
    \hline
      U & unexcited & 0 & white \\
      E & excited & 0 & yellow \\
      Z & forbidden zone / phase defect & 0 & grey \\
      F & wave front & 1 & cyan \\
      B & wave back & 1 & magenta \\
      h & head & 2 & blue \\
      t & tail & 2 & red \\
      p & pivot & 2 & green \\
      f & filament, tip & 2 & purple \\
      c & core, phase singularity &2 & orange \\
      g & growth particle & 2 & black \\
      s & shrinking particle & 2 & brown \\\hline
    \end{tabular}
\end{table}

\begin{figure*}[t]
    \centering
     \begin{tabular}{c c c }
     (a) & (b) & (c)\\
      \includegraphics[width=0.29\textwidth]{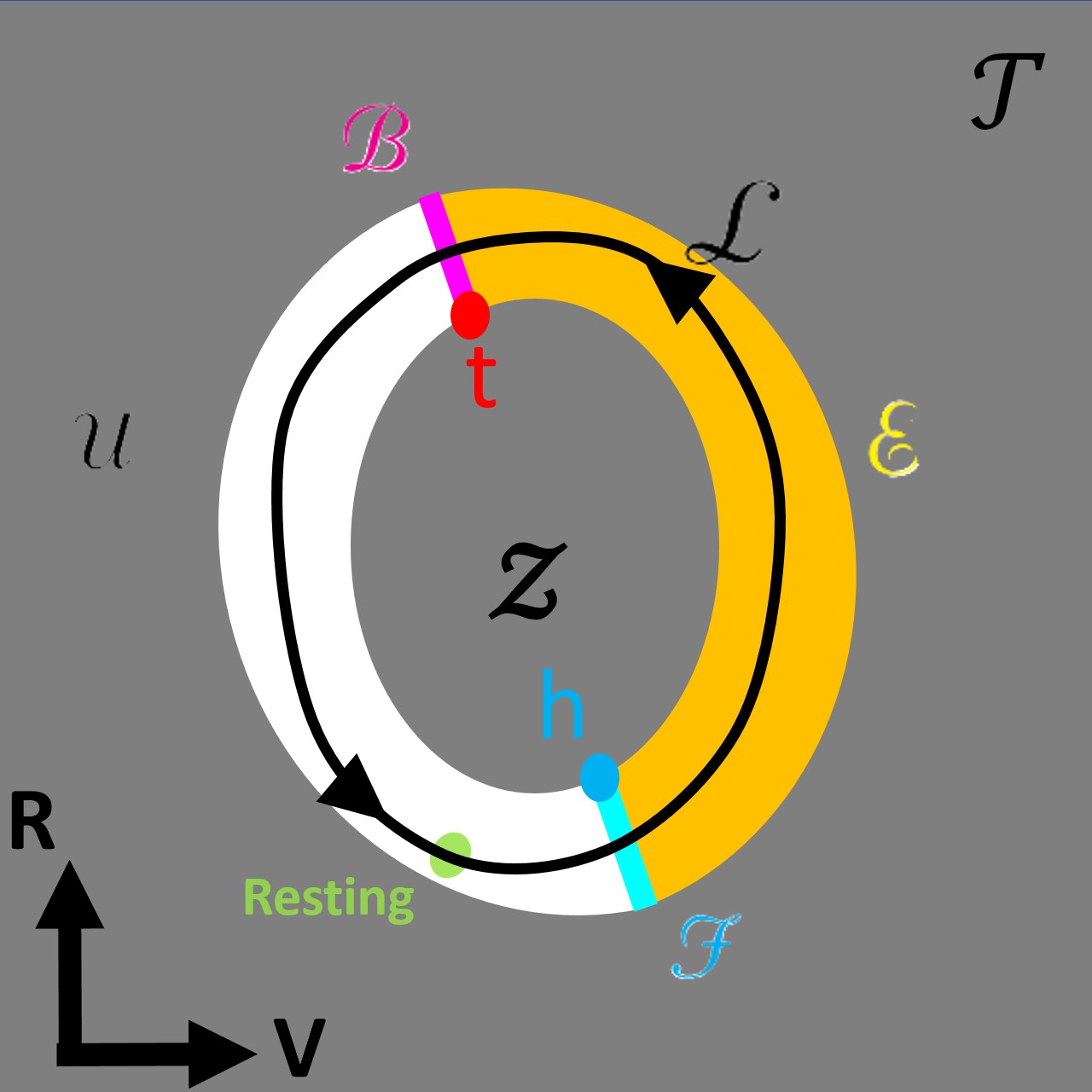}
& \includegraphics[trim={1.cm 0.3cm 2.cm 2.cm},clip, width=0.3\textwidth]{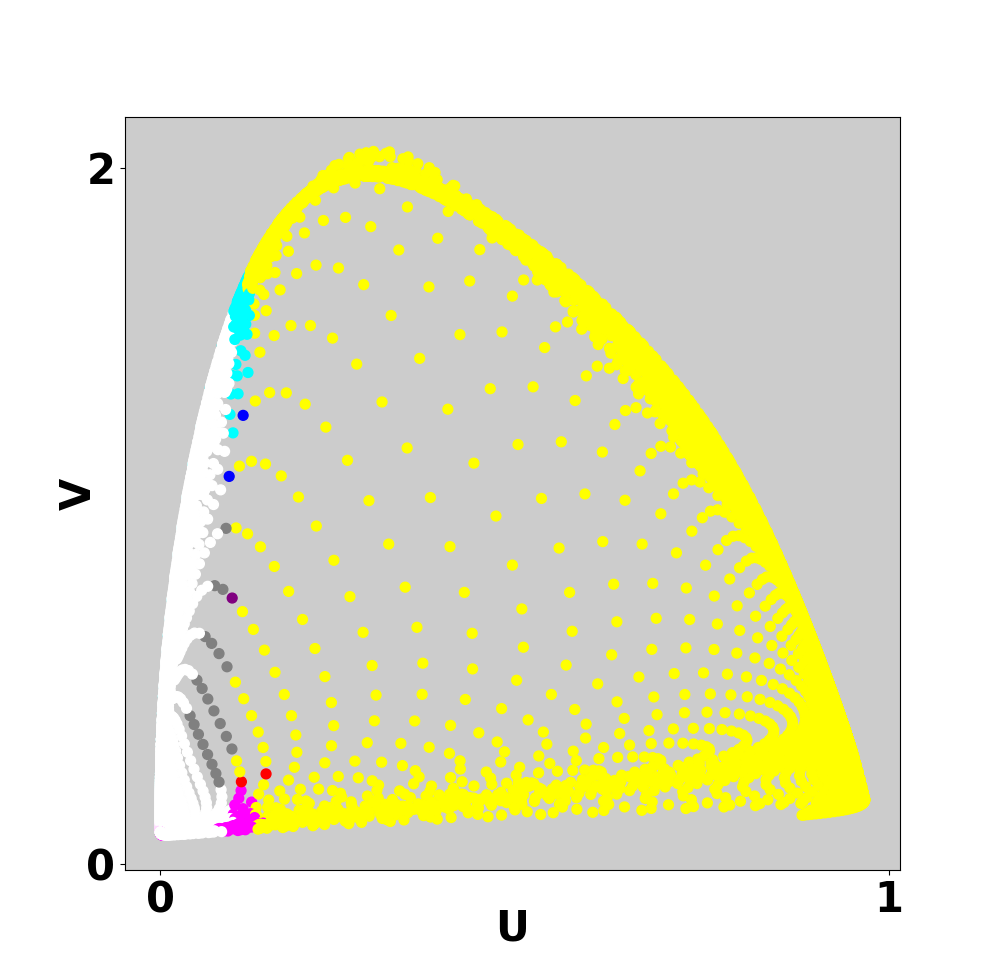} &
     \includegraphics[trim={10.cm 1.cm 13.cm 3.cm},clip, width=0.33\textwidth]{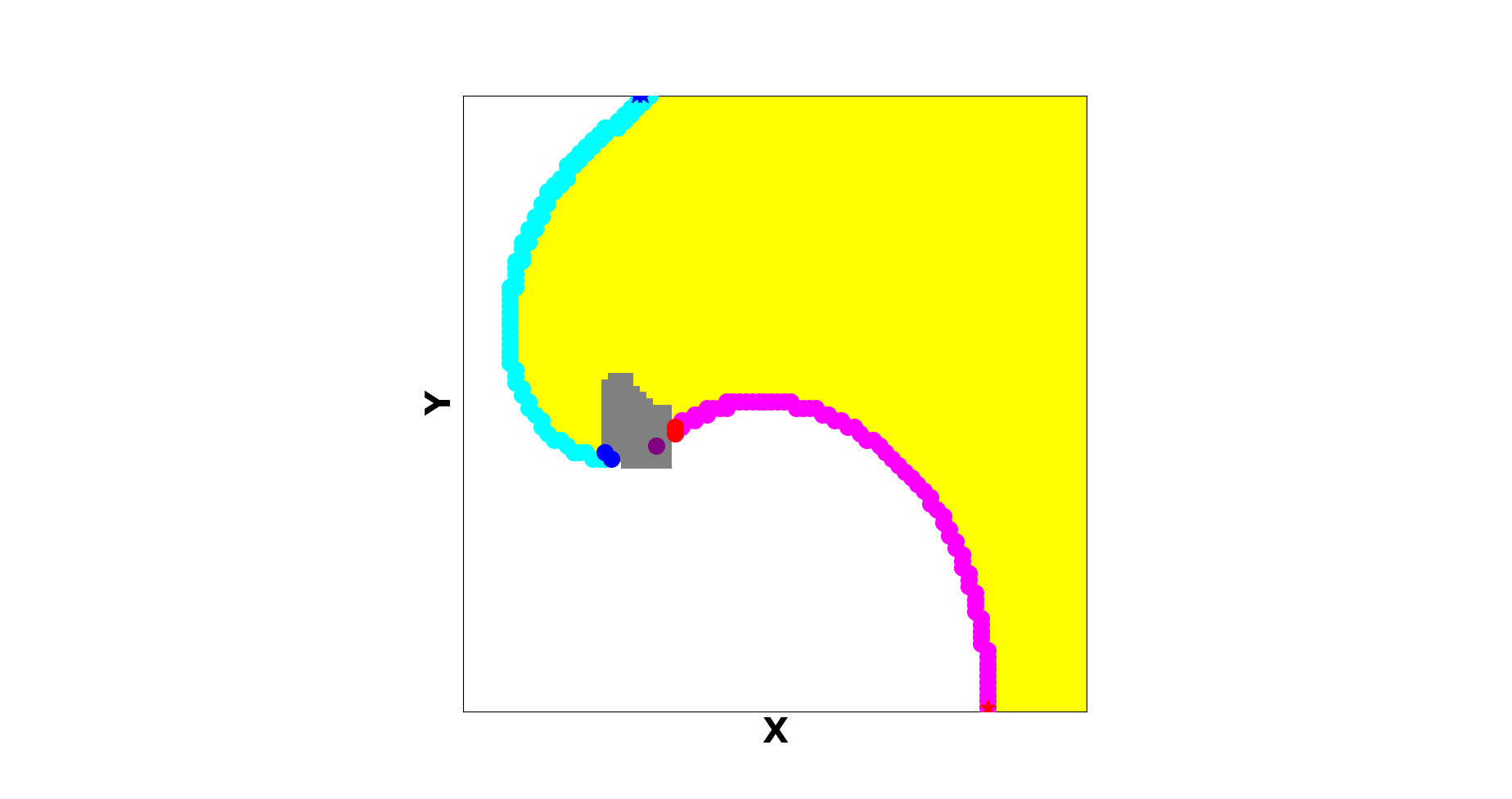}\\
  {\color{white}(d)} &   (d)& (e)  \\
  \includegraphics[width=0.29\textwidth]{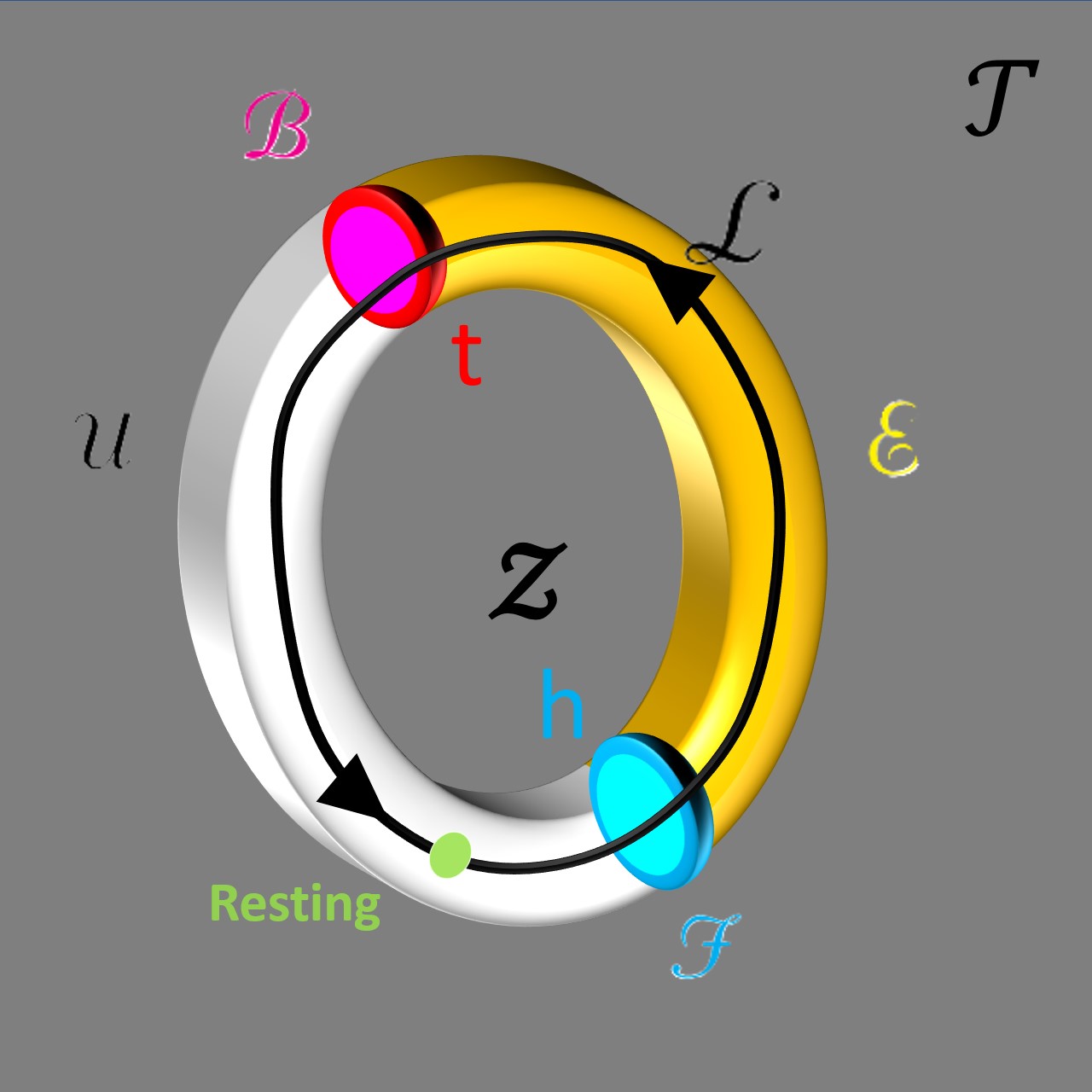}   &  \includegraphics[trim={2.5cm 1.cm 2cm 2.5cm},clip, width=0.29\textwidth]{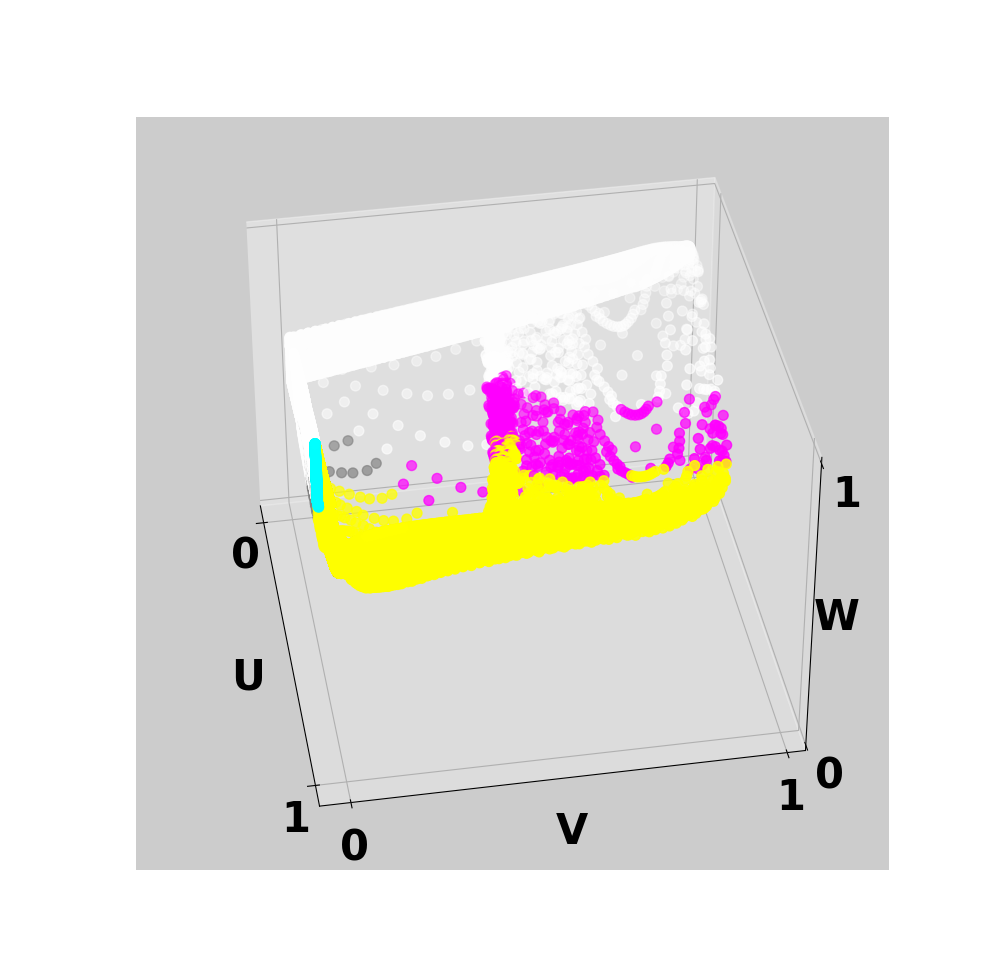}
      &
      \includegraphics[trim={10.cm 1.cm 13.cm 2.5cm},clip, width=0.33\textwidth]{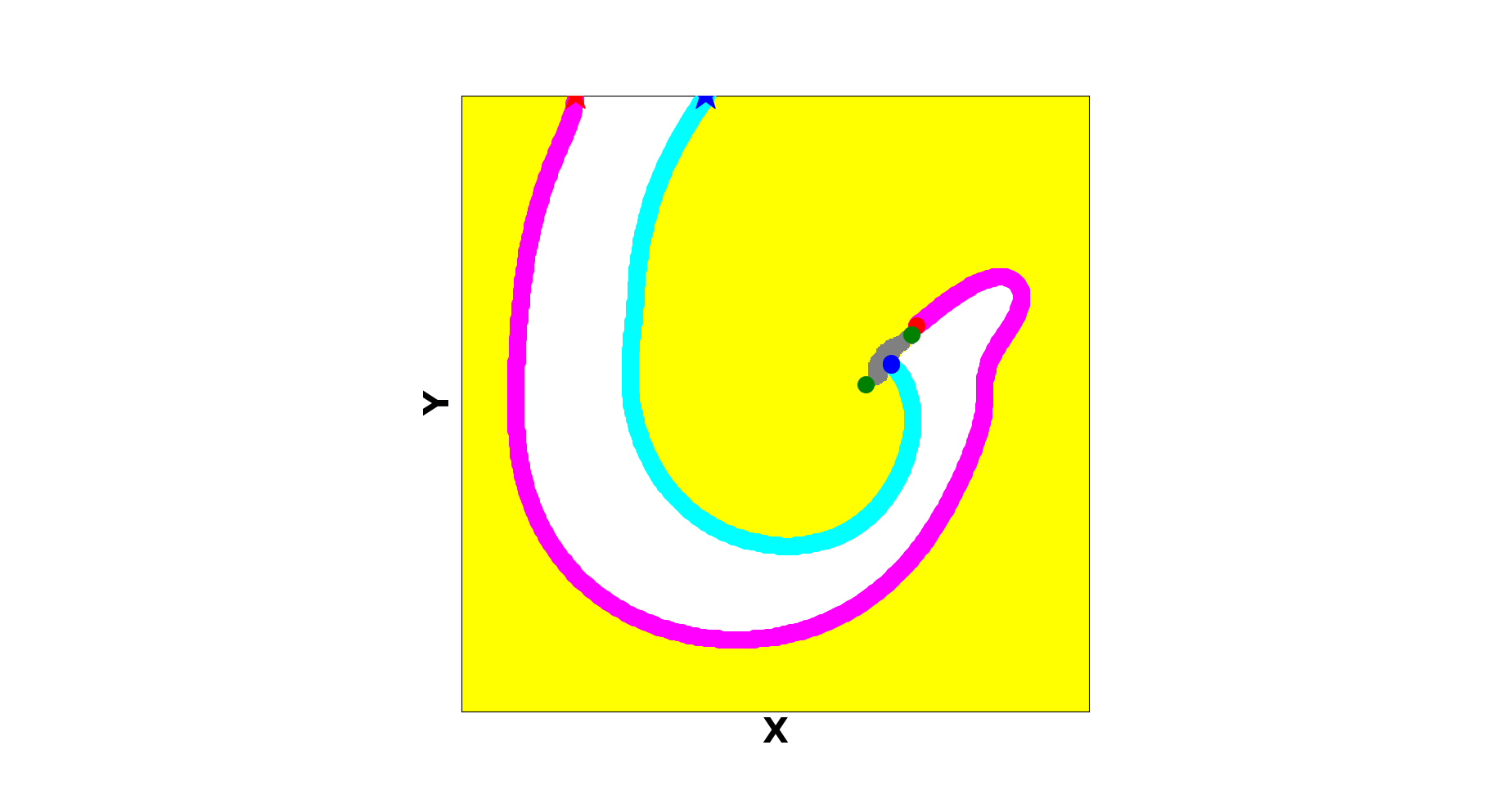}
    \end{tabular}
    
    \caption{
     Origin of phase defects from the geometric structure of state space in excitable media. (a) state space view in 2D and 3D. (b) Point cloud of states observed during the numerical simulation of a spiral wave with circular core (c) with Aliev-Panfilov kinetics \cite{Aliev:1996}, in an isotropic medium. Few points located at the rotor core lie in the forbidden zone of state space in panel b and are brought their by electrotonic effects, i.e. diffusion; they form the phase defect in physical space. (d-e) similar analysis for the three-variable Fenton-Karma (FK) model in an isotropic medium, supporting a linear-core rotor. Colors indicate the presence of fronts, backs, heads, tails and pivots and are used consistently as in Tab. \ref{tab:structures}. 
    }
    \label{fig:torus}
\end{figure*}

\subsubsection{Link between state space and excitation patterns\label{sec:mapping}}
The basic principle of excitability can be understood in the state space of the system (see Fig. \ref{fig:torus}a). Every element of an excitable medium goes through an excitation cycle, in a cardiac or neuronal context called an action potential. Given two linearly independent observables in the system, e.g. $V$ and $R$, one can plot the time traces $V(t)$ and $R(t)$ relative to each other to visualize the excitation cycle $\LL$, see the black line in Fig. \ref{fig:torus}a. 
A common choice for $V$ is a measure for the local excitation such as transmembrane voltage in electrically active media, and for $R$, the amount of recovery from excitation. 

Then, classical phase analysis introduces then the phase as a coordinate along the cycle. Here, we go further and will partition the state space into different domains. In a real medium, the cycle $\LL$ will occupy a ring of finite width in state space, due to factors such as inhomogeneity, statistical variations between excitable elements, diffusion, or electrotonic effects.

This broader version of $\LL$ forms a ring-shaped structure, i.e. an attracting portion of the state space which we denote as $\TT$. 

The states \textit{outside} $\TT$ form less probable zones $\ZZ$, which may physiologically be forbidden.
Note that for a 2D state space, $\ZZ$ consists of both the inner and outer regions shown in grey in Fig. \ref{fig:torus}. However, this theory can be generalized to include more state variables, and there it is seen that $\TT$ becomes a torus, and the `inner' and `outer' forbidden regions are connected. In practice, excitable systems have a fixed finite amplitude, which implies that the dynamics take place in the smaller subspace $\TT$ and most of the state space belongs to the forbidden zone $\ZZ$.  

Since the phase represents a polar angle, $\phiact$ (or $\phiarr$) varies periodically over $[0, 2\pi]$ when an excitation takes place. The reference point $(V,R) = (\Vstar, \Rstar)$ (see eq. \ref{eq:phase}) itself has no well-defined phase, and if in the spatial domain this point is found, it is classically called a phase singularity (PS). Therefore, the precise position of the classical PS in state space or real space depends sensibly on the chosen values $\Vstar$ and $\Rstar$.

Here, we present an alternative formulation that will yield no PSs and is applicable in the cases of a strongly repulsive `forbidden zone' $\ZZ$, occurring under the local system dynamics.

In the following, it will be useful to describe subsets according to their co-dimension: A subset of dimension $d$ in an ambient space of dimension $n$ has co-dimension $c=n-d$.\\

\subsubsection{Structures of co-dimension 0: unexcited, excited and forbidden regions \label{sec:codim0}}
We split the state-space attractor $\mathcal{T}$ in an unexcited region $\UU$ and excited $\EE$ part, see white and yellow regions in Fig. \ref{fig:torus}a. Since they have the same dimension as the state space itself, the co-dimension of these structures is $0$. An easy method to make such splitting is by defining $\UU$ as the part of $\TT$ with $V \leq \Vstar$ and $\EE$ as the portion of $\TT$ where $V \geq \Vstar$.

As a result, we can label excitation patterns in space, depending on whether their state belongs to $\EE, \UU$, or $\ZZ$. With a slight difference in notation, we label the corresponding spatial regions as $E$ (excited), $U$ (unexcited), and $Z$ (forbidden zone). 
In brief, $\vec{x} \in E \Leftrightarrow \vec{u}(\vec{x}) \in \EE$ and similarly for $U$ and $Z$. 

The result of such labeling is shown in Fig. \ref{fig:torus}, for a rigidly rotating spiral pattern (panels b-c) and a linear-core spiral wave (panels d-e).
Near the rotation center of spiral waves, also termed rotors, in the cardiac context, there exists a region with a state in the physiologically forbidden zone $\ZZ$. Since the phase is ill-defined away from $\TT$, we call this `gray' zone in the spatial domain a phase defect (PD). By the reasoning above, we have traced back the recently introduced concepts of phase discontinuity  \cite{Tomii:2021} or phase defect \cite{Arno:2021,Kabus:2022} to the shape of the dynamical attractor in state space. \\

\subsubsection{Structures of co-dimension 1: fronts, backs and PD boundaries \label{sec:codim1}}
The excited, unexcited, and forbidden regions in physical space have the same co-dimension as in state space, and so, this will also be true for their intersections. The oscillatory or excitable nature of the system gives rise to a torus structure of the attractor, such that the intersection of $\EE$ and $\UU$ consists of two disjunct sets of co-dimension 1 in state space, see Fig. \ref{fig:torus}a. Points in these sets can be discriminated based on the direction of the dynamics in the cycle $\LL$: The region where points excite is defined as $\FF$, and $\BB$ is defined as the points where points go from excited to unexcited. 

These letters are chosen since the corresponding regions map to the wave front and wave back in real space, respectively, see cyan and magenta lines in Fig. \ref{fig:torus}a. If a recovery variable can be observed, its value can also be used to discriminate the front region from the back region either in state space or in physical space: The wave front and wave back are the isosurfaces $V=\Vstar$ and distinguished by a second threshold $\Rstar$. If $V= \Vstar$ and $R\geq \Rstar$, this defines the waves front, if $V= \Vstar$ and $R< \Rstar$, this defines the waves back.

From the state space construction, we find that the wave front and wave back cannot intersect, as was assumed in classical theory. Instead, they both end on the phase defect boundary (PDB) $\partial Z = Z \bigcap T$. This border is divided into an excited PD boundary ($\partial Z_E = Z \bigcap E$) and an unexcited part ($\partial Z_U = Z \bigcap U$). \\

\subsubsection{Structures of co-dimension 2: heads, tails and pivots \label{sec:codim2}} 

Intersections of the $c=1$ structures will have a co-dimension $c=2$ and thus will be points on excitable surfaces, and curves in three-dimensional excitable media. 

From Fig. \ref{fig:torus}a, we can see there exist points that touch all three subdomains $\ZZ$, $\UU$, and $\EE$. We call these points lying on the outer rim of $\FF$ `heads' and points lying on the outer rim of $\BB$ `tails', denoted $\hh$ and $\ttt$, respectively:
\begin{align}
\hh &= \FF \cap \ZZ, &
\ttt &= \BB \cap \ZZ. \label{defheadtail_statespace}
\end{align}

The two-dimensional patterns in panels c. and e. of Fig. \ref{fig:torus}, show a head in blue and tail in red at the core of a spiral wave, where the wave front and wave back meet the PD. In previous work \cite{Arno:2023feynman}, we introduced heads and tails as intersections of respectively fronts and backs with a phase defect, in real space, which is consistent with eq. \eqref{defheadtail_statespace}. 

By construction, the head also lies at the intersection between the wave front and unexcited and excited PDB, and similarly for the tail.

Until here, our analysis in terms of heads and tails was valid for any excitable medium, regardless of the shape of the spiral wave core. E.g. the distinct cases of rigidly rotating and meandering cores were included in Fig. \ref{fig:torus}. 

To introduce the last building block, we will start from an excitable system that forms line defects of phase, i.e. where the PD contracts to a flattened region of abrupt phase variation, called a phase discontinuity \cite{Tomii:2021}. Our motivation comes from cardiac electrophysiology, where this phenomenon is known for a long time in simulations and experiments \cite{Fast:1990,Efimov:1999}, including our own \cite{Arno:2021, Kabus:2022}. Such linear cores were experimentally observed in myocardial tissue \cite{Efimov:1999} and recently in optical voltage mapping of human atrial myocytes \cite{Kabus:2022}. Moreover, they appear in almost all ionic models of cardiac electrophysiology \cite{Fenton:1998, BuenoOrovio:2008, Clayton:2011}. 

We attribute the spontaneous flattening of PDs to the dynamics in state space, see Fig. \ref{fig:torus}a. The middle of the forbidden zone is, in general, repulsive, pushing all excitable elements toward the attractor $\TT$. We hypothesize that this repulsion property leads to a minimization of the volume occupied by the PD in physical space, while the circumference of the PD must remain finite to allow re-excitation along the PD boundary. In systems that have a long excitation duration $\tau$, the boundary of a sustained PD needs to have a length equal to $v \tau$, with $v$ the propagation speed of a front in the medium, leading to the so-called linear core regime (see also Fig. \ref{fig:headcases}d). One method to reduce the phase defection region $Z$ of co-dimension $c=0$ to a line defect of co-dimension $c=1$ is to first calculate the PD density using an order parameter \cite{Tomii:2021,Kabus:2022} and then fitting a spline to the elongated defects to obtain phase defect lines. 

In this case, linear rotor cores have two end points, around which the wave front will turn. They are known as `pivots' in cardiac electrophysiology \cite{seitz_af_2017}. We take over this name for the end points of a flattened PD, even if the wave has not made a pivoting turn there yet. 

Pivots, heads, and tails are the fundamental building blocks of the theory exposed here. An introduction with less mathematical detail was given in a separate paper \cite{Arno:2023feynman}.

\subsubsection{Structures of co-dimension 3: twistons\label{sec:codim3}}

Heads and tails cannot intersect in our framework, as they are separated by the physiologically forbidden region $\mathcal{Z}$ in state space, and hence by a phase defect in real space. However, in 3D, a head can intersect with a pivot curve, for which we suggest this intersection occurs classically as a twistons coined by Fenton and Karma \cite{Fenton:1998}, see Sec. \ref{sec:twistons}. 

\subsection{Topological constraints}
\subsubsection{Half-integer topological charge of heads, tails, and pivots \label{sec:topoheadtail}}

The classical topological charge in an excitable or oscillatory medium with activation phase $\phiact$ is given by: 
\begin{align}
    Q = \frac{1}{2\pi} \oint_{\mathcal{C}} \vec\nabla \phiact \cdot \vec{d\ell} \label{classicalcharge}
\end{align}
where $\mathcal{C}$ is a closed contour. Hence, this quantity evaluates to an integer number. Note that for a phase defect that does not touch the medium boundary, one can compute this charge as well, using the PD boundary as the contour $\mathcal{C}$. As the phase is well defined on the PD boundary, we also find an integer topological charge $Q$.

\begin{figure*}[ht]
     \centering
    \begin{tabular}{c}
    (a) \\
 head, $Q=+\frac{1}{2}$ \\
\frame{\includegraphics[width=0.1\textwidth]{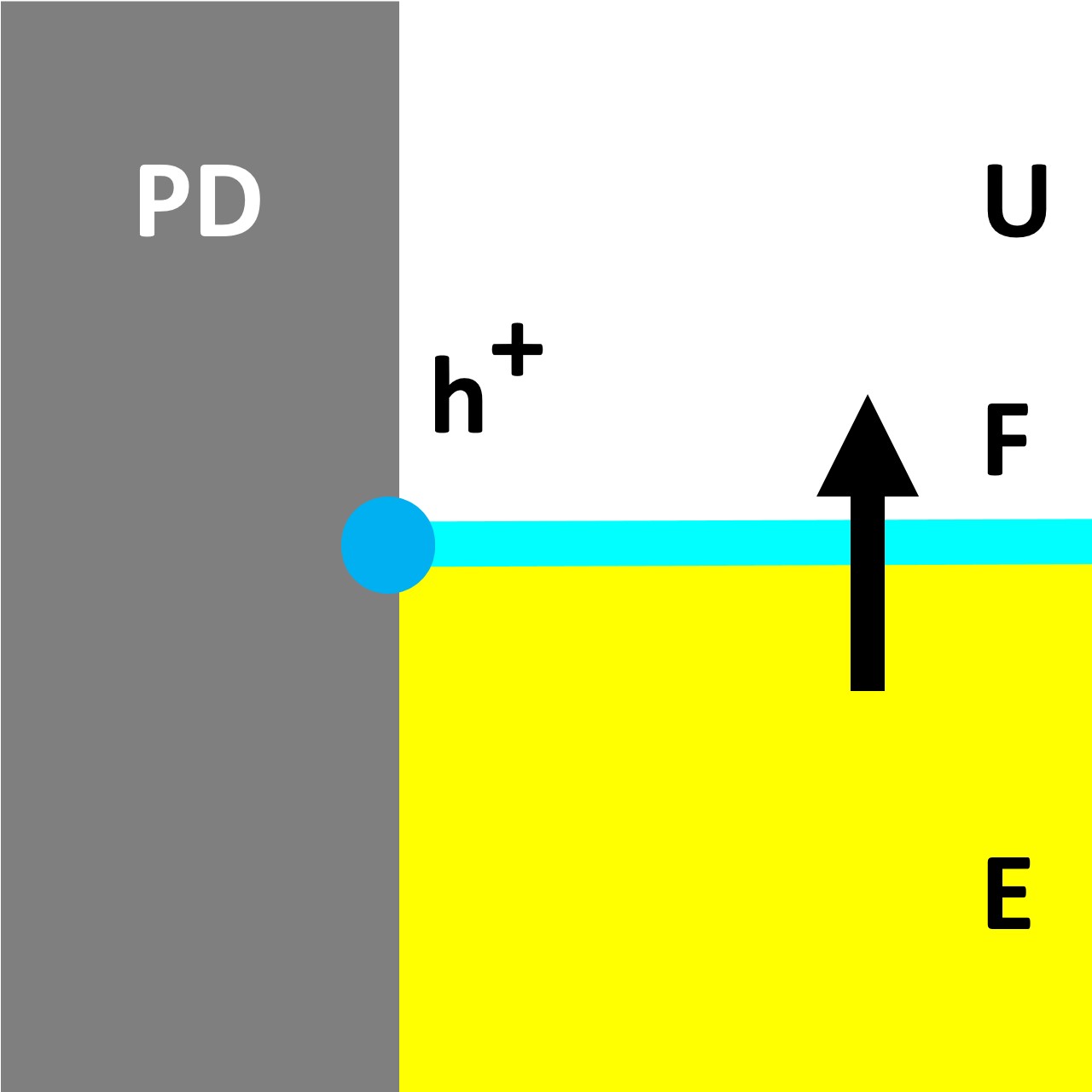}}\\
{\large{\textbf{}} }\\
tail, $Q=-\frac{1}{2}$\\
\frame{\includegraphics[width=0.1\textwidth]{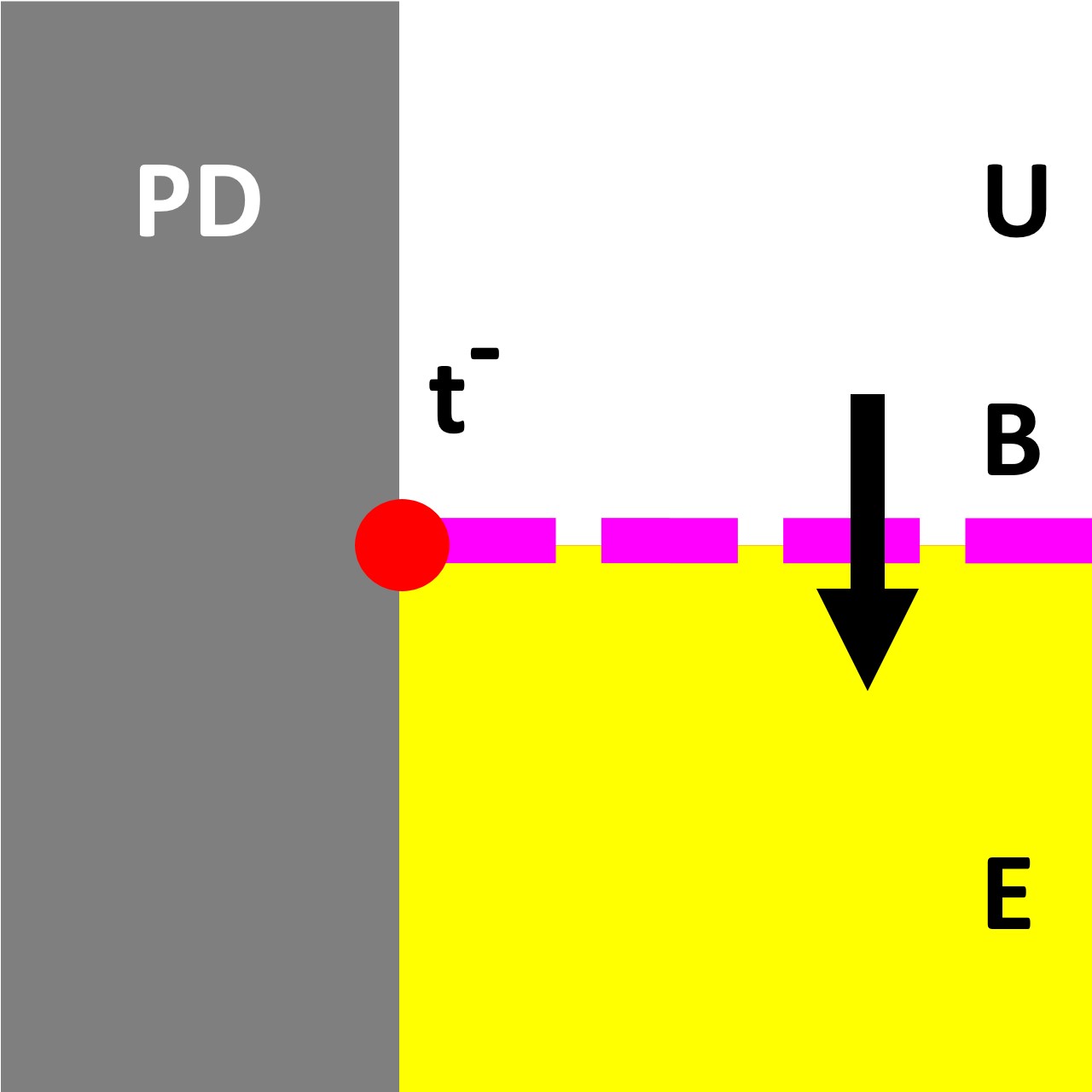}}\\
{\large{\textbf{}} }\\
tail, $Q=+\frac{1}{2}$\\ 
\frame{\includegraphics[width=0.1\textwidth]{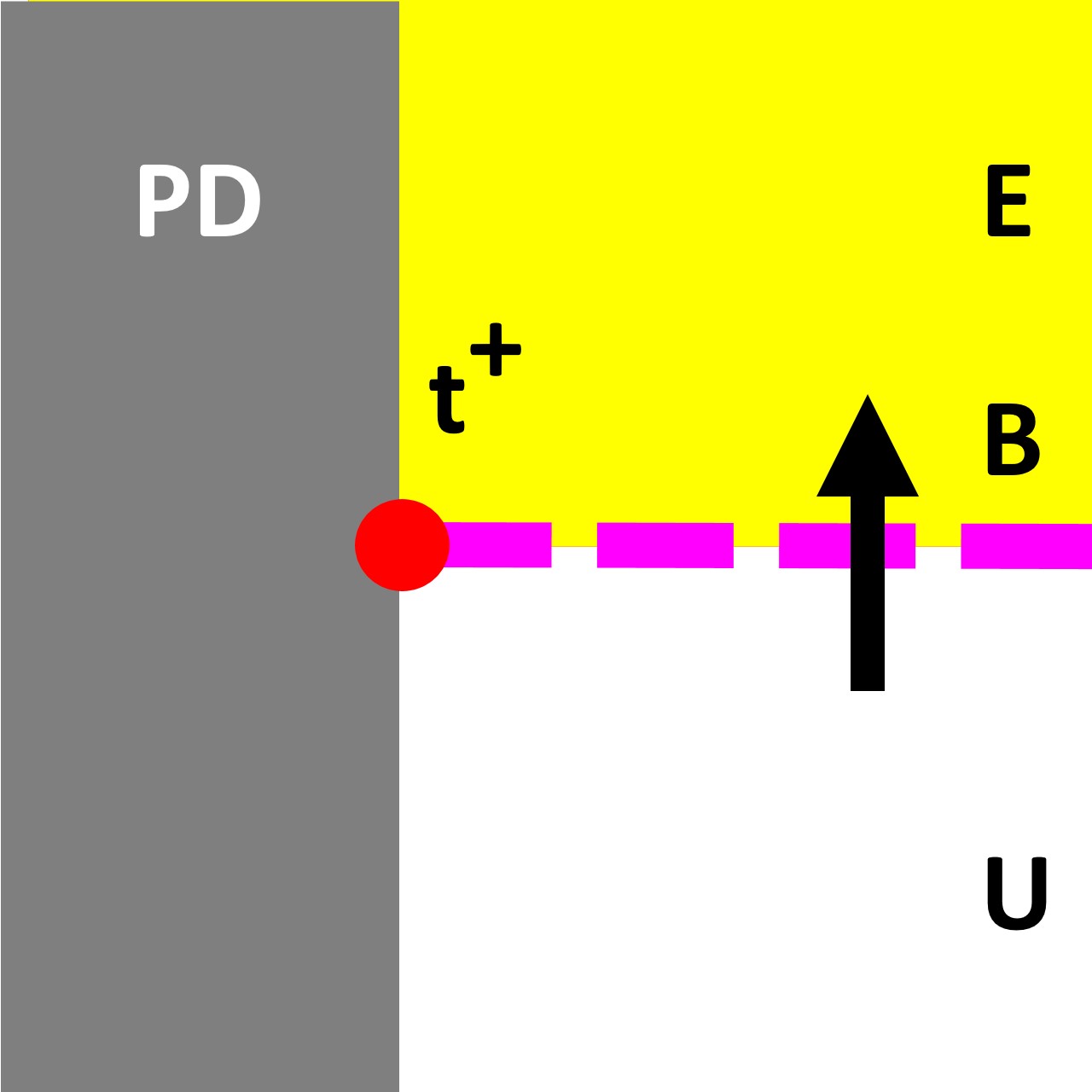}}\\
{\large{\textbf{}} }\\
head, $Q=-\frac{1}{2}$ \\
\frame{\includegraphics[width=0.1\textwidth]{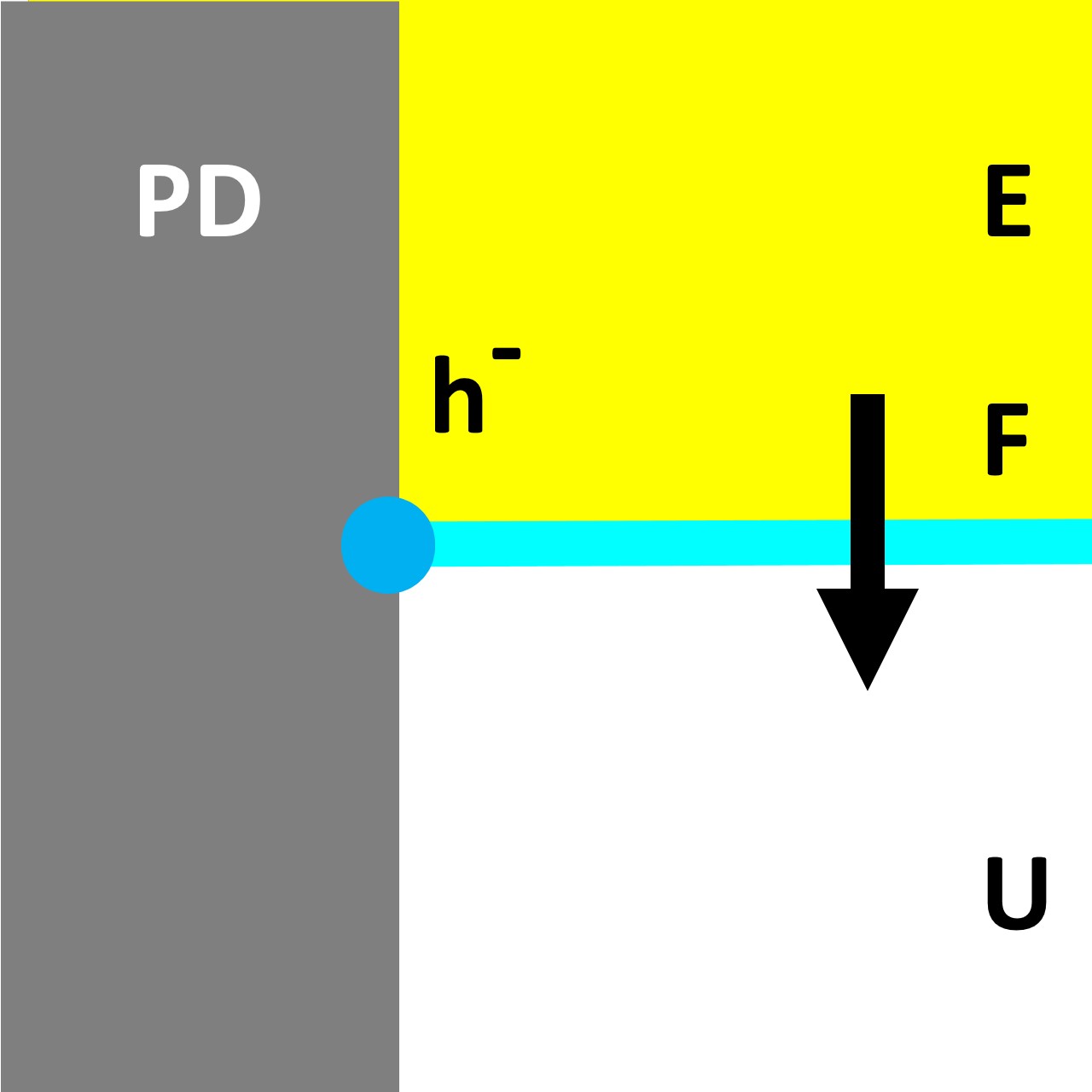}}
{\large{\textbf{}} }\\
\end{tabular}
\begin{tabular}{c c c}
 \multicolumn{3}{c}{(b) }\\

\multicolumn{3}{c}{wave front collision}\\
\frame{\includegraphics[width=0.1\textwidth]{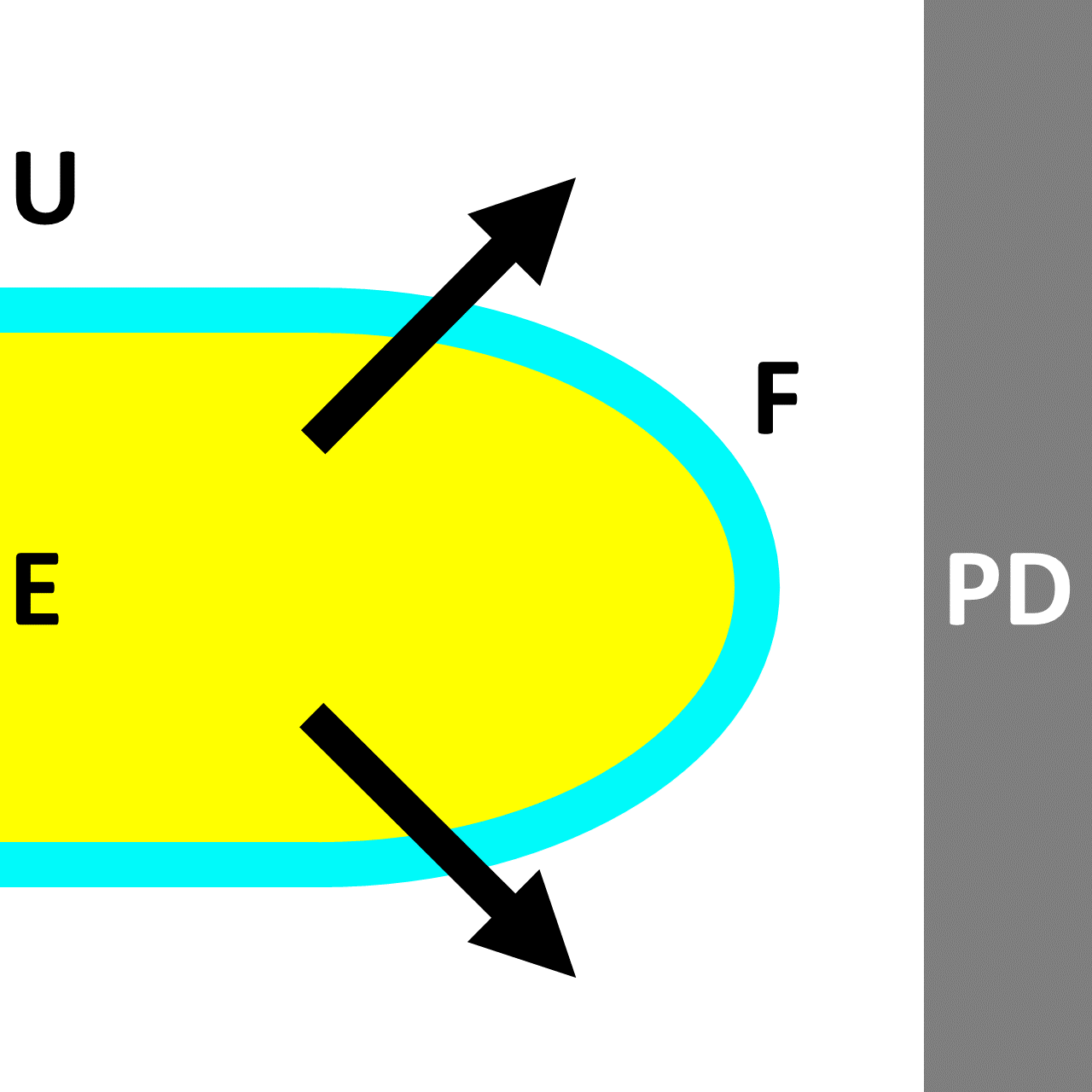}}&
{\Large $\rightarrow$} & 
\frame{\includegraphics[width=0.1\textwidth]{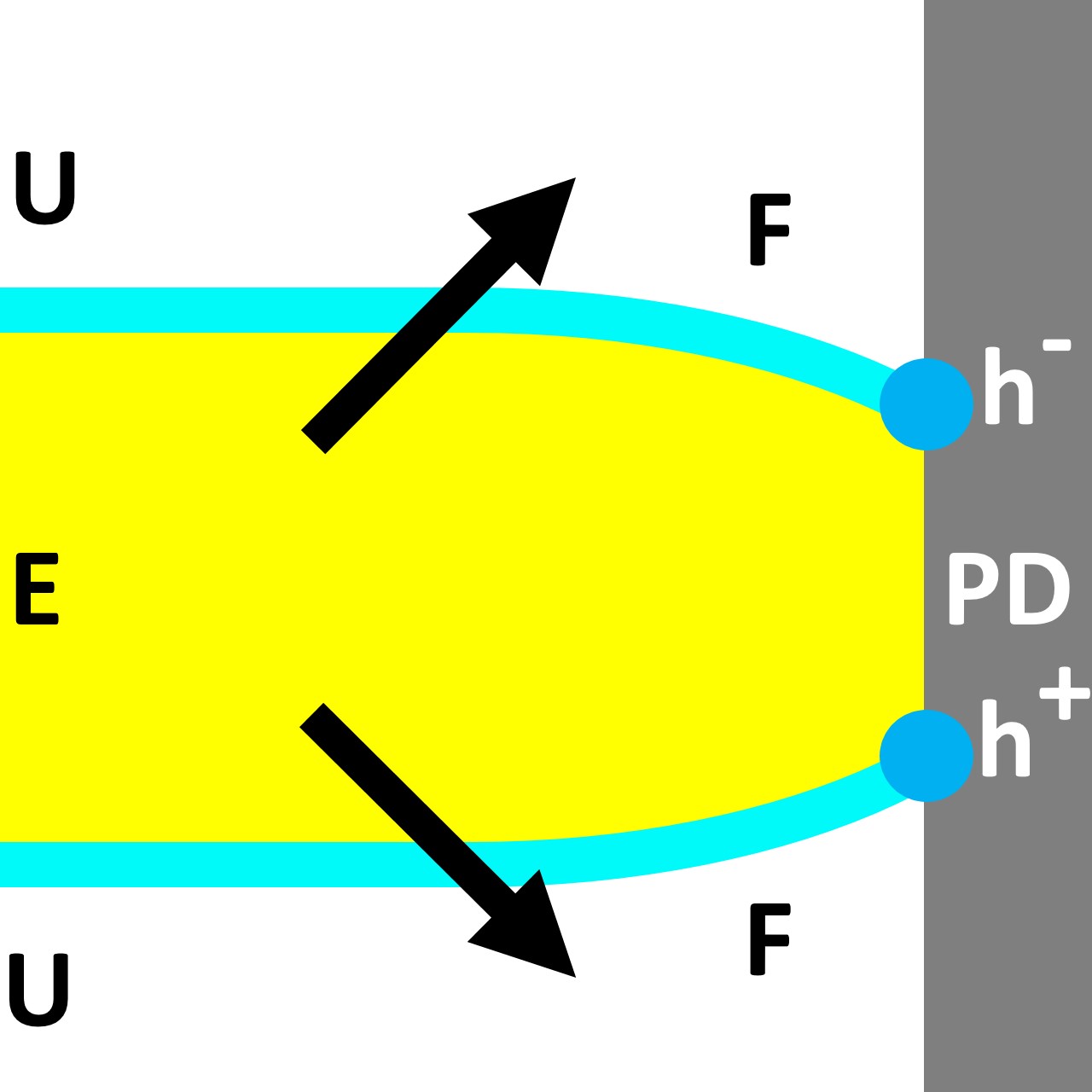}} \\
 \multicolumn{3}{c}{\large{\textbf{}} }\\

\multicolumn{3}{c}{wave front detachment}\\
\frame{\includegraphics[width=0.1\textwidth]{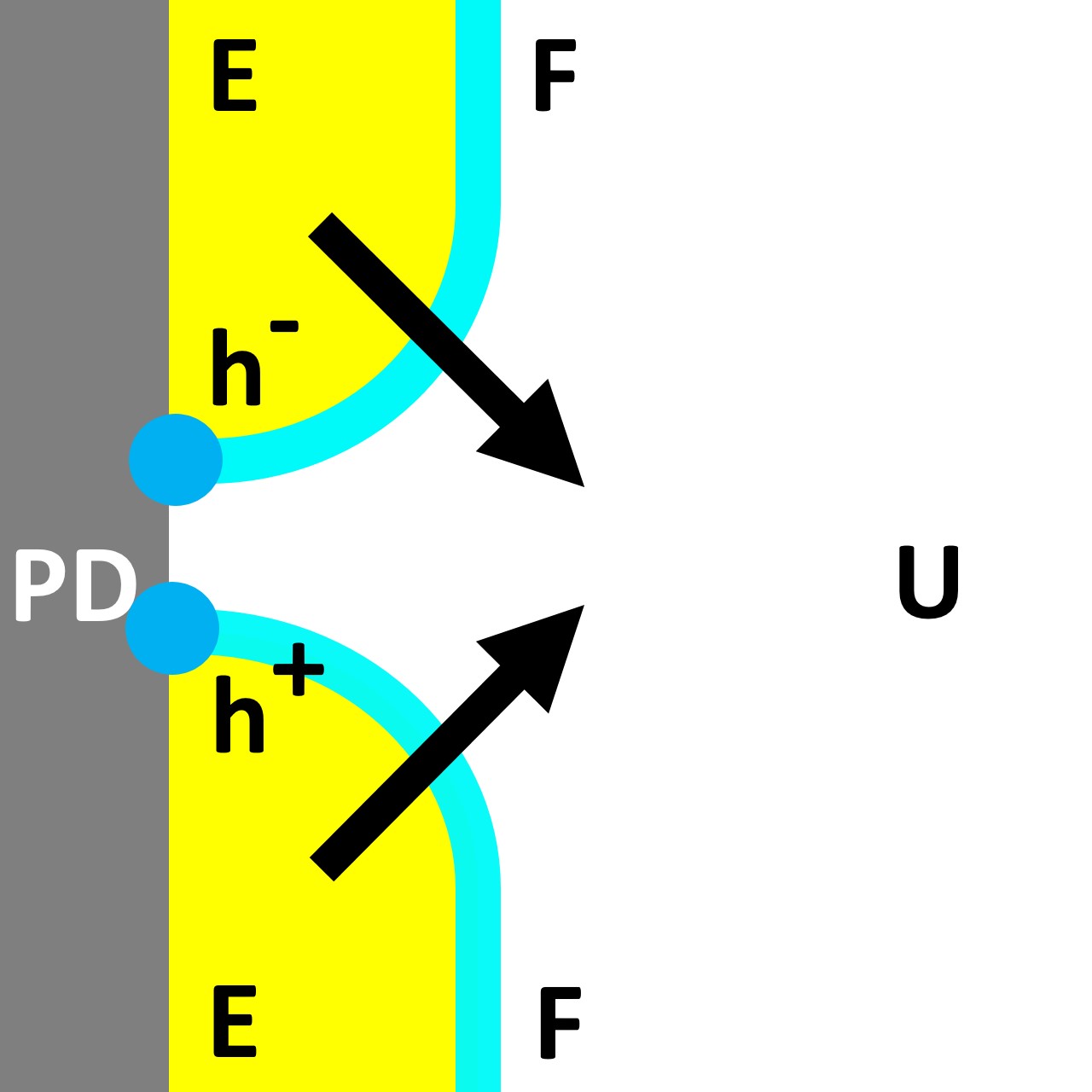}}&
{\Large $\rightarrow$} & 
\frame{\includegraphics[width=0.1\textwidth]{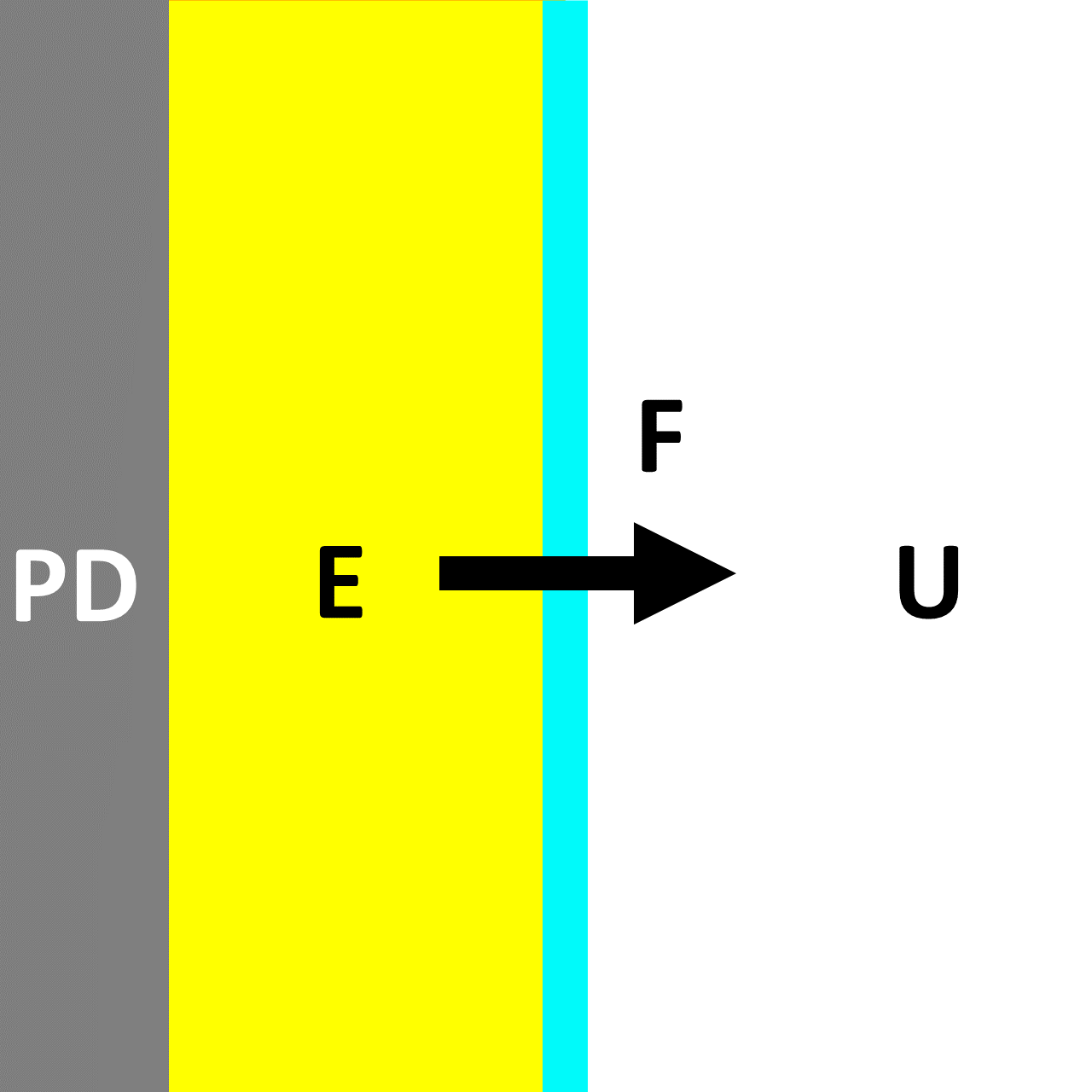}}\\
 \multicolumn{3}{c}{\large{\textbf{}} }\\

\multicolumn{3}{c}{wave front birth}\\
\frame{\includegraphics[width=0.1\textwidth]{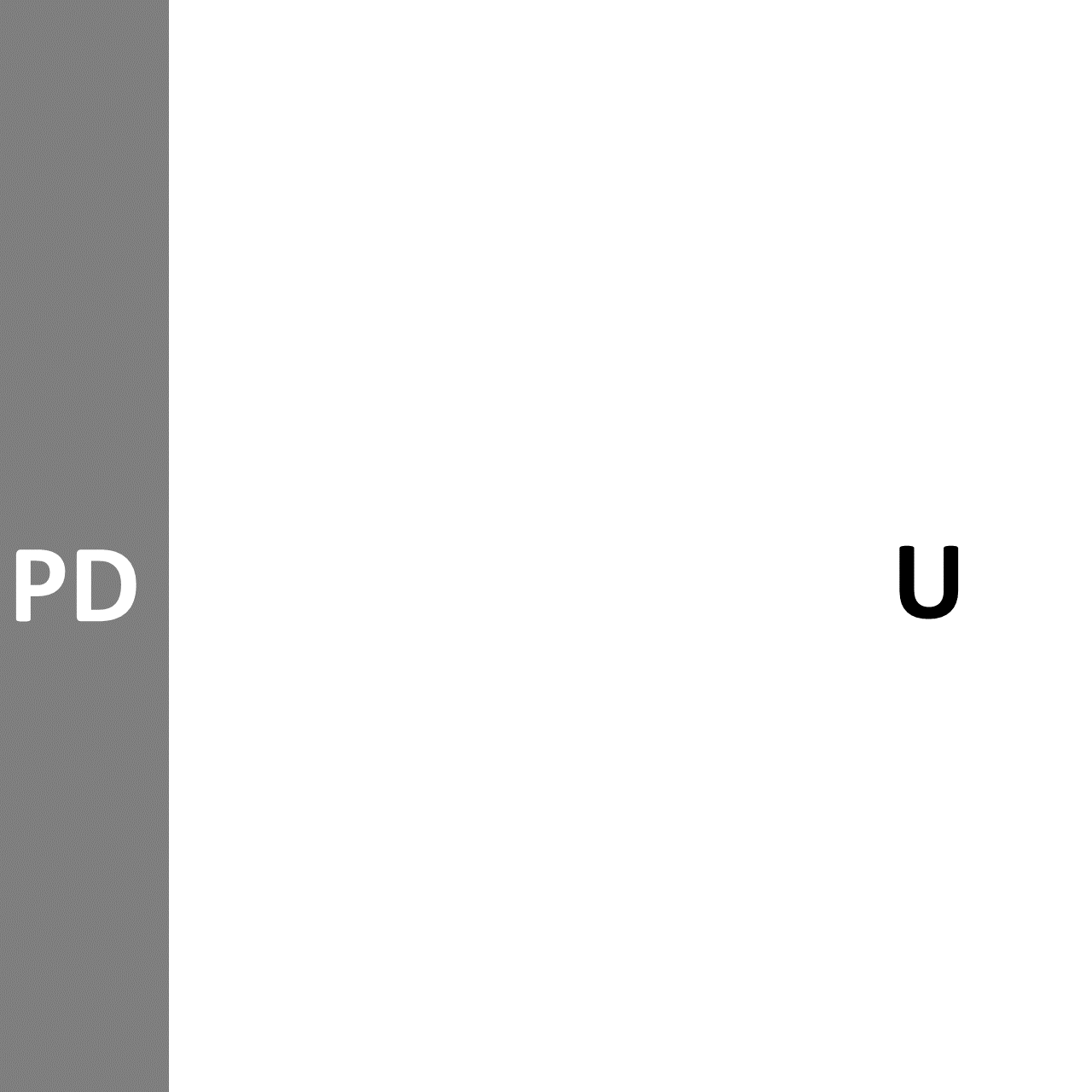}}&
{\Large $\rightarrow$} & 
\frame{\includegraphics[width=0.1\textwidth]{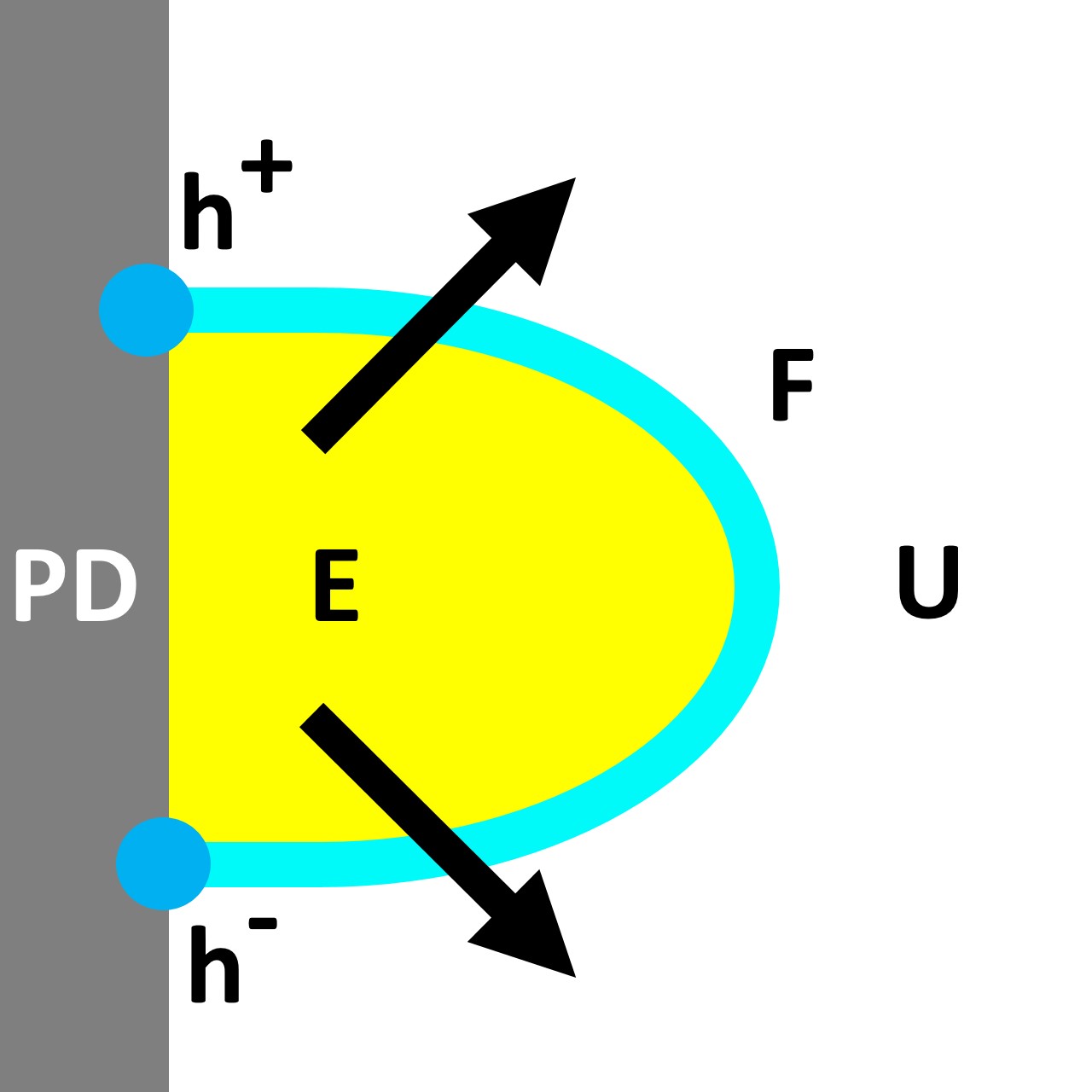}} \\
 \multicolumn{3}{c}{\large{\textbf{}} }\\

\multicolumn{3}{c}{wave front death}\\
\frame{\includegraphics[width=0.1\textwidth]{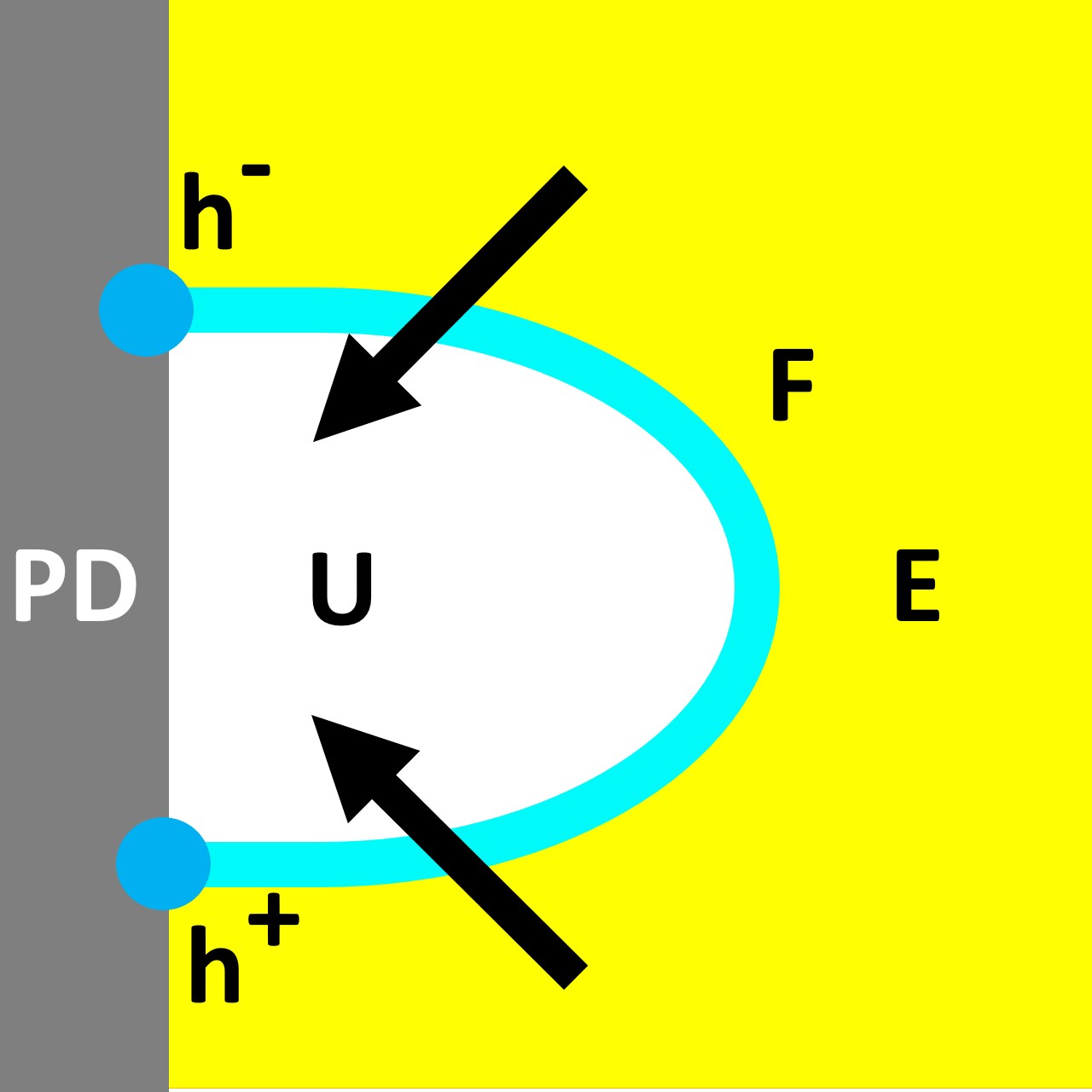}}&
{\Large $\rightarrow$} & 
\frame{\includegraphics[width=0.1\textwidth]{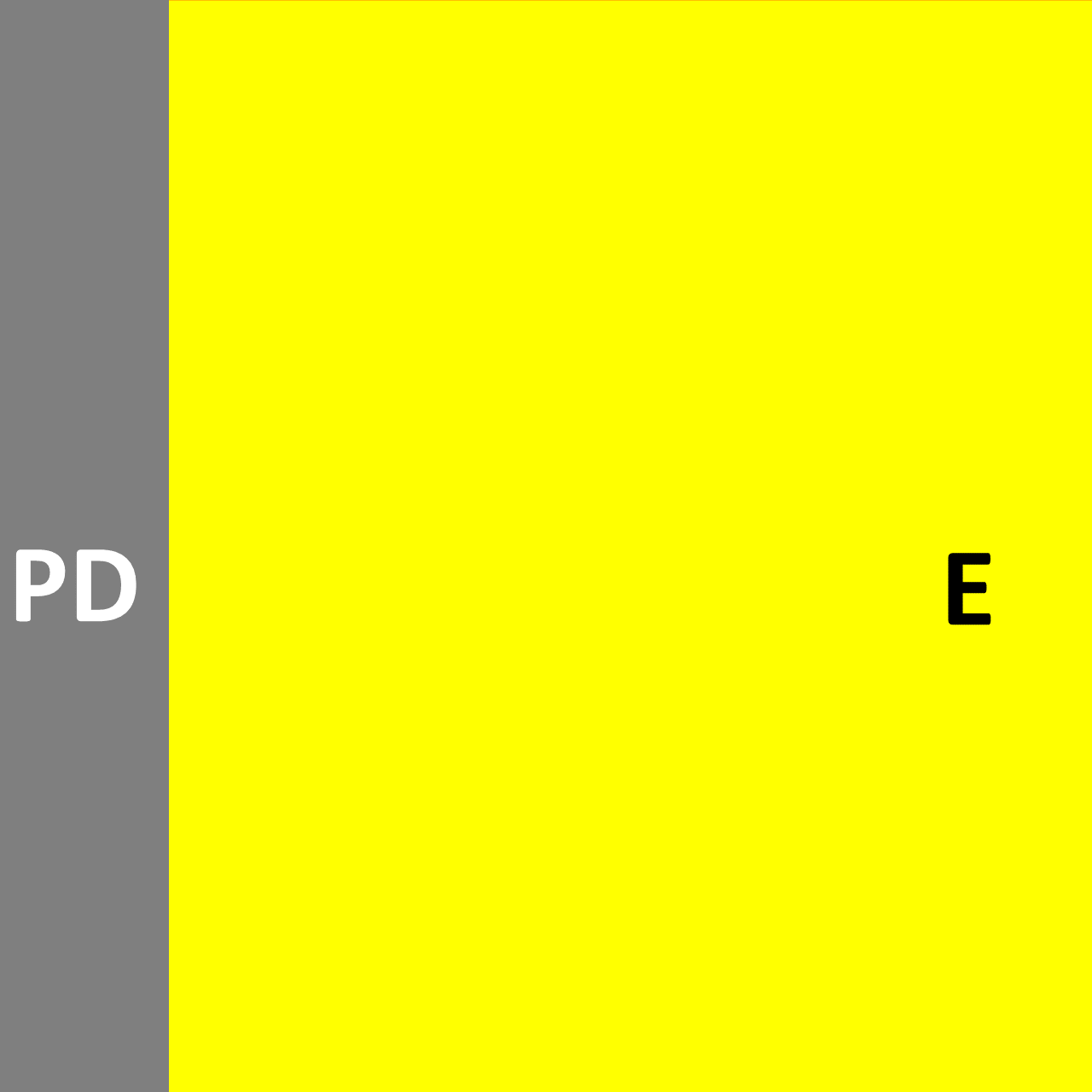}}
\end{tabular}
\begin{tabular}{c c c}
 \multicolumn{3}{c}{(c) }\\

\multicolumn{3}{c}{wave back collision }\\
\frame{\includegraphics[width=0.1\textwidth]{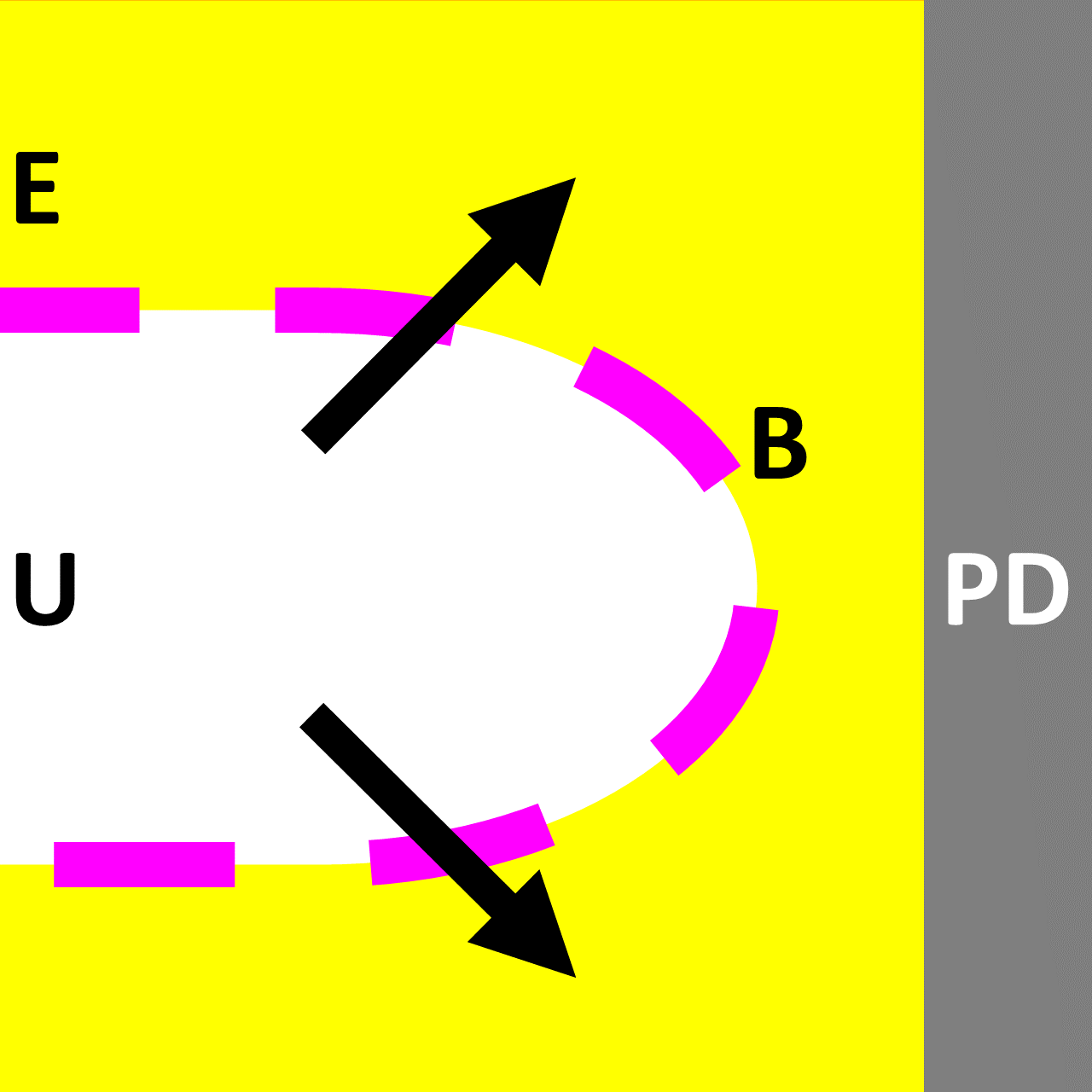}}&
{\Large $\rightarrow$} & 

\frame{\includegraphics[width=0.1\textwidth]{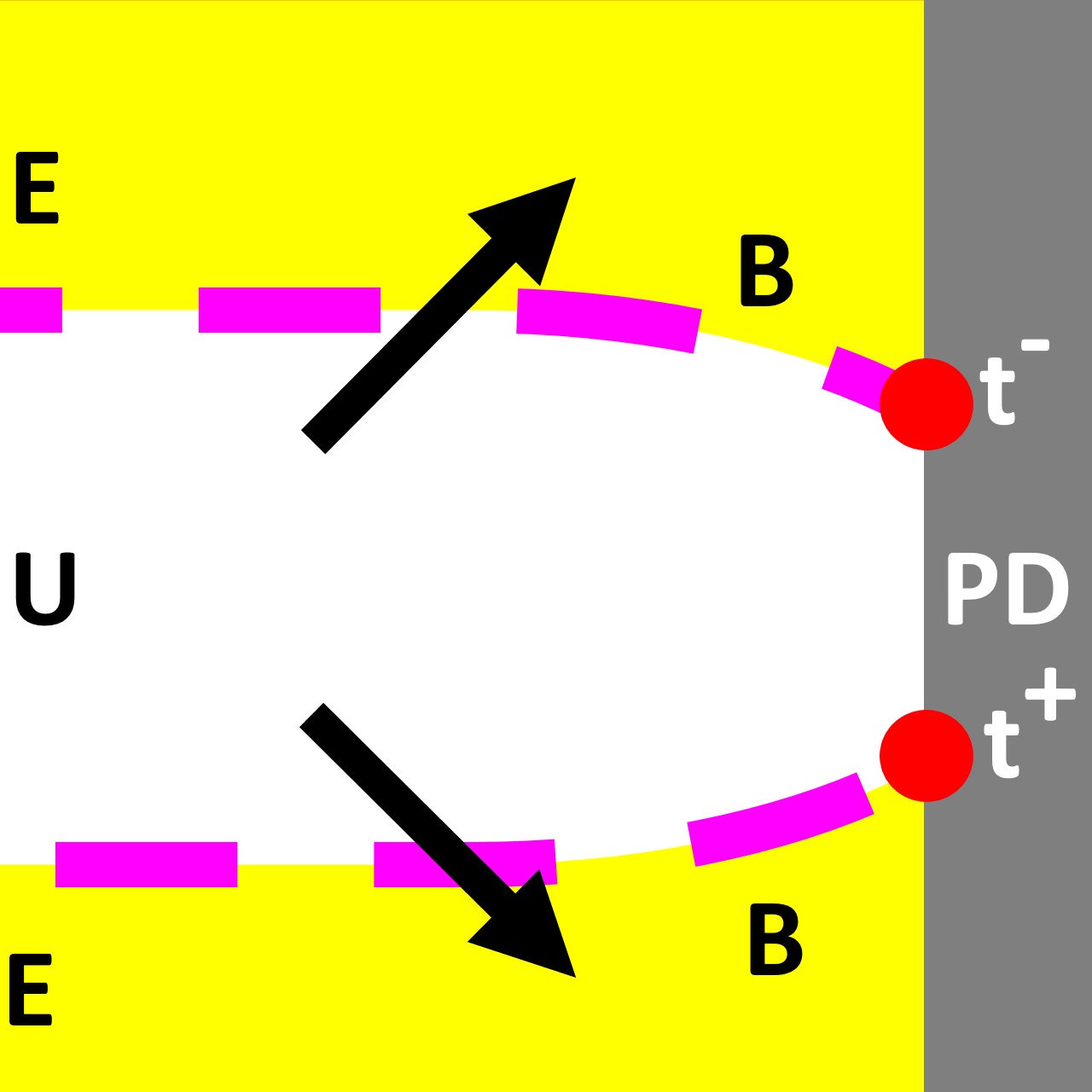}} \\
 \multicolumn{3}{c}{\large{\textbf{}} }\\

\multicolumn{3}{c}{wave back detachment}\\
\frame{\includegraphics[width=0.1\textwidth]{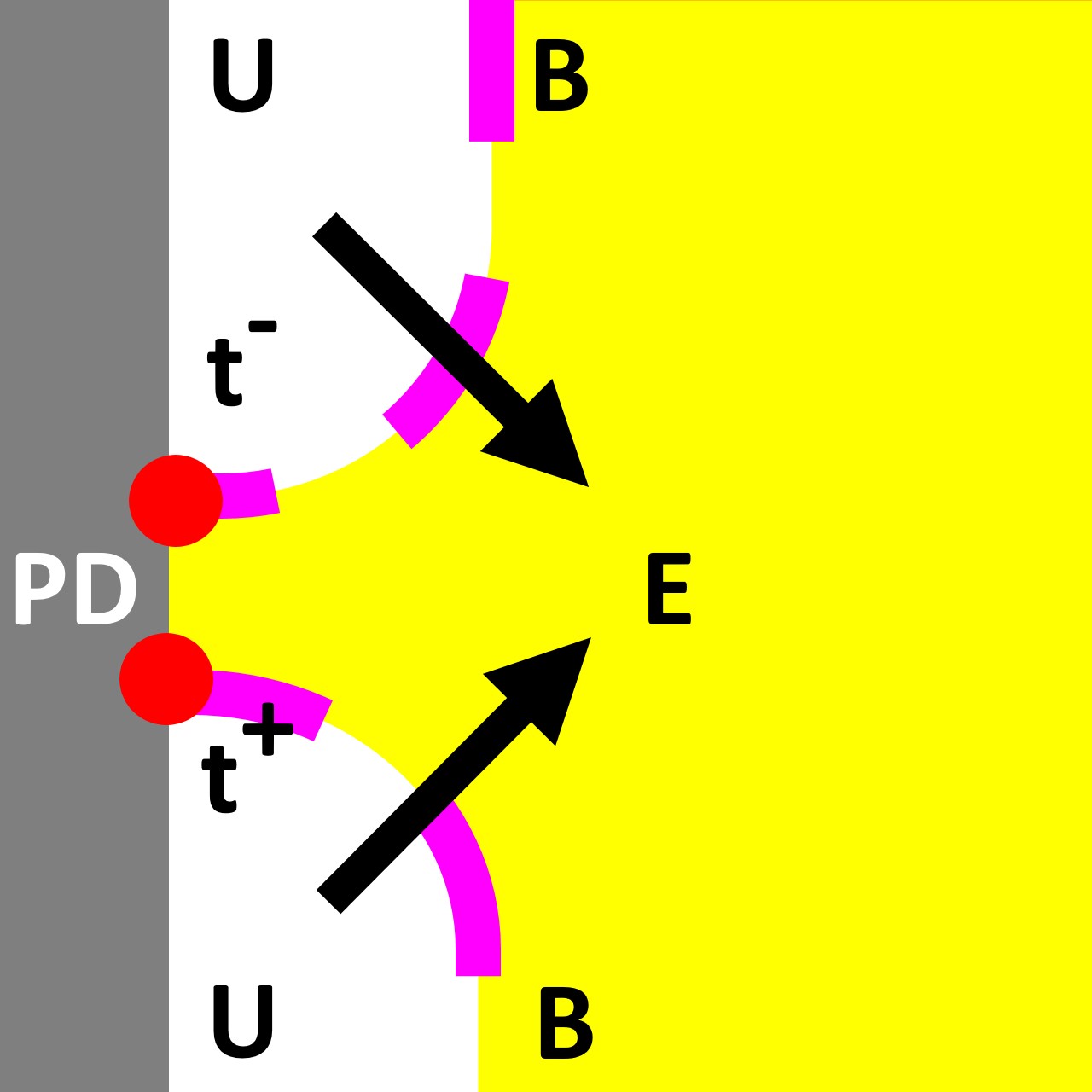}}&
{\Large $\rightarrow$} & 

\frame{\includegraphics[width=0.1\textwidth]{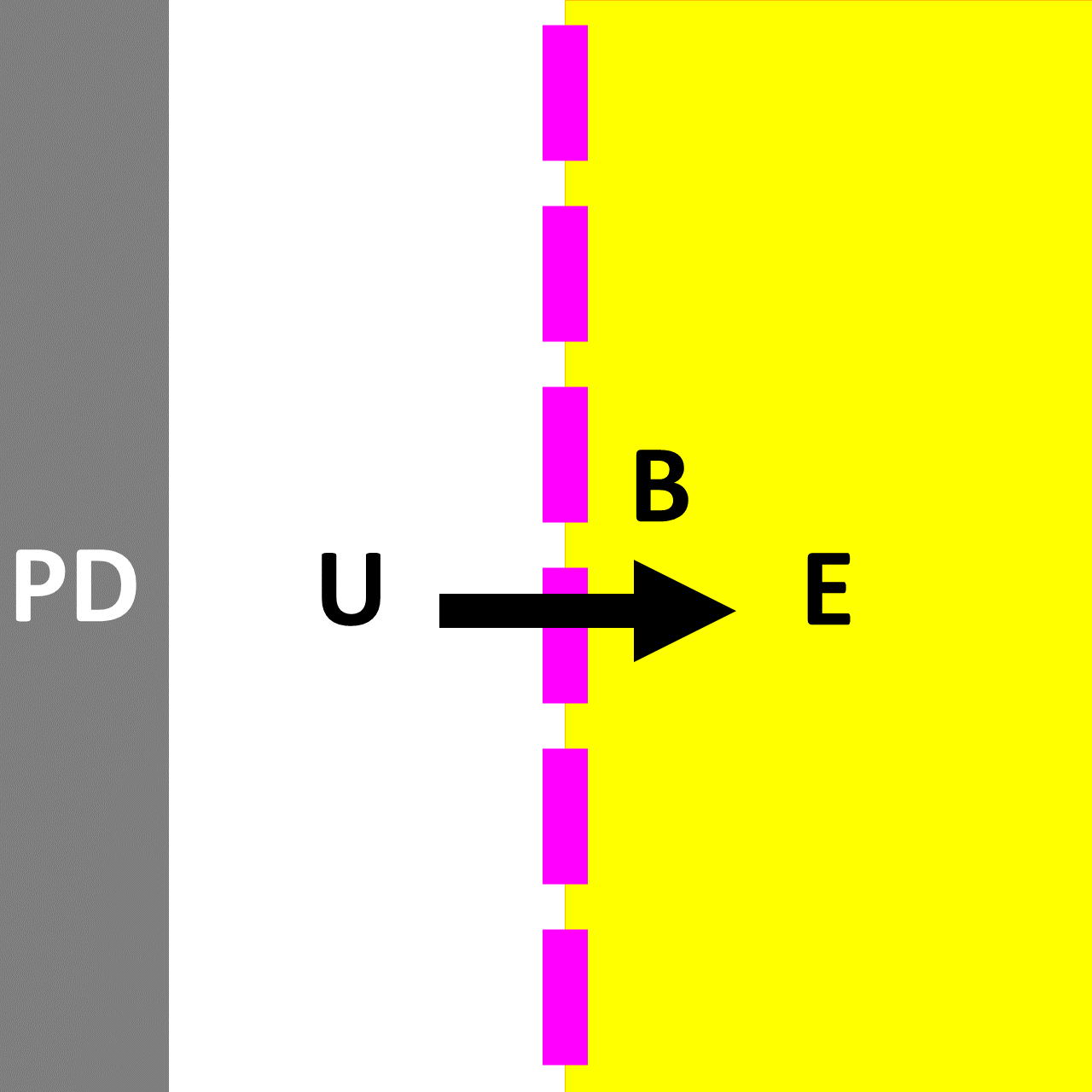}} \\
 \multicolumn{3}{c}{\large{\textbf{}} }\\

\multicolumn{3}{c}{wave back birth}\\
\frame{\includegraphics[width=0.1\textwidth]{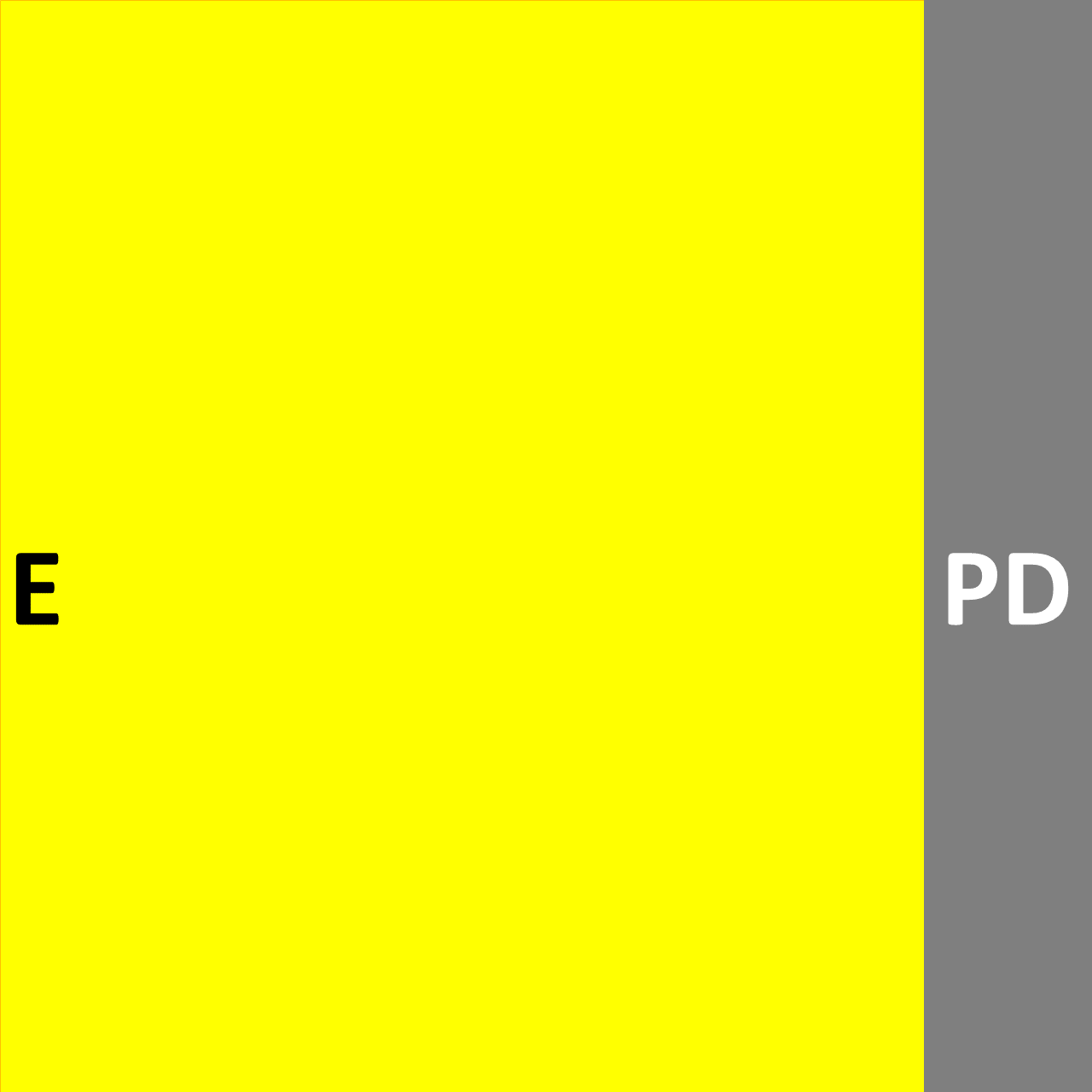}}&
{\Large $\rightarrow$} & 

\frame{\includegraphics[width=0.1\textwidth]{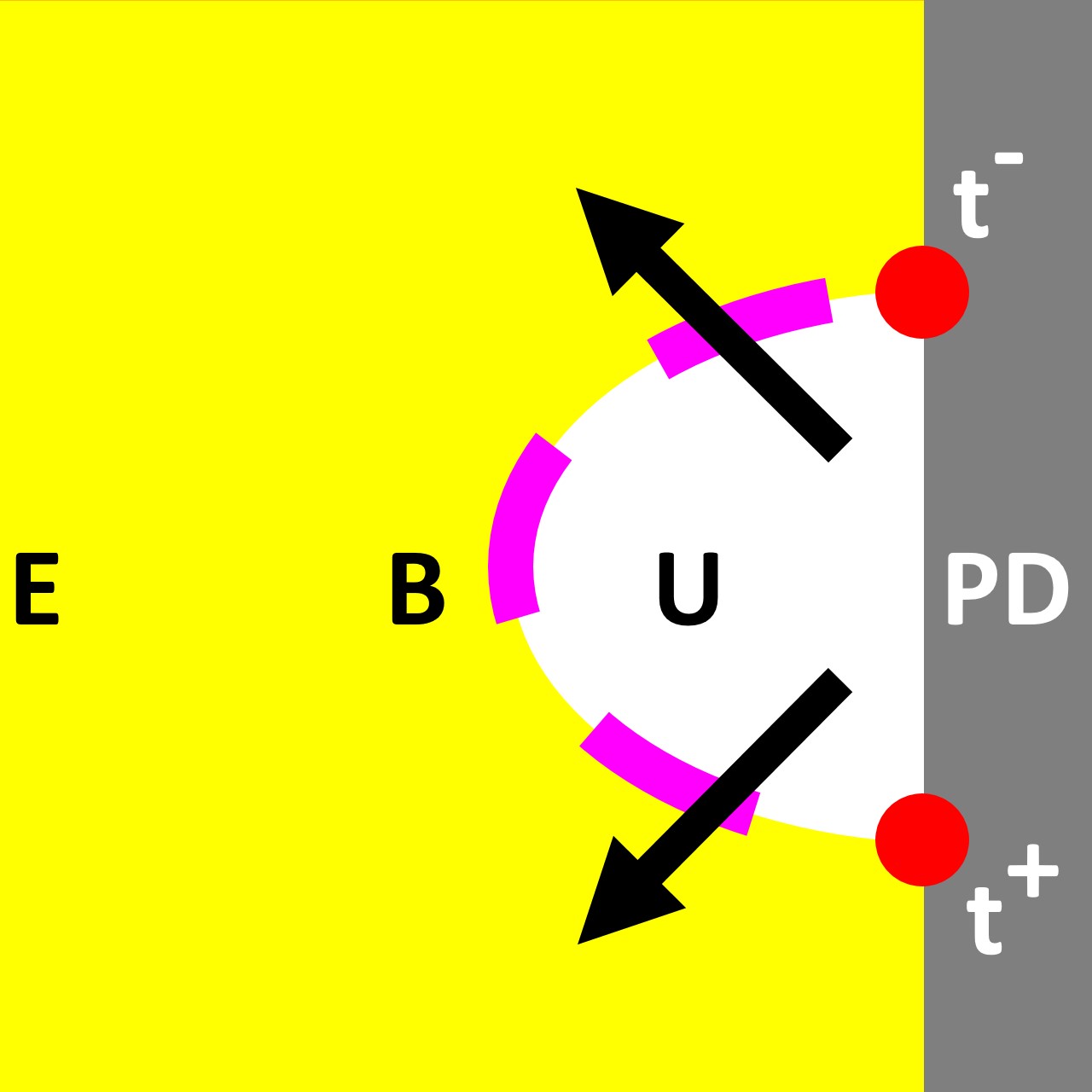}}\\
 \multicolumn{3}{c}{\large{\textbf{}} }\\

\multicolumn{3}{c}{wave back death}\\
\frame{\includegraphics[width=0.1\textwidth]{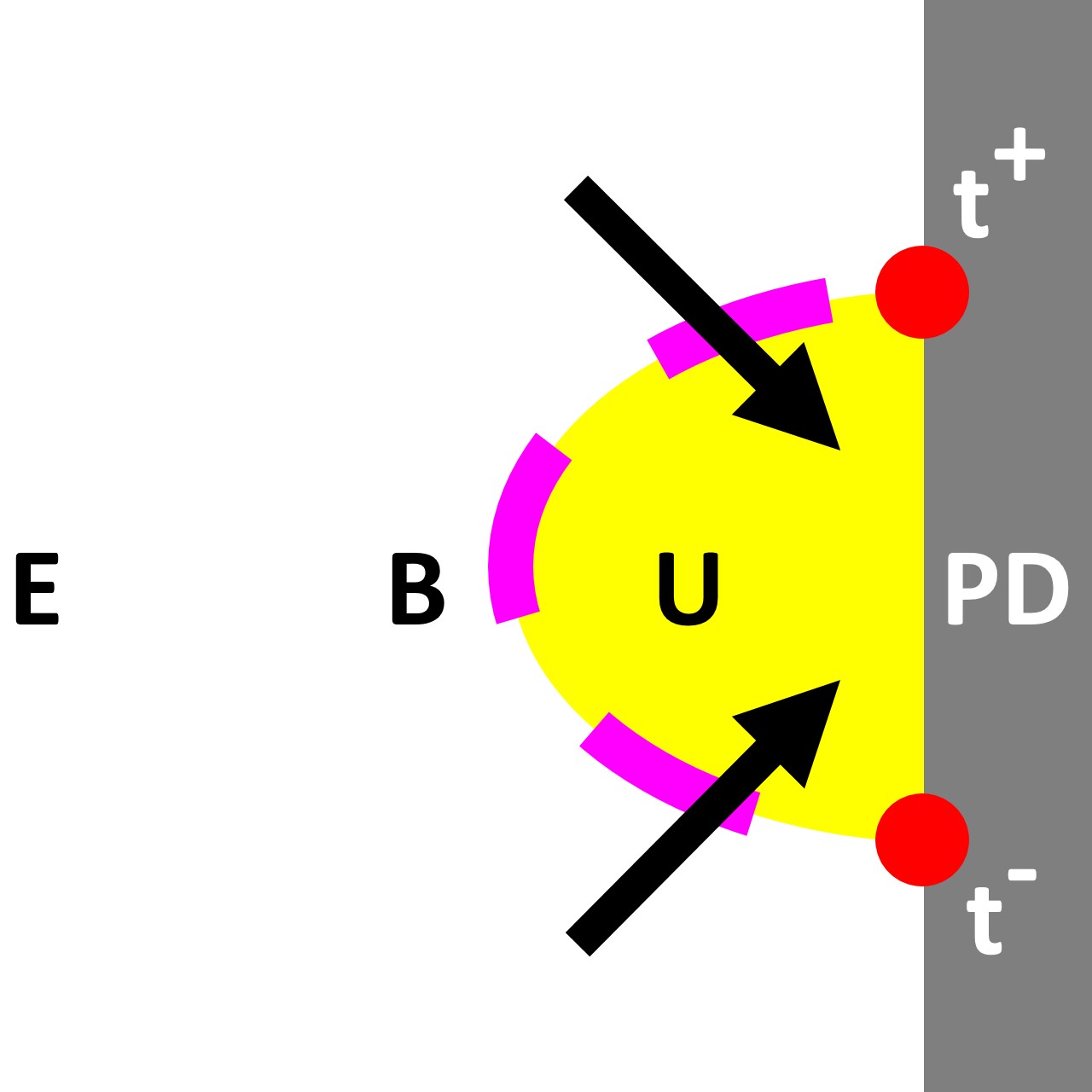}}&
{\Large $\rightarrow$} & 

\frame{\includegraphics[width=0.1\textwidth]{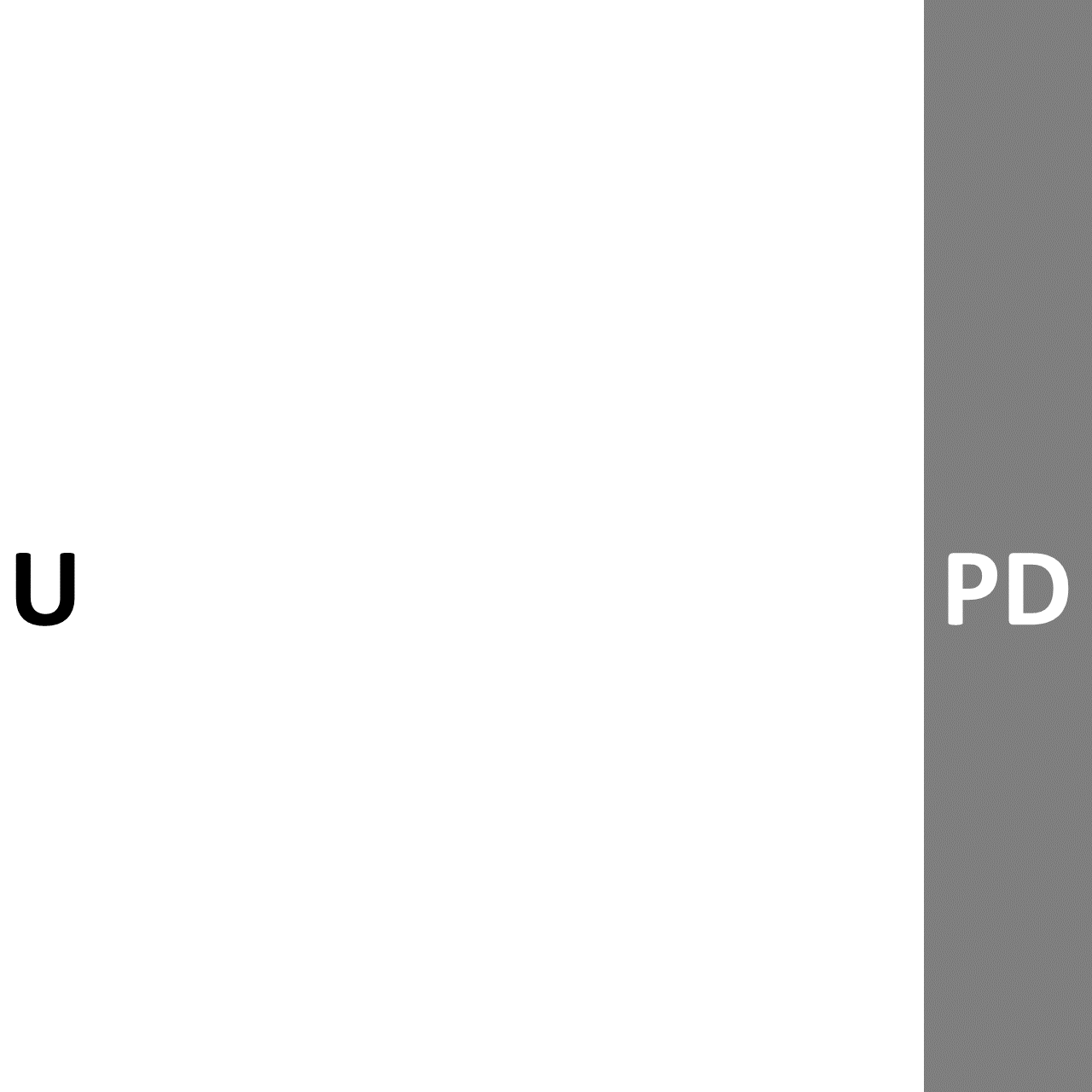}}
\end{tabular}
\begin{tabular}{c c c}
 \multicolumn{3}{c}{(d)}\\
 \multicolumn{3}{c}{ PD creation } \\
\frame{\includegraphics[width=0.1\textwidth]{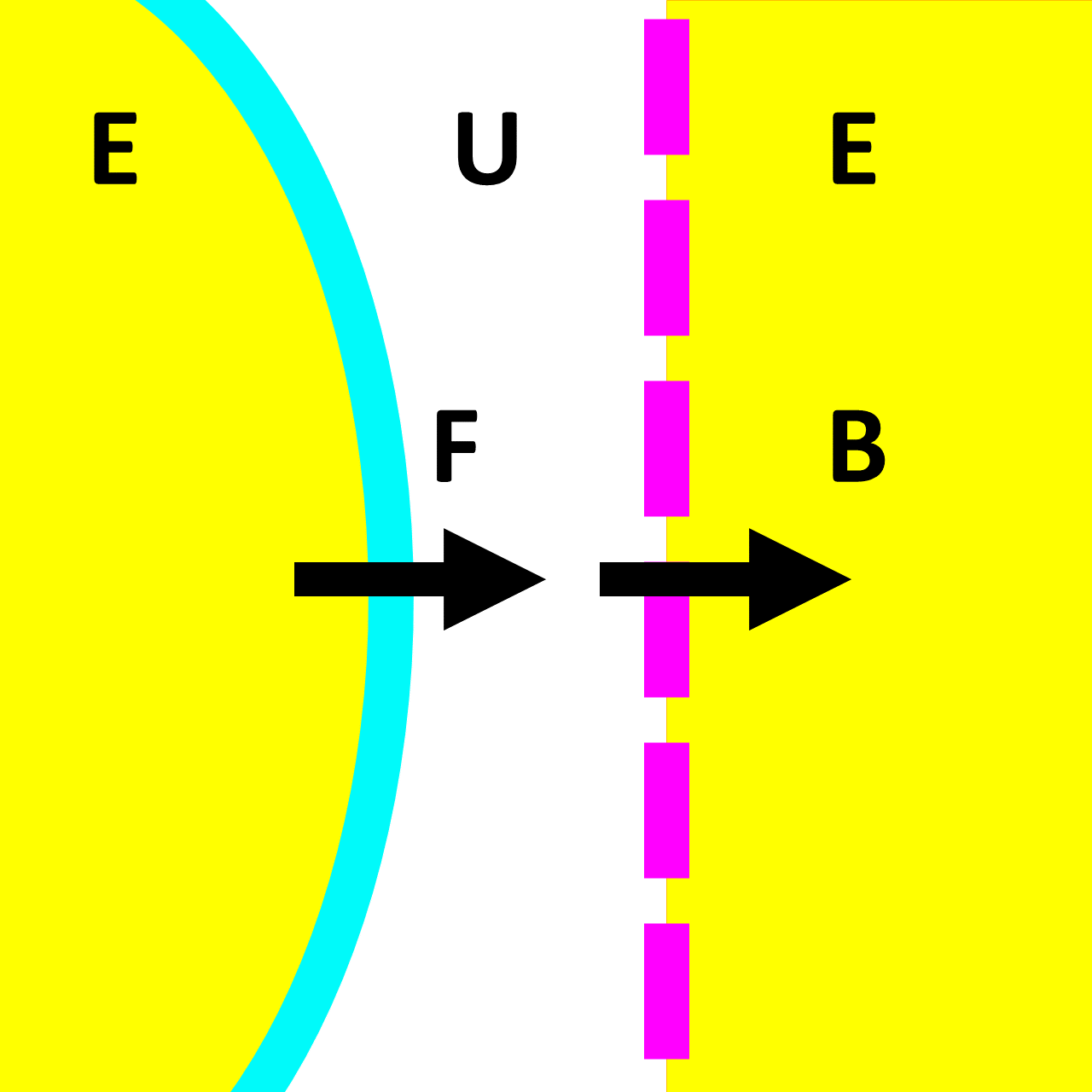}}  & 
{\Large $\rightarrow$} & 
\frame{\includegraphics[width=0.1\textwidth]{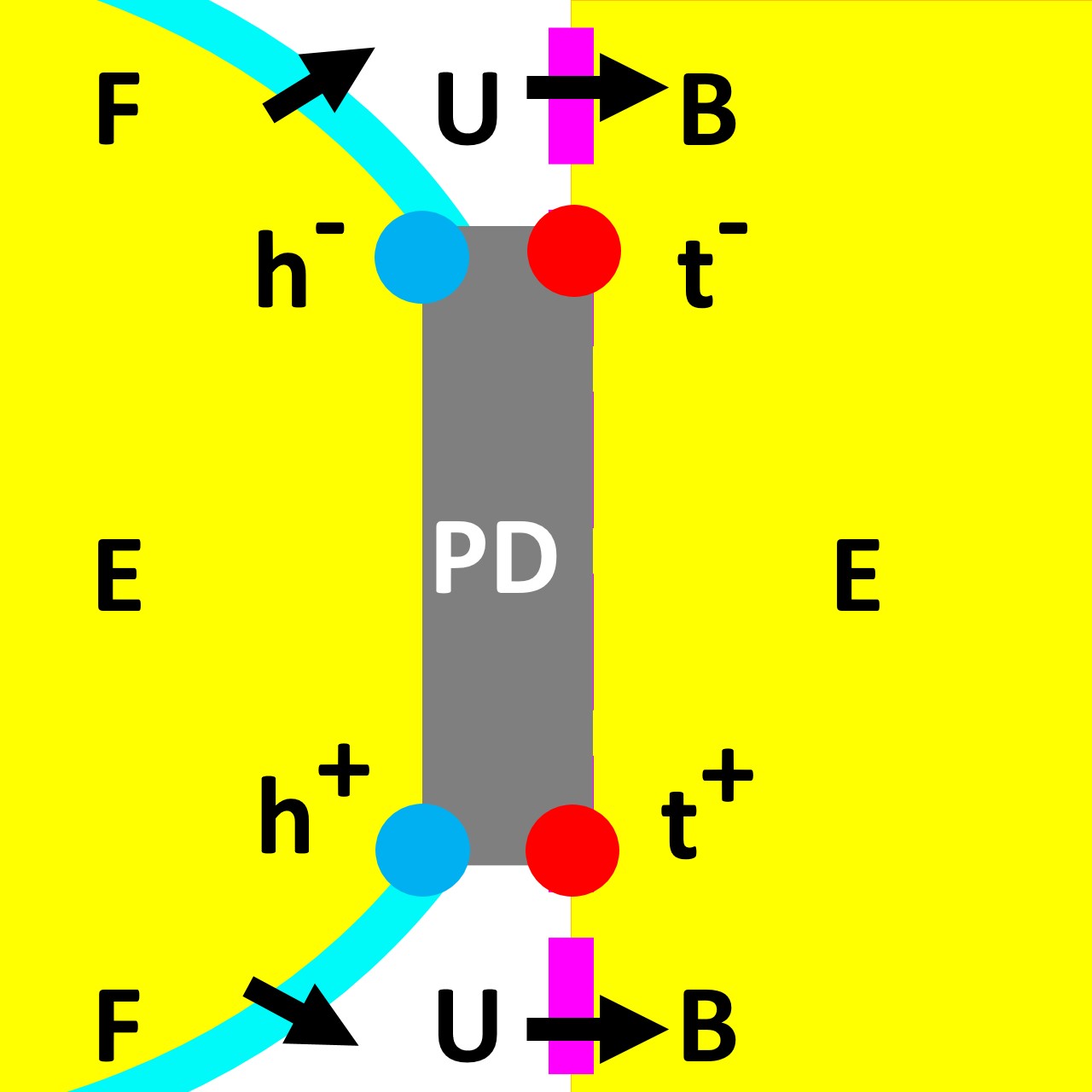}} \\
 \multicolumn{3}{c}{\large{\textbf{}} }\\
 \multicolumn{3}{c}{ PD splitting } \\
\frame{\includegraphics[width=0.1\textwidth]{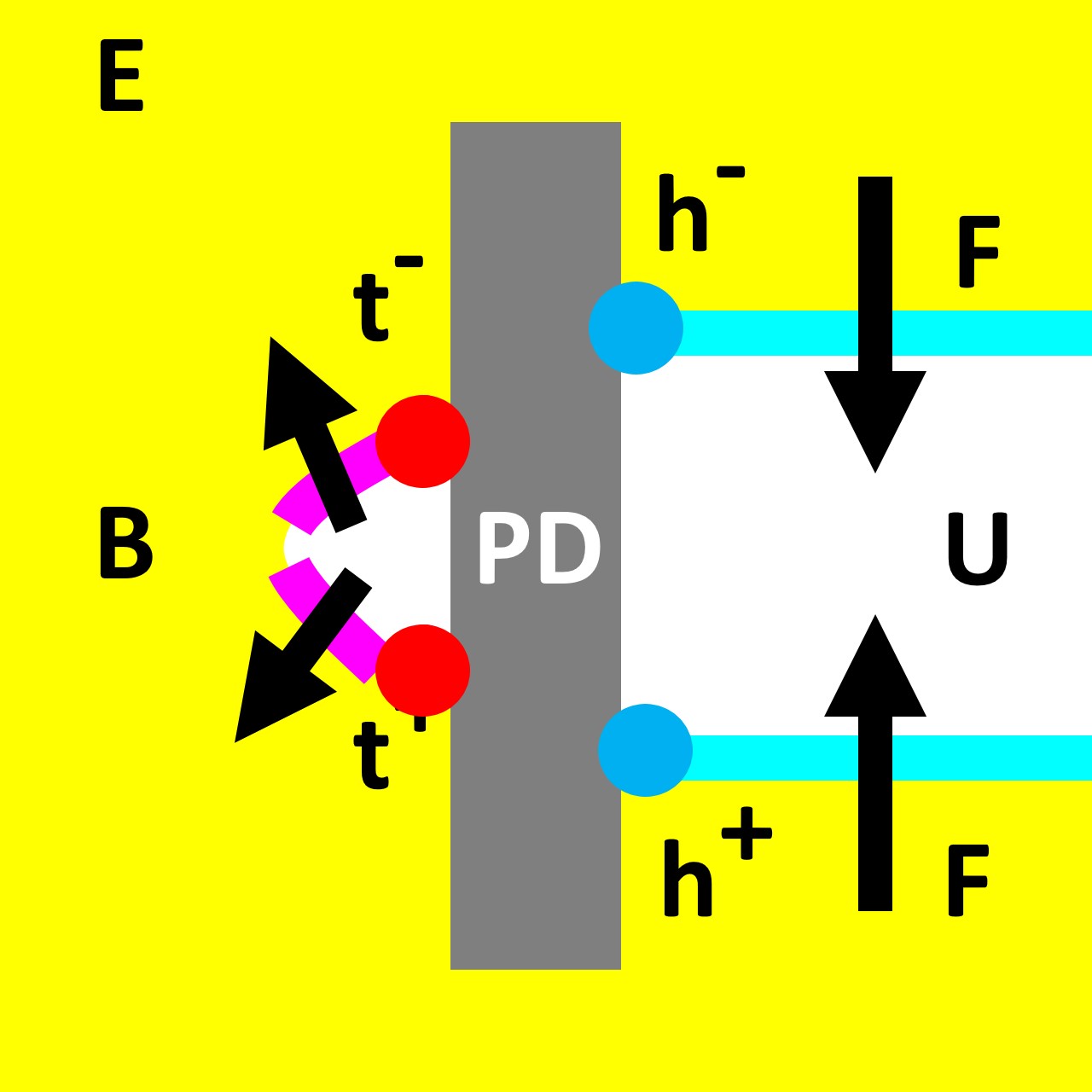}} &
{\Large $\rightarrow$} & 
\frame{\includegraphics[width=0.1\textwidth]{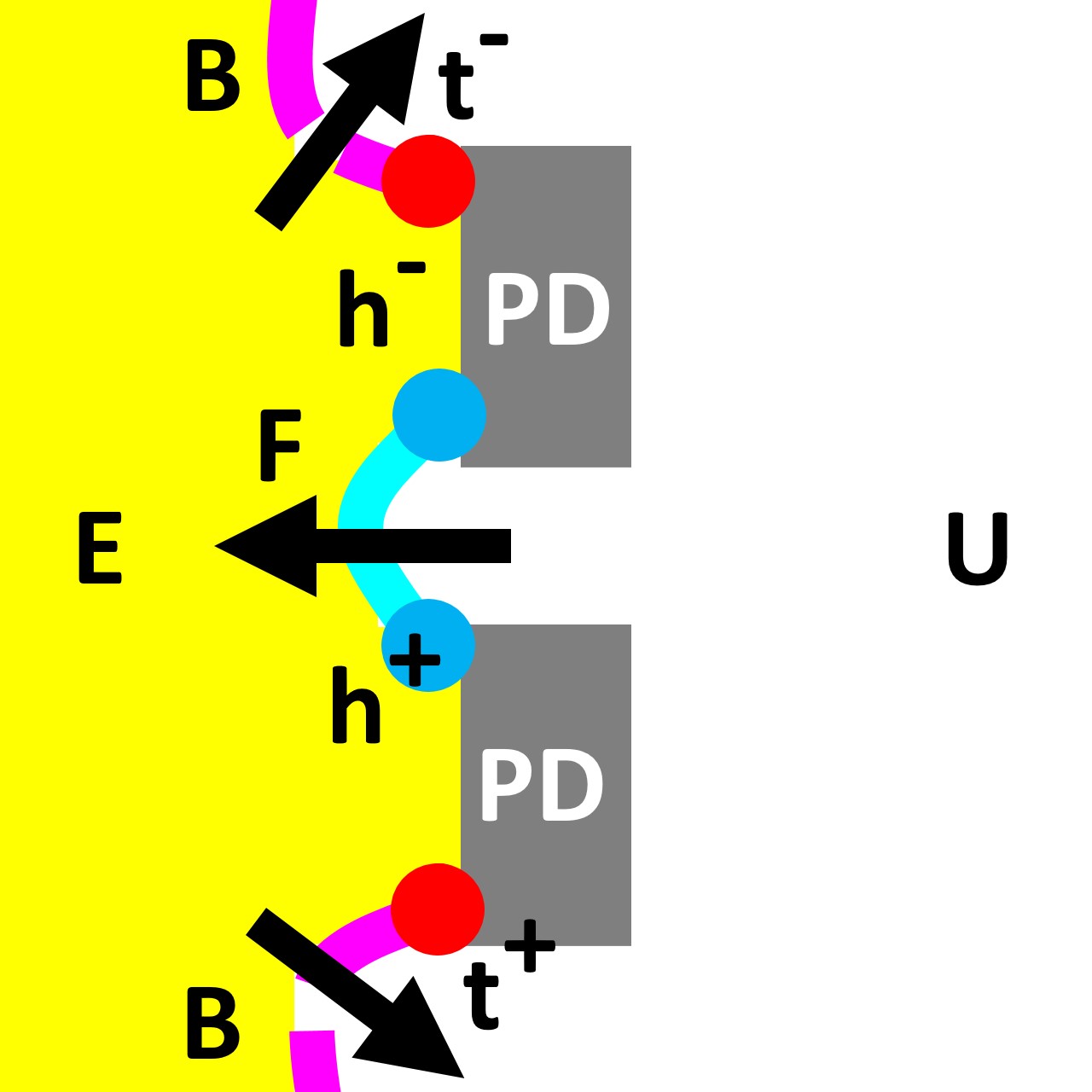}} \\
 \multicolumn{3}{c}{(e) }\\
tip, $Q=+1$ && tip, $Q=-1$ \\
\frame{\includegraphics[width=0.1\textwidth]{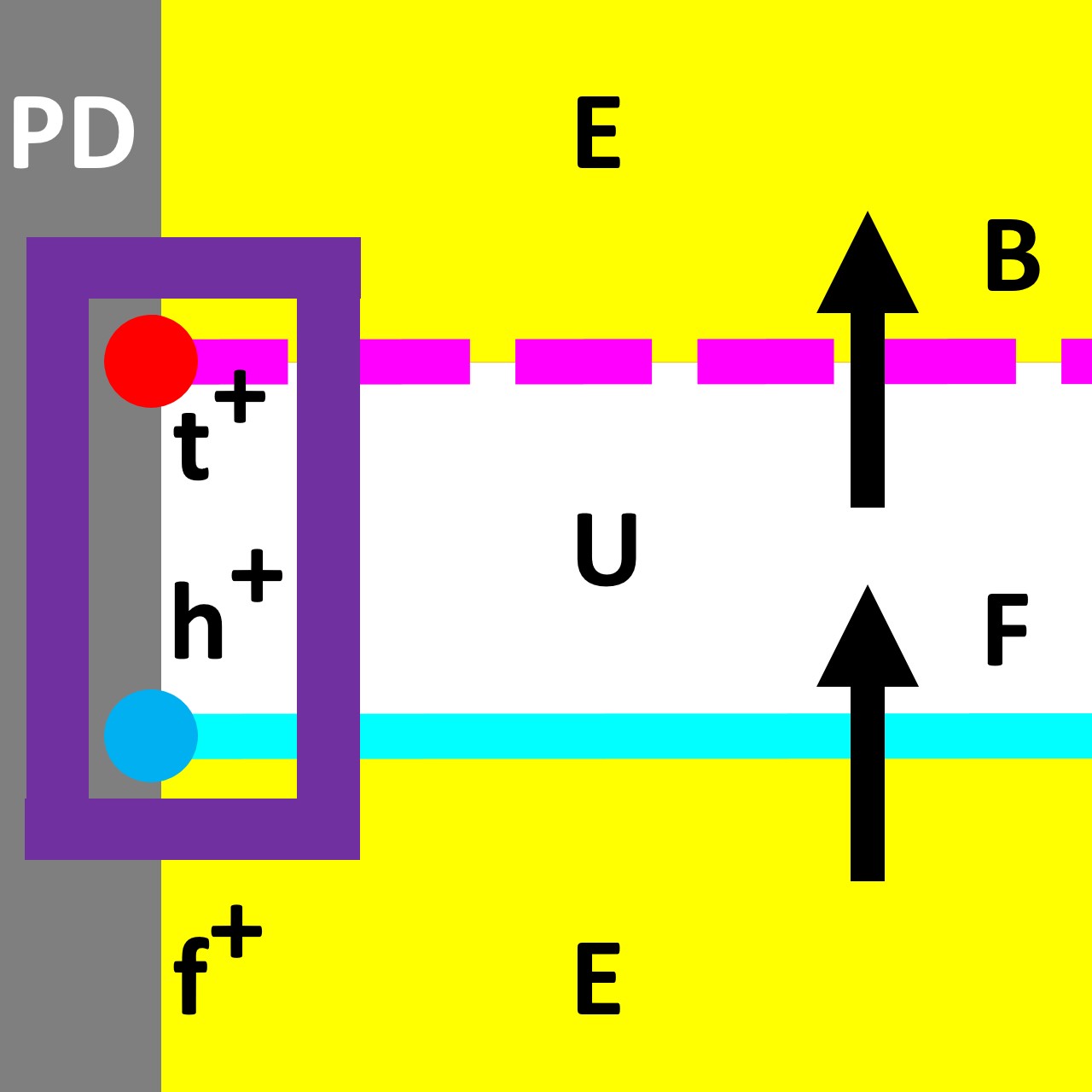}}&
{\Large $\rightarrow$} & 
\frame{\includegraphics[width=0.1\textwidth]{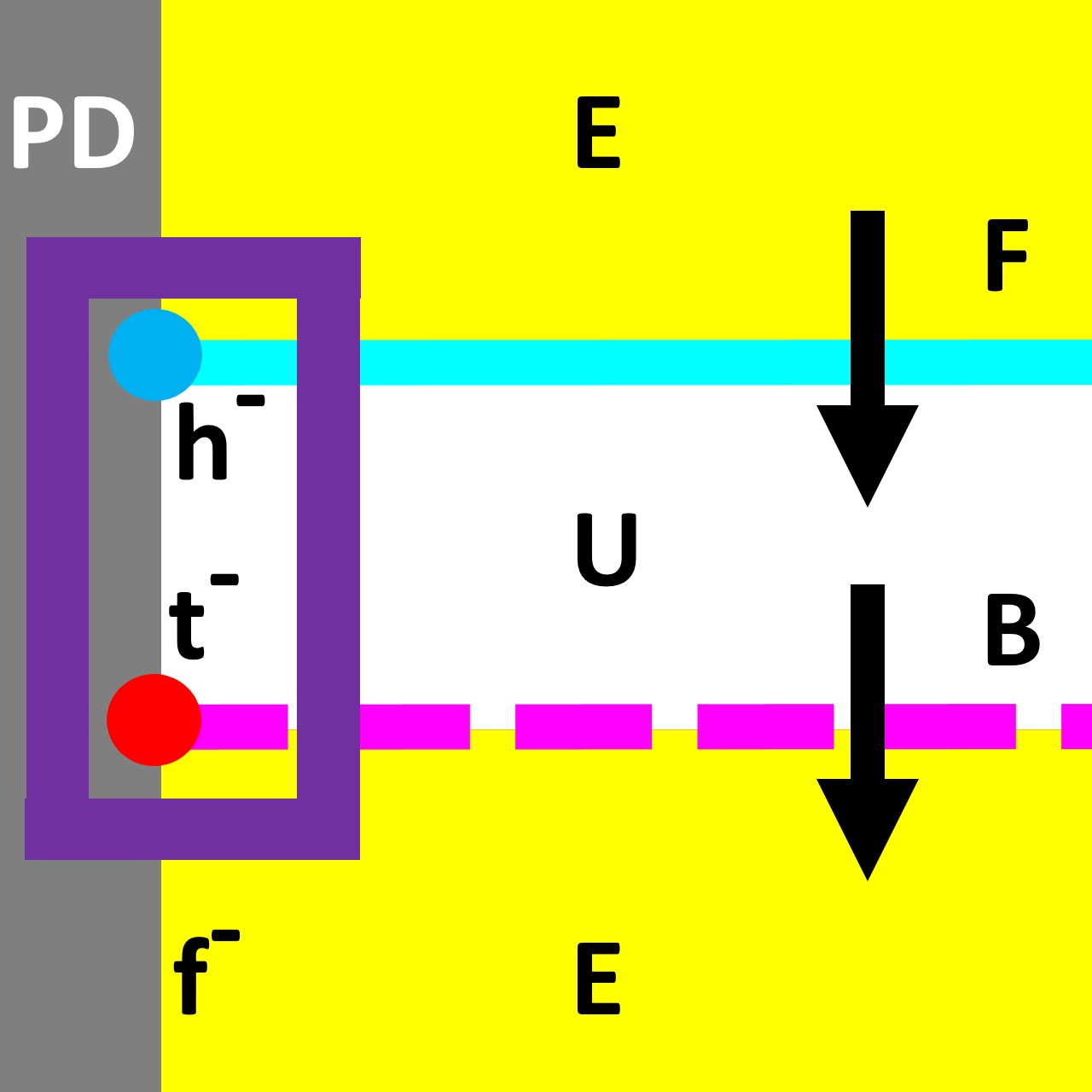}} \\
 \multicolumn{3}{c}{}\\
\multicolumn{3}{c}{ \frame{\includegraphics[width=0.215\textwidth]{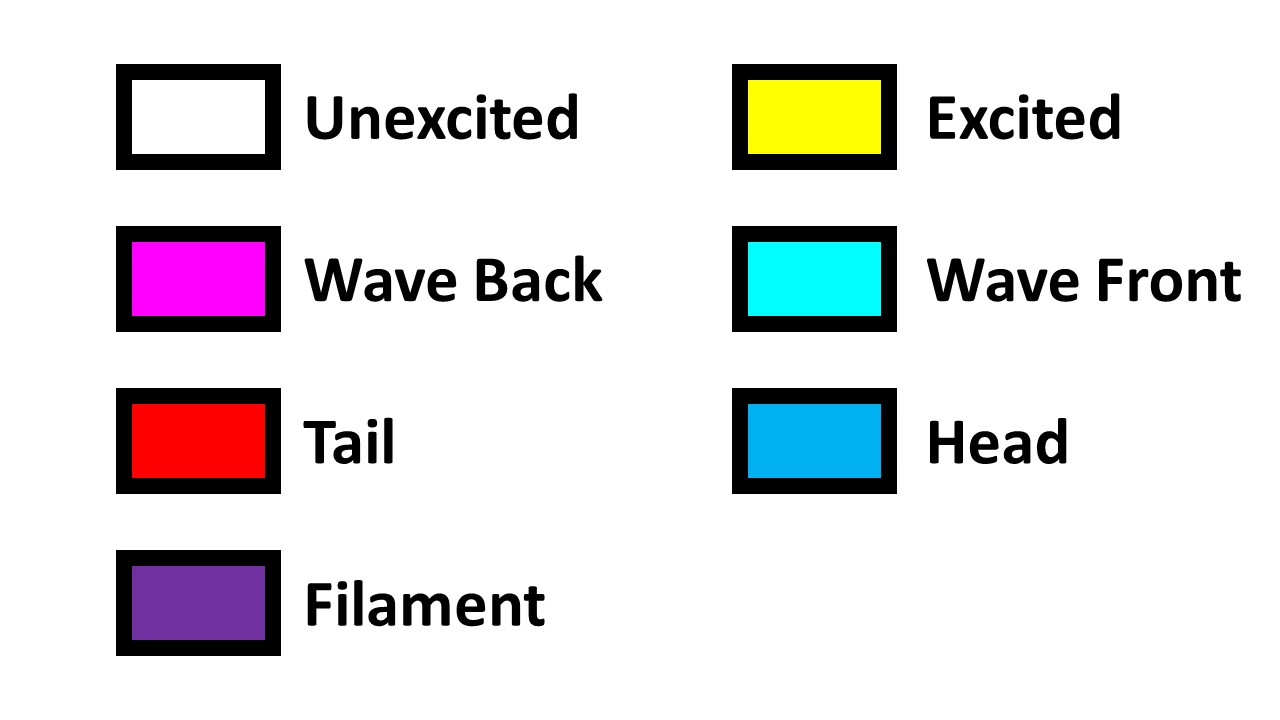} }}
\end{tabular}
  \caption{Overview of spatial transitions and temporal processes involving heads and tails in 2D. (a) A topological charge $Q=\pm \frac{1}{2}$ is associated to heads (h) and tails (t). (b) Illustration of the only four processes that can occur at the PD boundary changing the number of heads. Heads are created or annihilated in a pair of opposite charge. (c) Same as b. for wave back. (d) PD creation by collision of a wave front and back, introducing a quadruple $(\text{h}^+,\text{t}^+,\text{h}^-,\text{t}^-)$. (e) Example of the classical tip, indicated in purple,  with charge $Q=\pm 1$, appearing as a bound state of heads and tails with equal charge $Q=\pm \frac{1}{2}$.
   }
    \label{fig:headcases}
\end{figure*}

In principle, one could consider looking at which parts of the contour the topological charge originates from, to find a charge density. However, such a notion of charge density sensibly depends on the precise definition of phase, for which we previously showed that this phase can be reparameterised \cite{Kabus:2022}, as it is essentially a coordinate on the dynamical attractor $\TT$ in state space. 
Nonetheless, we note that on the PD boundary, a head or tail marks the transition between an excited PD boundary ($\partial Z_E$) and an unexcited part ($\partial Z_U$). Since the entire PD boundary is closed, i.e. the PD is a subset of the domain, it should, in two spatial dimensions, always contain an even number of heads and tails, and for this reason, we \textit{define} the topological charge of heads and tails to be $\pm \frac{1}{2}$, as in Fig. \ref{fig:headcases}a. For example, heads that circumvent a PD counterclockwise are given $Q=+\frac{1}{2}$, and are denoted as h$^+$. Clockwise heads have $Q=-\frac{1}{2}$ and are denoted as h$^-$. In a loose sense, these definitions correspond to the notion of the quantized angular momentum of a scroll, and the sign indicates the chirality. Alternatively, in the limit of a steep wave front and wave back, all dynamics take place there, and a wave front and wave back will give rise to a phase change of $\pi$, such that they attribute for half of the total rotation around the attractor, yielding a topological charge $\pm \frac{1}{2}$. 

Like heads and tails in 2D, pivot points also exist in two chiralities, to which a different topological charge $P=\pm \frac{1}{2}$ was associated \cite{Arno:2023feynman}. A pivot point has positive charge $P=+\frac{1}{2}$, denoted p$_+$ if the wave encircles it counterclockwise, and negative charge $P=-\frac{1}{2}$ if the wave encircles it clockwise. We use a different symbol for the pivot charge since it is intrinsically different from the head and tail charges. 

Moreover, pivots cannot be interpreted as a point in the state space: During activation, a pivot point will go through all phases. The emergence of a pivot is purely due to the mapping from phase space to physical space. The half-integer topological charge of a pivot corresponds to the rotation of $\pi$ in physical space: When the front goes from one side of the phase defect to the other side, it rotates around the pivot over $\pi$. For this reason, we believe the charge of the head and tail are intrinsically different from the charge of a pivot, which explains the distinct notations h$^{\pm}, \text{t}^{\pm}$, and p$_{\pm}$. In general, we call the charge of a pivot a $P$-charge and the charge of a head or tail a $Q$-charge.

\subsubsection{Constraints on head and tail creation, and annihilation  \label{sec:transitions}}

Heads and tails grant an interpretation of excitable media reminiscent of particle physics. Fig. \ref{fig:headcases}b lists all four possibilities to create or annihilate heads at the PD boundary. We conclude that heads are always created or annihilated in a pair of opposite charge 

In particle-physics terminology, h$^+$ and h$^-$ would correspond to a particle anti-particle pair. Fig. \ref{fig:headcases}c shows the same property for tail pairs. Since local excitation duration can be inhomogeneous in the medium, the arrows indicating motion in the case of wave backs and tails are only indicative. 

\subsubsection{Heads and tails at the medium boundary \label{sec:boundary}}

We have so far only considered intersections of wave fronts and backs with PDs. However, fronts and backs can also end at the medium boundary. In some cases, e.g. 3D patterns or inexcitable obstacles, see below, it is useful to consider the ambient space of the excitable medium also as a PD, and the medium boundary as a PD boundary. Moreover, the local activation state, `phase', of cells inside and outside the medium is different, such that also at the medium boundary, there is an abrupt change in phase. 
This way, we introduce `boundary heads' and `boundary tails', with the same topological charges of $\pm \frac{1}{2}$: see Fig. \ref{fig:headcases}a with the grey zone being a non-excitable domain. These boundary heads and tails were already shown in Figs. \ref{fig:torus}b. and e. 
With this extension, we can prove that in 2D, the total topological charge carried by heads and tails is zero: Every head or tail touches exactly one excitable region. Then, consider the boundary of every excited region $E$ in the domain. This border consists alternatively of a PD boundary ($E\cap Z$) and a wave front or wave back ($E\cap U$). At both end points of a wave front or wave back, heads or tails of opposite charge are located, so the total charge is zero. 

\subsection{Bound states \label{sec:BoundStates}}

In previous work \cite{Arno:2023feynman} we mentioned already four cases in which heads, tails, and pivots may group and move together for a certain amount of time. For completeness, we now revise these bound states before going to 3D. A complete overview of all regions, and fundamental and bound particles can be found in Tab. \ref{tab:structures}. 

\subsubsection{Tips}
Once a linear-core rotor has formed in 2D, the head point will generally follow the tail around the phase defect boundary, see Fig. \ref{fig:headcases}e. This bound state can be detected by tip tracking algorithms \cite{Fenton:1998}, and so we call it a tip in 2D. In three dimensions, this will form an extended curve of tips, which is classically called a rotor filament. 

\subsubsection{Cores}
If a spiral wave with no linear core is created via a conduction block, the phase defect region will generally shrink to a small region, i.e. the core, around which the spiral wave rotates. This core particle then also coincides with the classical PS. Since during its formation, it contains a head, tail, and two pivots, all of the same charge, we call this bound state a core particle. 

\subsubsection{Growth sites}

As mentioned above, a PD typically arises from a wave front hitting an incompletely recovered region, see Fig. \ref{fig:headcases}d. From the point of first contact, the wave front will travel along the wave back in all directions to create a PDL in 2D. During this process, a head, tail, and pivot of the same topological charge coincide. We call this bound particle a PD growth site, see Fig. \ref{fig:PDgrowth}a. Later on, the wave back retreats from the PD, such that the head curls around it and forms a rotor, see Fig. \ref{fig:PDgrowth}b. 

\subsubsection{Shrink sites}

\begin{figure*}
\begin{tabular}{c c c }
   (a) & (b) & (c)  \\
       \frame{\includegraphics[height=3.5cm]{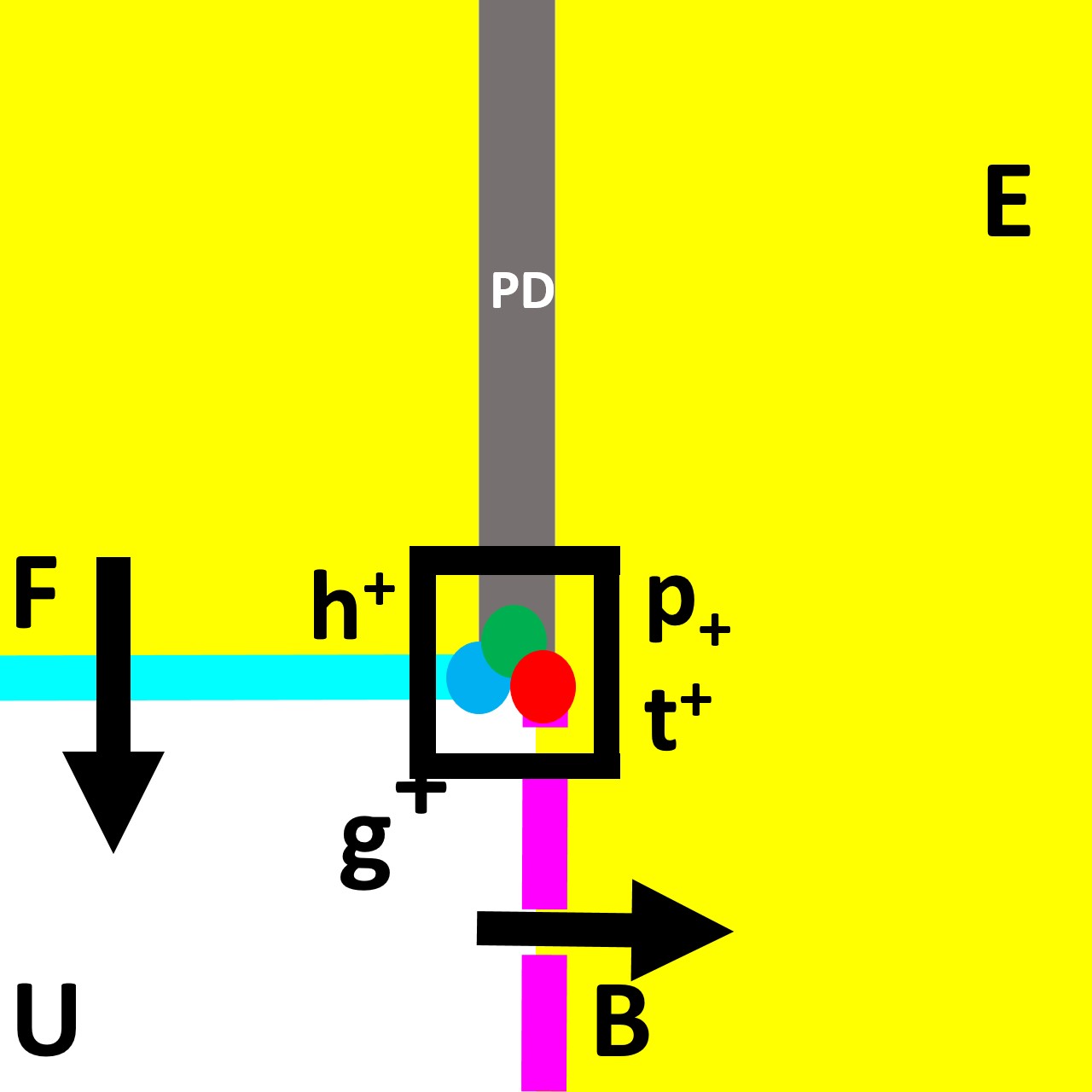}}
  &     \frame{\includegraphics[height=3.5cm]{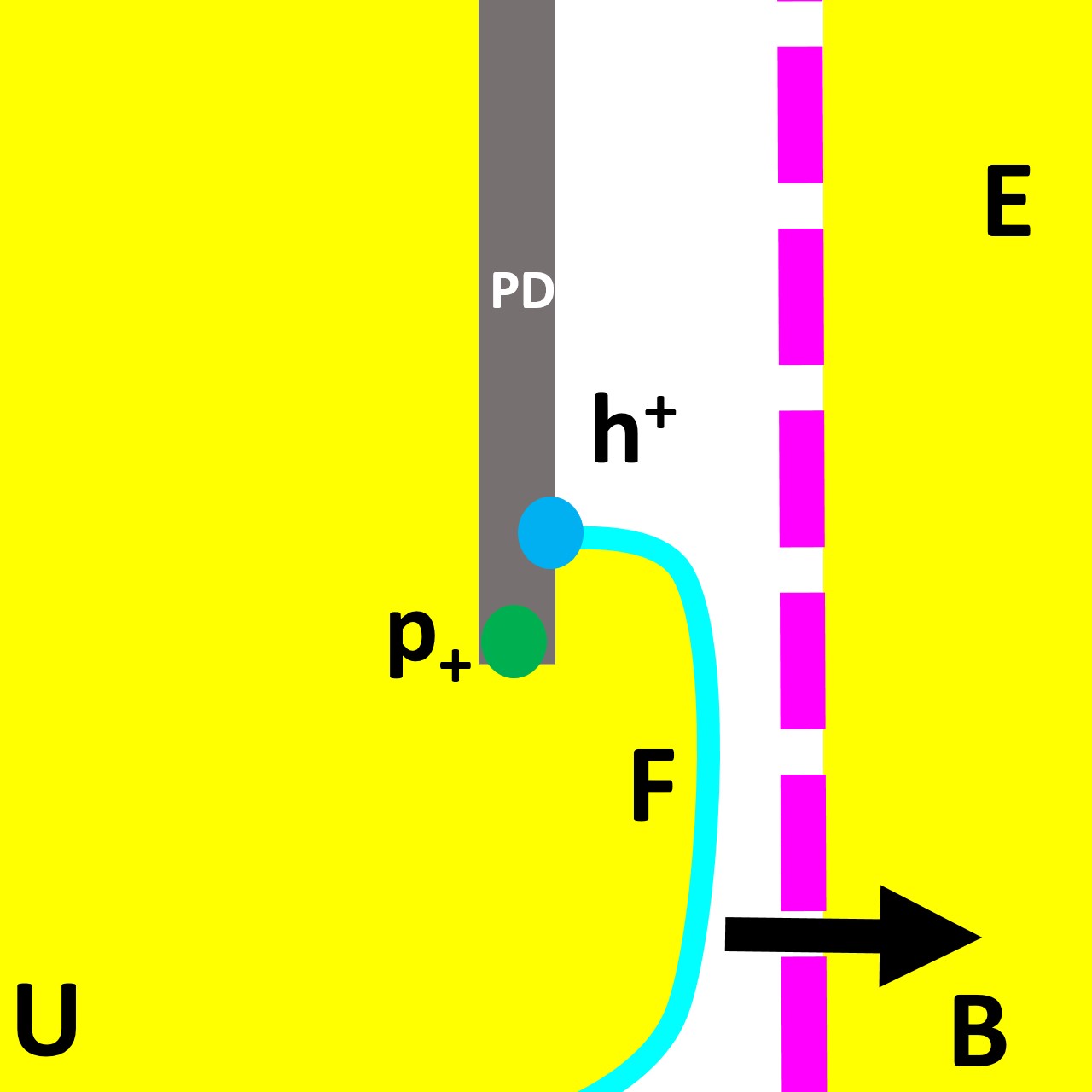}}
&     \frame{ \includegraphics[height=3.5cm]{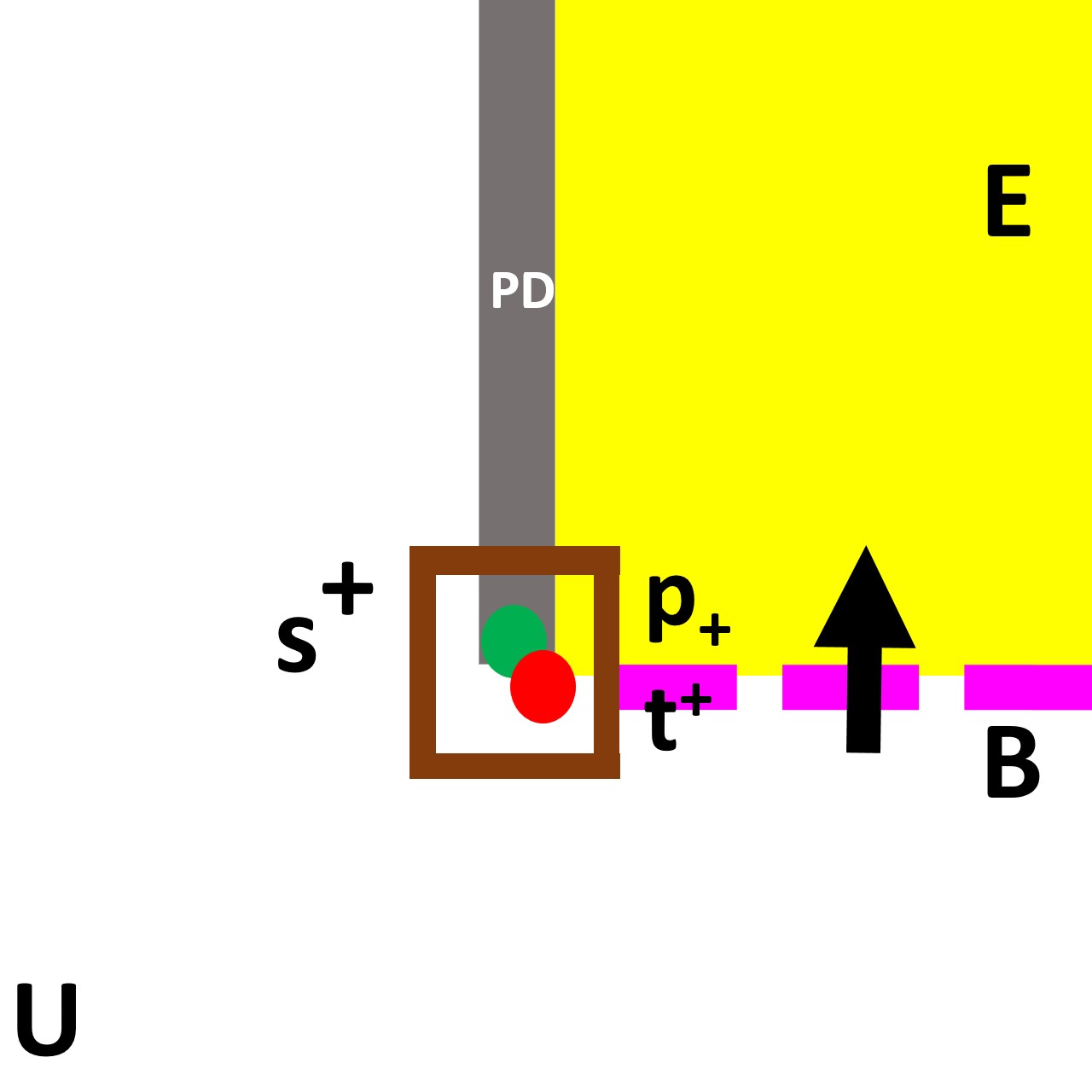}}

\end{tabular}
    \caption{Mutual relation of head, tails and pivots during growth and shrinking of a phase defect. (a) initial PD formation via a wave block since an impeding wave (top left) touches an incompletely recovered region (right). A growth point is formed, consisting of a head, tail, and pivot of the same chirality and topological charge. (b) During further evolution, the pivot will generally not coincide with head and tail. (c) Retraction of the oldest PD end (pivot), as a bound state with a tail point. In 3D, similar stages of development occur. Only case (a) is covered by the classical theory of phase singularities and filaments. 
    \label{fig:PDgrowth}}
\end{figure*}

The wave back of the resulting spiral will at some point touch the pivot site, and in the approximation of a toroidal state-space attractor, the tissue recovers to the resting state at both sides of the phase defect line. At these points, the PD will locally cease to exist, and the pivot retreats together with the tail point, see Fig. \ref{fig:PDgrowth}c. We call this bound state of a tail and pivot a `PD shrinking site', denoted as $s$. 

\section{Topological description of 3D excitation patterns\label{sec:3D}}

\subsection{Theory}
\subsubsection{Heads and tails are closed curves in 3D \label{sec:headtail3D}} 
Just like the concept of a PS extends to filament curves in three spatial dimensions \cite{Clayton:2005}, heads and tails will become curves in a 3D excitable medium. Since head curves and tail curves are intersections of the $E,U,$ and $Z$ regions, they cannot end inside the medium, i.e. they are topologically preserved in space as well as under time evolution. 

\begin{figure*}
    \centering
    \begin{tabular}{c c c c }
      (a) & (b) & (c) & (d)   \\
      \includegraphics[height=2.5cm]{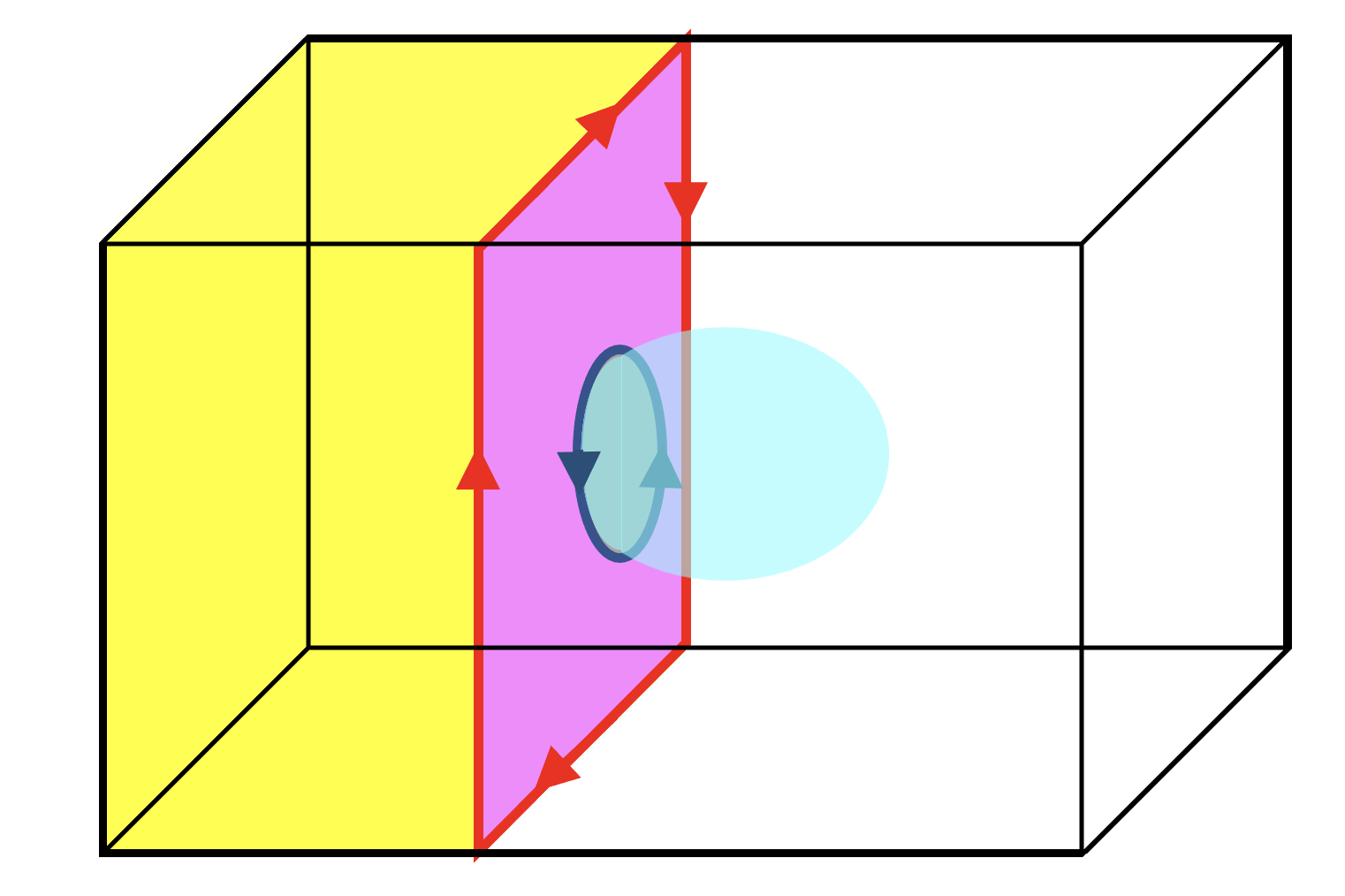}   & 
      \includegraphics[height=2.5cm]{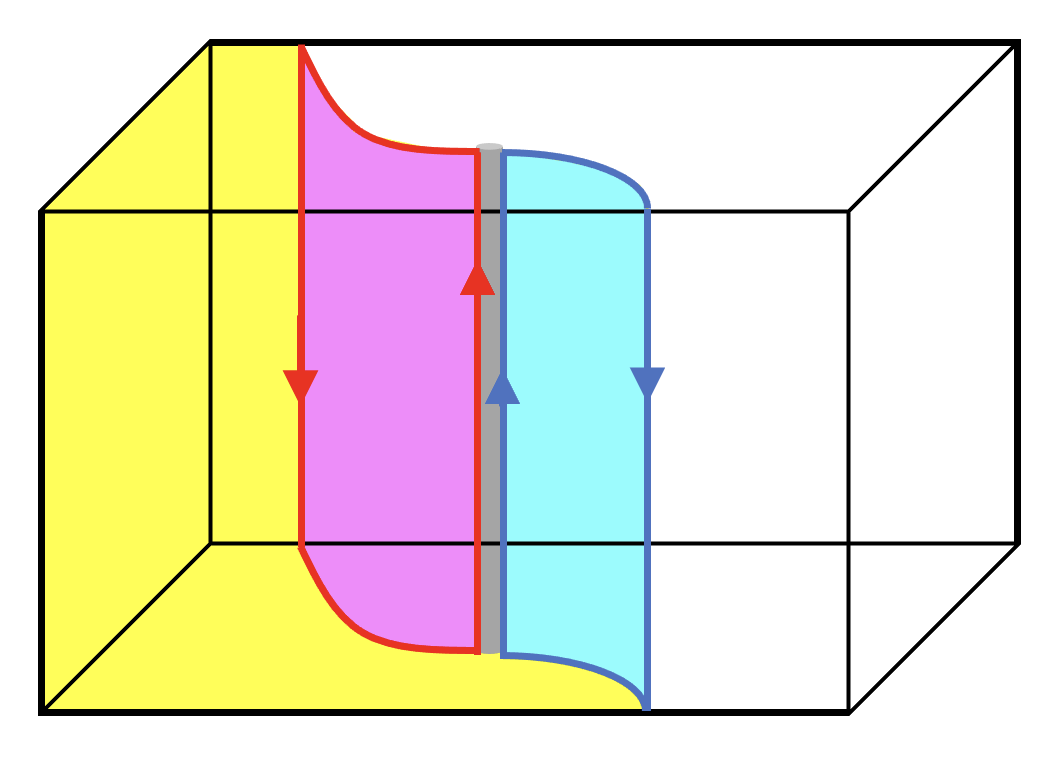} & 
       \includegraphics[trim={4.cm 2.cm 8.cm 1.cm},clip,height=2.5cm]{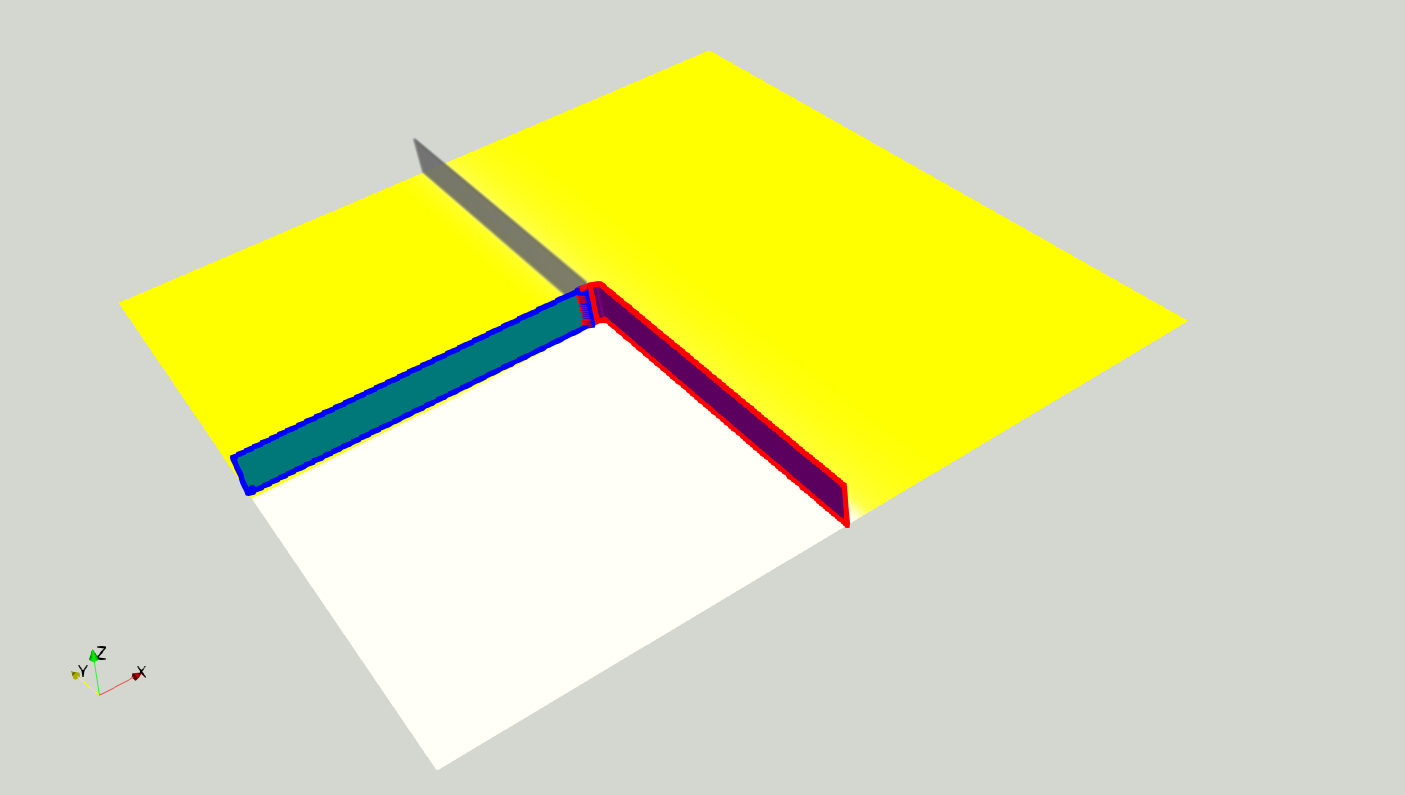} & 
       \includegraphics[trim={15.cm 6.cm 15.cm 10.cm},clip,height=2.5cm]{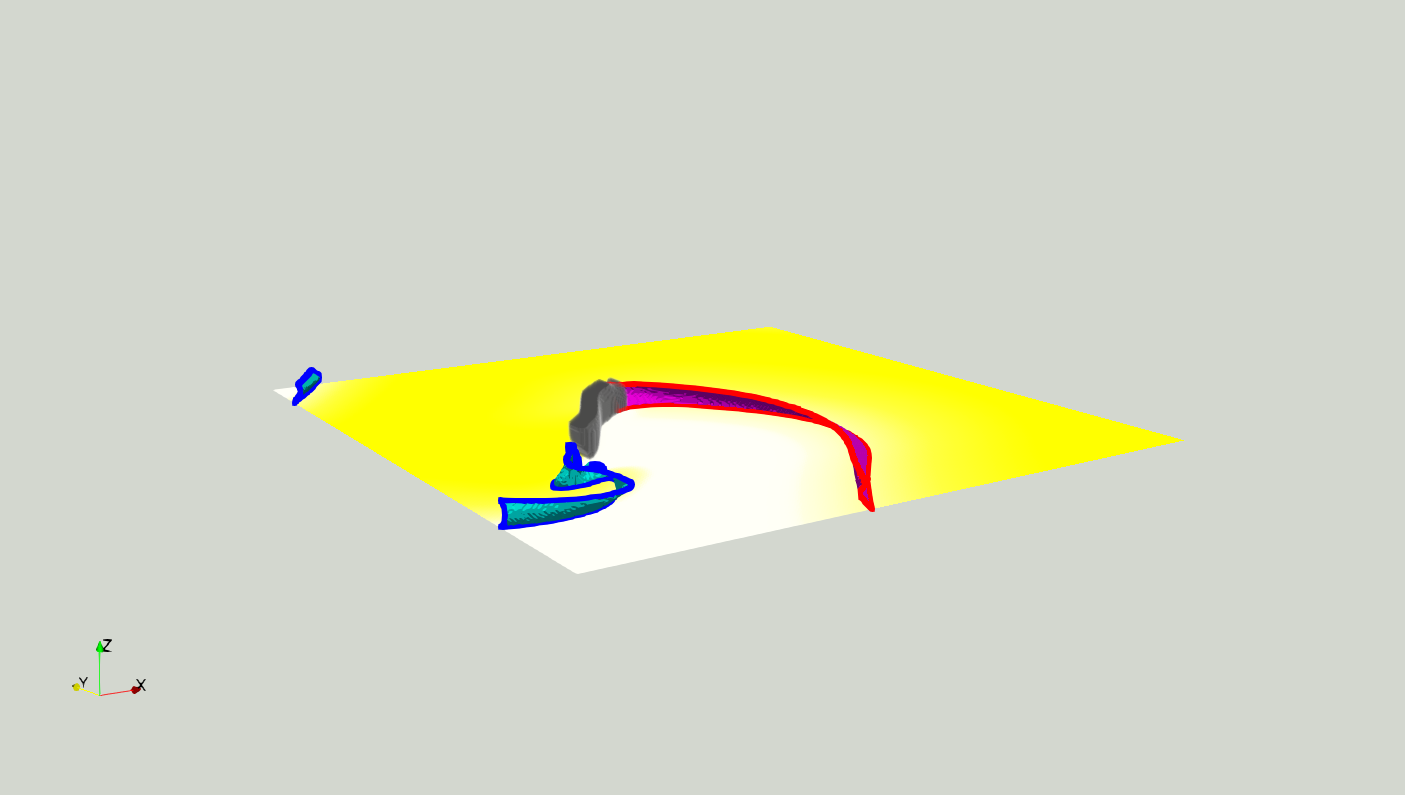} \\
    \end{tabular}
    \caption{Heads and tails form closed strings in 3D. (a) Creation of a closed head loop (blue) around a PDS (grey) if a wave front (cyan) hits an unrecovered region on the left. (b) A straight scroll wave. 
    (c) Initial condition of an S1-S2 scroll wave initiation in an isotropic medium with FK reaction-kinetics \cite{Fenton:1998}, showing consistently closed head and tail curves, together with the phase defect on the left. 
(d) Head and tail curves for a linear-core scroll in an anisotropic medium $D_1= 1, D_2=0.25$ for which the fibre angle varies from $-60 ^{\circ}$ at the bottom to $60 ^{\circ}$ at the top with BOCF reaction-kinetics \cite{BuenoOrovio:2008}. Topological conservation of head and tail strings follows as they are the boundary of wave fronts and wave backs.}
    \label{fig:headtail3D}
\end{figure*}

When generalizing the processes in Fig. \ref{fig:headcases} to three dimensions, we see that when a wave front touches a PD, it creates a circular head curve, see Fig. \ref{fig:headtail3D}a. A head line can be considered to carry a `head current' or `head field' $\vec{H}$, following the right-hand rule: When moving the fingers of your right hand in the direction of front propagation and the palm of your hand towards the PD, your thumb points in the direction of $\vec{H}$. When such current intersects a plane, pointing out of it, it will yield a charge $Q=+\frac{1}{2}$ and when pointing into the plane, one sees a head point with $Q=-\frac{1}{2}$. This corresponds to the half-integer topological charge in two dimensions, see Sec. \ref{sec:topoheadtail}. A similar process occurs for the other cases in Fig. \ref{fig:headcases}b-c. For tail curves, the fingers of your right hand should point towards the excited part and the palm of your hand to the PD, then your thumb points in the direction of the `tail current' or `tail field' $\vec{T}$, see Fig. \ref{fig:headtail3D}a.

We still follow the convention that the medium boundary can be seen as a phase defect. In this case, the head and tail curves must form closed loops without any flips in the tangential fields $\vec{H}$ and $\vec{T}$. 

Note in Fig. \ref{fig:headtail3D}a that the tail curve lying on the medium boundary is a mere wave back observed on the surface of the medium. Likewise, in Fig. \ref{fig:headtail3D}b, the part of the head curve lying on the medium boundary is seen as a wave front.

Fig. \ref{fig:headtail3D}b shows a sketch of a circular-core scroll wave, where the head and tail curves form together the filament or rotation axis of the scroll. Where the filament touches the medium boundary, however, the head and tail curves disentangle and run back as boundary heads and tails over the medium boundary to the other free end of the filament. Initiating a scroll wave via the S1-S2 protocol already sets a topological equivalent situation, see Fig. \ref{fig:headtail3D}c. A first stimulus (S1) initiated on the left has swept the medium, and is followed by a second stimulus (S2) in a quarter of the medium. The second stimulus instantaneously creates a planar phase defect and a planar wavefront, circled by a head curve. 

\begin{figure*}
        \includegraphics[width= 0.6\textwidth]{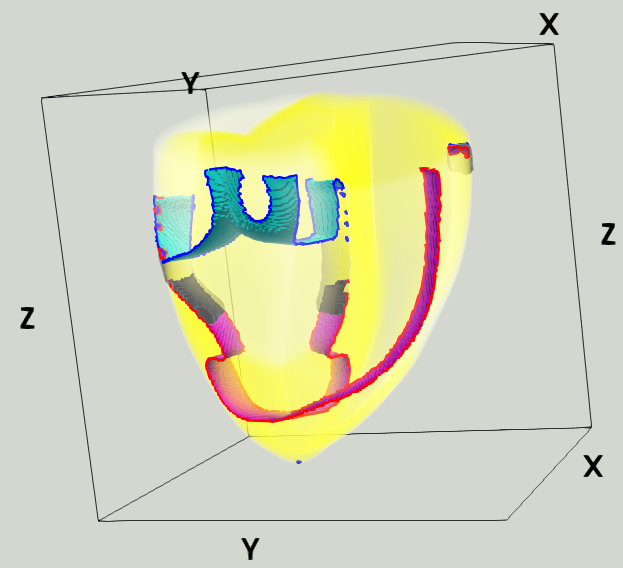} 
    \caption{Excited (E), unexcited (U), and PD regions, together with front (F), back (B) and corresponding heads and tails detected in a human biventricular mesh, with uniaxial anisotropy for which the the local fiber orientation was measured by Hren \textit{et. al.} \cite{Hren:1995}. Simulation was done in the BOCF model \cite{BuenoOrovio:2008}. \label{fig:Hren}} 
\end{figure*}

Fig. \ref{fig:headtail3D}d shows a linear-core scroll wave in the BOCF model in a medium with rotational anisotropy $D_1=1, D_2=0.25$, where also inside the medium, the head and tail curves are separated, yet connected by a PD surface. Fig. \ref{fig:Hren} shows an example of three phase defects, together with the detected fronts and backs, and the corresponding heads, tails in a human biventricular mesh, with uniaxial anisotropy.



\subsubsection{Pivot curves and phase defect lines at the boundary of phase defect surfaces \label{sec:pivot3D}}

In the regime of linear cores, phase defect lines become phase defect surfaces (PDSs) in three dimensions \cite{Arno:2021}. Like spiral tips, heads, and tails, the pivot points from 2D now become pivot curves in 3D. As with head curves and tail curves, we can associate a direction to pivot curves using the right-hand rule: If your fingers curl like the wave turning around the pivot, your thumb indicates the direction of the pivot curve's vector charge $ \vec{P}$, which one can think of as a vector field on the rim of the PDS. 

\begin{figure*}
\begin{tabular}{c c c c }
       (a) & (b) & (c) & (d)   \\
    \includegraphics[height=2.2cm]{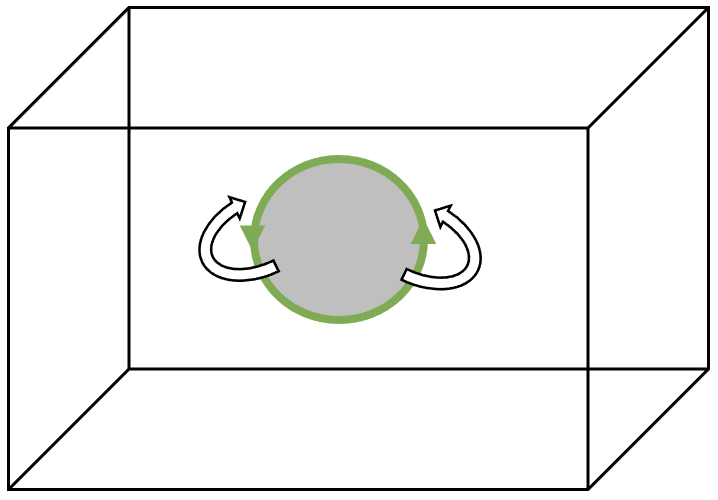} & 
        \includegraphics[height=2.2cm]{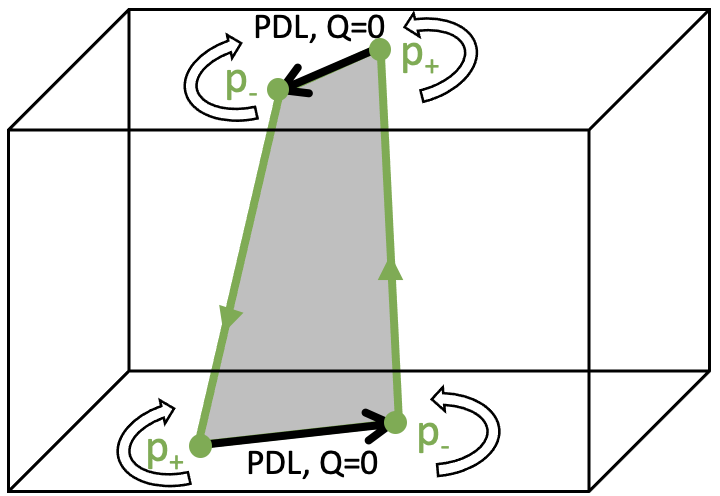} &
            \includegraphics[height=2.2cm]{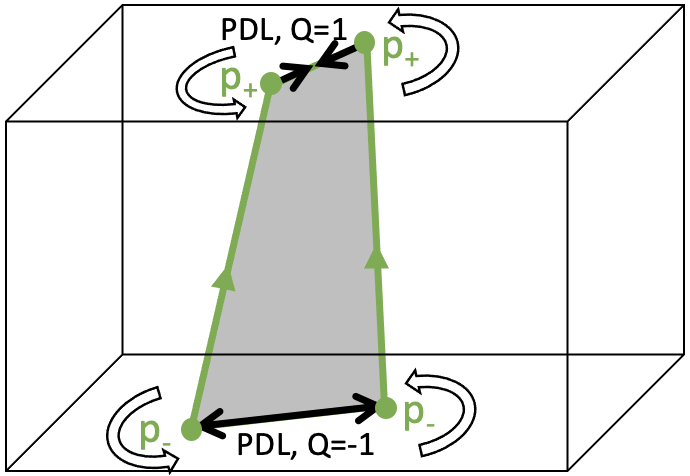} &
                \includegraphics[height=2.2cm]{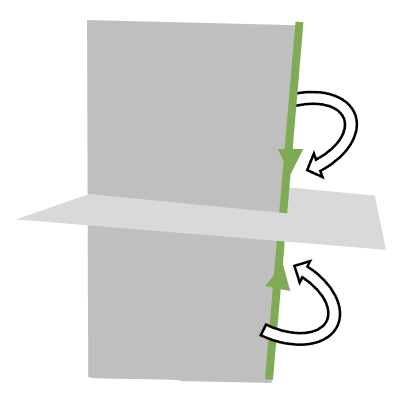}

\end{tabular}
    \caption{Pivot curves at the rim of a PD surface in 3D. Double arrows indicate the direction in which the scroll wave travels around the PDS. Green arrows indicate the pivot field $\vec{P}$, which can originate from pivot charges on the bounding surfaces. (a) A local wave block has a closed circular pivot curve around it. (b) A wave block extending to the medium boundaries, where two PDLs are seen. Note that the pivot charge on both surfaces is assigned while looking at the pattern from outside the medium. (c) A circular-core rotor, with two similarly oriented pivot fields (semi-filaments). This situation cannot lead to a closed pivot curve and we conclude that they emerge from pivot charges on the PDLs on the bounding surface. (d) Proof that the direction of $\vec{P}$ is continuous along a pivot curve. Otherwise a transverse PDS (light gray) would be needed.  \label{fig:pivotcurves}} 
\end{figure*}

Fig. \ref{fig:pivotcurves} presents some examples of pivot vector charges. For simplicity, the head and tail curves are not drawn here. Panel a. shows a local wave block inside the medium: A propagation wave from the front of the medium to the back reaches an unrecovered region and needs to travel around it, as indicated by the double arrows. As a result, a closed pivot curve emerges. When this wave block grows and collides with the medium boundaries, a transmural block can form (panel b.), which is bound by PDLs on either side with a total charge $Q=0$. In the classical viewpoint, there is no filament, but we find two pivot lines of opposite chirality traversing the medium, that one could call `semi-filaments'. These semi-filaments end in the pivot points of the PDLs that are seen on the medium boundary. It is interesting to see that both conduction blocks on either surface (e.g. the endo- and epicardium of the heart) are connected via a phase defect surface.  

Note that the end points of the pivot curves in the bulk are precisely the pivot points of the 2D theory. The sign of those surface pivot charges is chosen as if viewed from outside the medium. Thereby, we are consistent with the 2D theory, which was developed for patterns on the cardiac surface, seen from outside the heart. Note further, that when the $\vec{P}$ field touches the medium boundary, one finds a p$_+$ and when this field leaves the medium boundary, one finds a p$_-$. Thus, in analogy to the closed head and tail curves, we may choose to make the pivot curves continuous by assigning a pivot vector charge to a phase defect line with $Q=0$ on the medium surface, flowing from p$_+$ to p$_-$. 

The case of a linear-core scroll wave is different, see Fig. \ref{fig:pivotcurves}c. Here, the wave circumscribes both pivot lines in the same direction, and on the outer medium boundary, PDLs of topological charge $Q=1$ and $Q=-1$ are seen. 
In the case drawn, one has two p$_-$ on the bottom medium boundary and two p$_+$ on the upper medium boundary. 

Along both pivot curves, the vector charge $\vec{P}$ points upward. It is therefore recommended to think of $\vec{P}$ as an electrical field, which has net sources or sinks in electrical charges, rather than a current, which must be conserved in space. The field lines of $\vec{P}$ are organized in two flux tubes, i.e. the pivot curves, containing $P=\frac{1}{2}$, starting at the topological charges of the pivots on the bottom surface, with total charge $P=1$, and point towards the other medium boundary with total charge $P=-1$. 

Nonetheless, we can close the pivot curves at the medium boundary, by convening that the classical topological charge $Q=\pm 1$ can act as a net source or sink of the pivot field $\vec{P}$. That is, between pivot pairs on the surface we draw 
\begin{align}
    \text{p}_- &\leftrightarrow \text{p}_-,  & \text{p}_+ &\rightarrow \leftarrow \text{p}_+, & \text{p}_+ & \rightarrow \text{p}_-. 
\end{align}

Finally, we argue that the pivot field cannot suddenly change direction on a pivot curve. The dark-gray PDS in Fig. \ref{fig:pivotcurves}d shows such a hypothetical example. However, in the region where there could be a scroll wave encircling the pivot curve in opposite directions, the phases of the excitable element will not be continuous. Hence, there must be another PDS (light grey) that intersects the first PDS. We conclude that on a non-branching piece of the PDS border, the pivot charge is continuous. 

To summarize, we see that, when considering the boundary as a PD, the rim of a PDS is a continuous curve, called the pivot curve, along which a `pivot field' can be seen to circulate in a continuous manner. This field can have net sources or sinks on the medium boundary if there are phase defect lines with a net topological charge present. 

\subsection{Applications to excitation patterns}

\subsubsection{Isthmus-mediated ventricular tachycardia \label{sec:loops}}

We recently showed that the recombination of quasi-particles can provide additional geometric insights in 2D pattern evolution via Feynman-like diagrams \cite{Arno:2023feynman}. In 3D, even more complex dynamics are revealed, as phase defect surfaces that end both within the medium and on its boundary may become arbitrarily complex. We argue that one way to keep track of this complexity and reason about transitions in 3D is to use the continuity of pivot curves and the $\vec{P}$-charge they carry. Here, we apply this reasoning to the question under which an initial conduction block can grow and form a 3D scroll wave with a U-shaped filament.

\begin{figure}
\begin{tabular}{c c }
   (a)  & (b) \\
\includegraphics[width=0.2\textwidth]{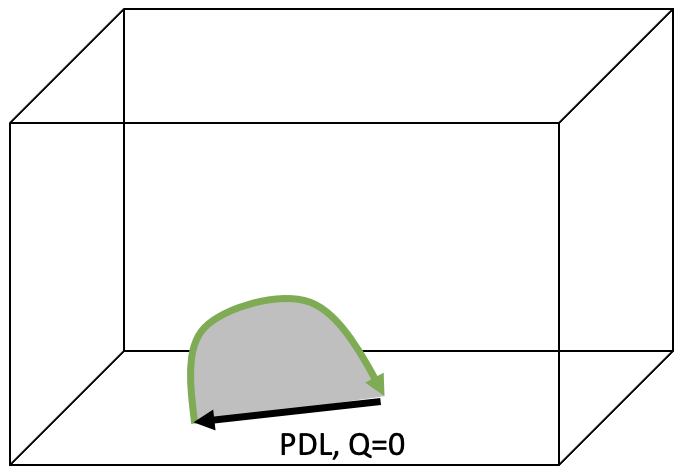}
    &
    \includegraphics[width=0.2\textwidth]{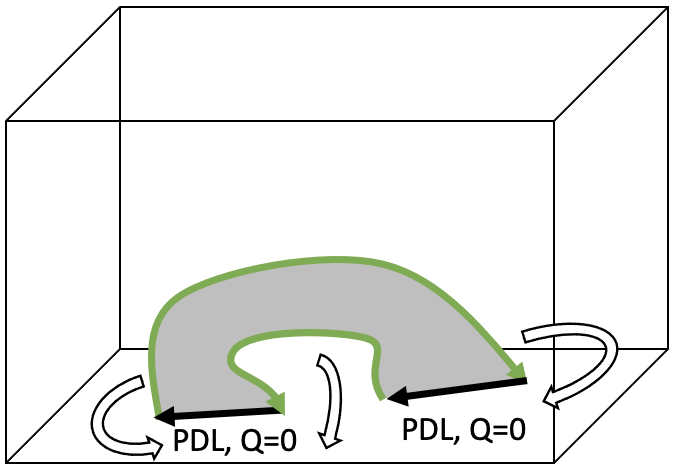}\\
     (c)  & (d) \\
    \includegraphics[width=0.2\textwidth]{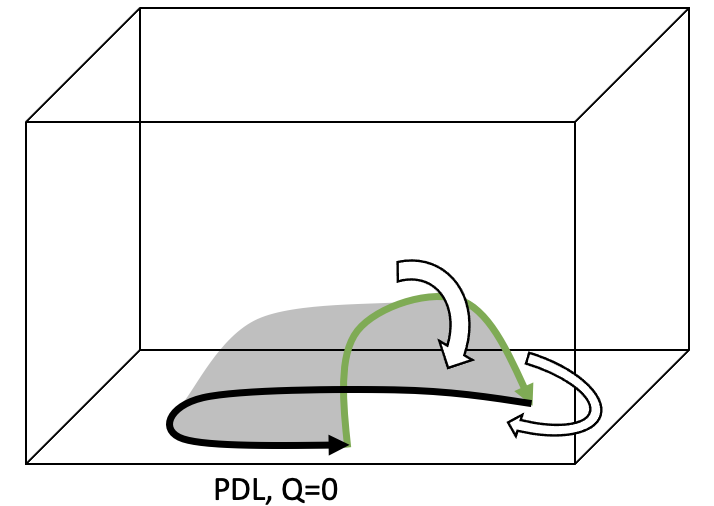}
    &
    \includegraphics[width=0.2\textwidth]{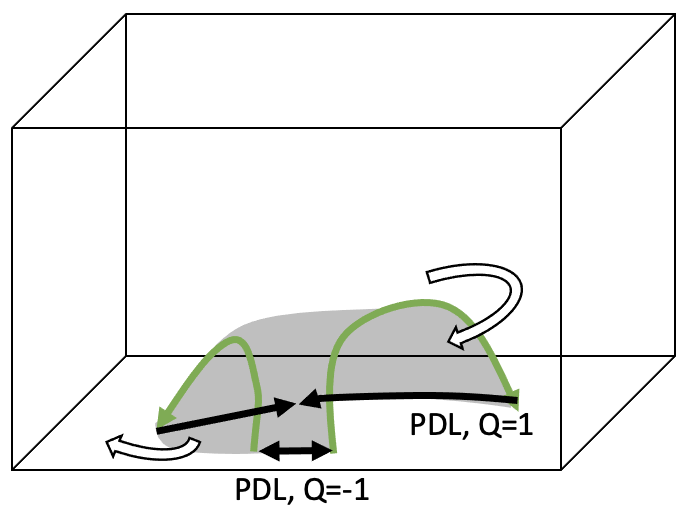}
    \end{tabular}
    
\caption{Example of pivot loop analysis of rotor formation. Double arrows indicate travel direction of the excitation fronts. (a) A wave block loop is formed near the medium boundary. (b) Detachment of the pivot from the boundary will not yield a rotor pair, since the created PDLs have no net topological charge. (c) A stretching of the PDS can yield an istmus through which the activation propagates (d), the sides of which are PDSs with opposite charge. 
    \label{fig:pivotloops}}
\end{figure}

When a wave block occurs, it forms a closed loop with a continuous direction of the pivot charge, as shown in Fig. \ref{fig:pivotcurves}a and Fig. \ref{fig:pivotloops}a illustrates this for a wave block near the medium boundary. Since in this case, the pivot field, indicated by the green and black arrows, is continuous, it can start from or can shrink to a single point. Therefore, this conduction block is easy to create or destroy, and will generally not sustain an arrhythmia on its own, but rather show a transient conduction block. 

To create sustained scroll waves, this simple loop needs to be converted to a loop with the property that it has a flip of the $\vec{P}$-field. If the conduction block line is locally detached from the medium boundary, there is still a simple closed pivot curve, see panel b. Such a case could happen due to a crescent-shaped inhomogeneity causing the initial block. To create a sustained rotor, the PDL on the surface first needs to grow sufficiently, as shown in panel c. Then, it can split into two parts, via the surface process in the right panel of Fig. \ref{fig:headcases}d.

Hence, two nearly parallel pivot curves are created, each with a non-zero topological charge. This situation is a bent version of situation in Fig. \ref{fig:pivotcurves}c, corresponding to a linear-core rotor. The sequence of panels a-c-d describes the onset of isthmus-mediated ventricular tachycardia in the heart and therefore shows why, in our opinion, pivot loop analysis can serve as a tool to understand three-dimensional cardiac arrhythmia patterns and their onset. 

\subsubsection{Twistons \label{sec:twistons}}
It was found in numerical simulations that, if a linear-core scroll wave is initiated in a medium with locally varying anisotropy, the rotor filament can exhibit local concentrations of twist, which travels along the filament. Fenton and Karma were the first to detect this behavior and called this traveling twist a `twiston' \cite{Fenton:1998}. 

We are now in the position to shed new light on their observation. Due to the local anisotropy in the medium, the scroll wave becomes twisted, and so the head curve itself becomes curved. For this reason, all points of the head curve do not simultaneously reach the pivot curve, when moving around the PDS. Where the head points do, it turns from one side of the PDS, intersecting the pivot, to the other. During this turn, a local twist is produced. Since the head described in the new viewpoint can coincide with the filament in the classical viewpoint,this intersection of the head with the pivot is also known as the twiston.

\begin{figure}
\begin{tabular}{c c}
    (a) & (b) \\
    \includegraphics[height=3cm]{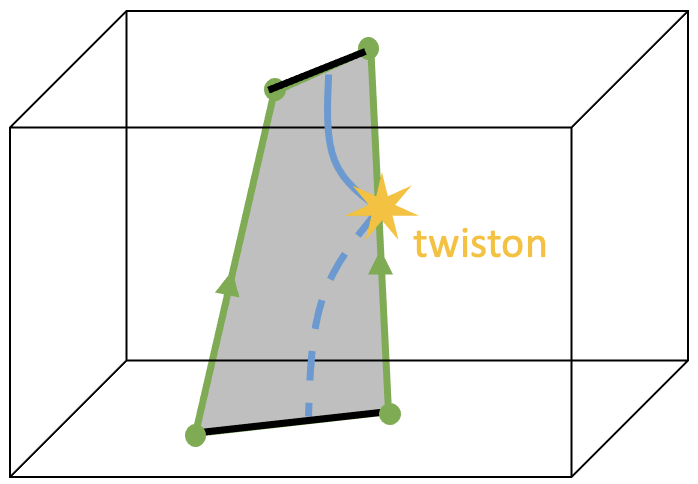} & \includegraphics[trim={2.cm 0.cm 20.cm 0.cm},clip,height=3cm]{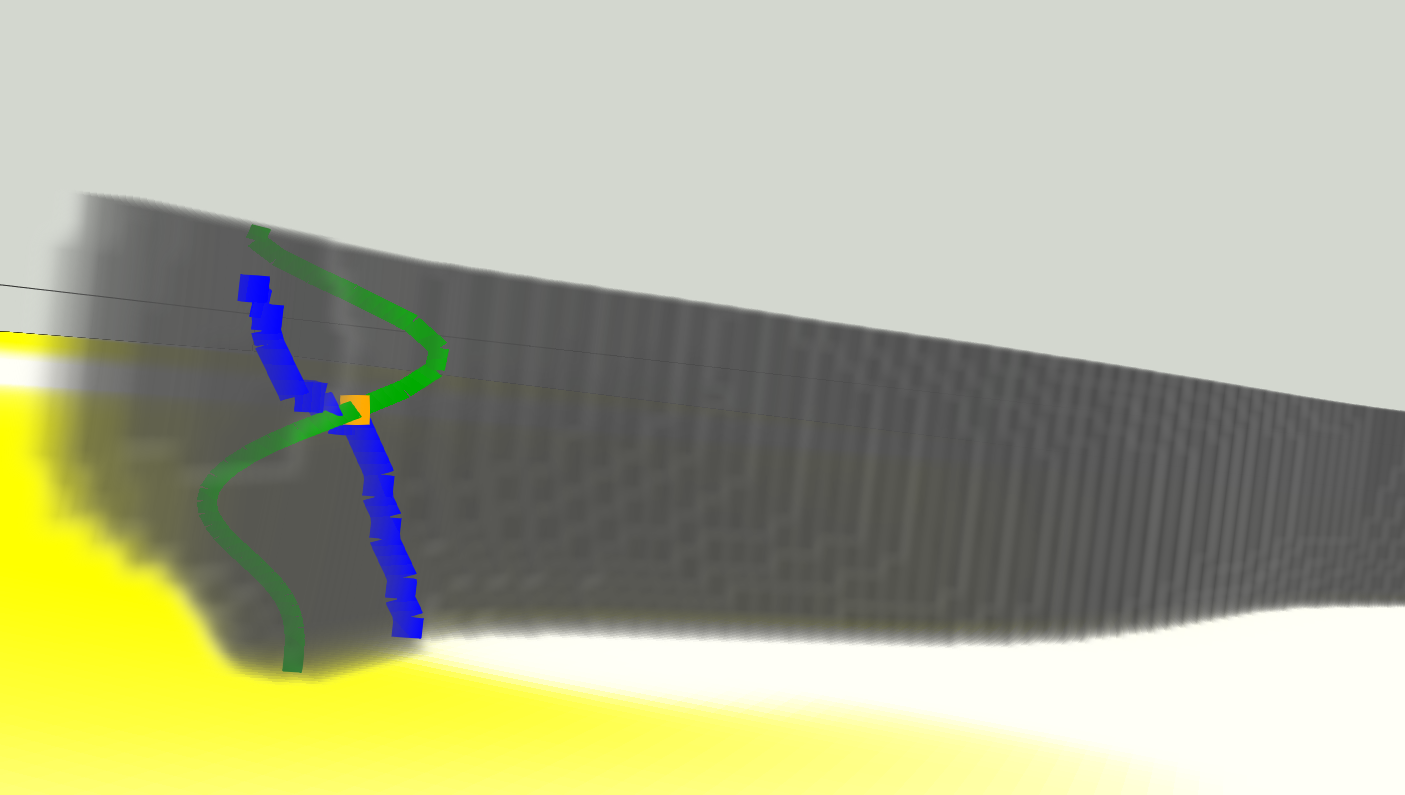}
\end{tabular}
    
    \caption{Explanation of the twiston phenomenon, as a scheme (a) and in an anisotropic simulation ($D_1= 1, D_2=0.25$) for which the fibre angle varies from $-60 ^{\circ}$ (bottom) to $60 ^{\circ}$ (top), with BOCF reaction-kinetics \cite{BuenoOrovio:2008} (b). Here, the head (blue), pivots (green), and twiston (orange) are annotated as well.  
    \label{fig:twiston}}
\end{figure}

This process is schematically rendered in Fig. \ref{fig:twiston}a and illustrated by a simulation in Fig. \ref{fig:twiston}b. In short, the twiston is a co-dimension three structure where a head curve intersects a pivot curve. Therefore, it can only occur in more than two spatial dimensions. From this observation, we conclude that twistons may be a common feature in complex arrhythmia patterns since every PDS with a head curve on it, both conduction block lines, $Q=0$ and rotors, $Q=\pm 1$, can exhibit the non-synchronous turning of this head around a pivot curve. 

The wave back is also expected to intersect a pivot curve, which would present a second type of twiston, which we, however, did not pursue in simulations yet.

\subsubsection{Hybrid re-entry \label{sec:hybrid}}

\begin{figure}
    \centering
 \includegraphics[trim={10.cm 4.cm 10.cm 5.cm},clip,width=0.4\textwidth]{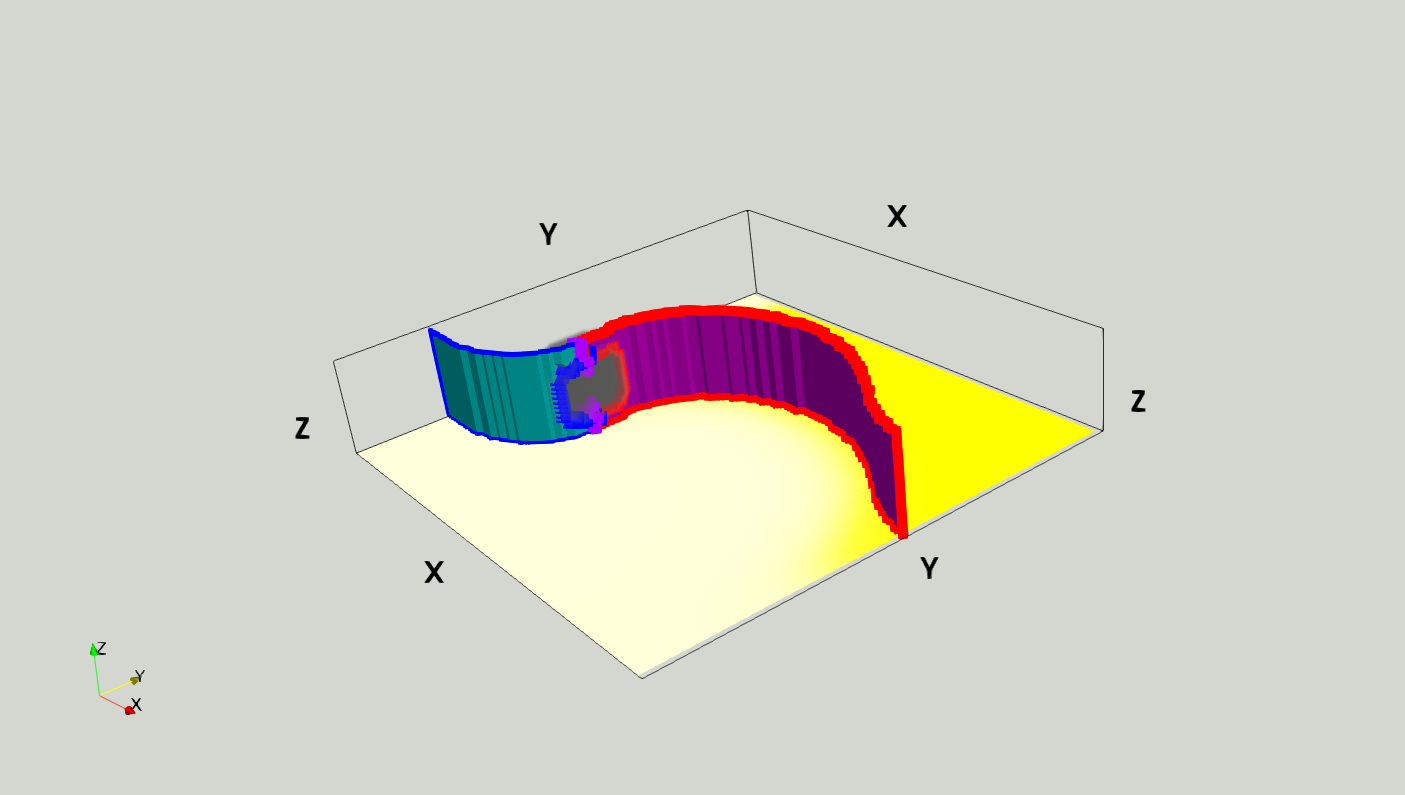}
    \caption{A case of hybrid re-entry, where a PD joints an inexcitable obstacle in a 3D case, where a PDS intersects an obstacle. The purple curve indicates the filament. Simulation was done in an isotropic medium with AP reaction-kinetics \cite{Aliev:1996}, for which an unexcitable sphere (black) is present at the core of the created spiral.  \label{fig:hybrid}}
\end{figure}

Since inexcitable obstacles can also be regarded as phase defects, PDs in the same excitation structure may span both the obstacle and a PD formed by a conduction block. Fig. \ref{fig:hybrid} shows a circular rotor intersecting an inexcitable obstacle. Since realistic excitable media, such as cardiac tissue, include a lot of inhomogeneities, it can be expected that most re-entrant patterns are of the hybrid type. Note that the filament indicated in purple is interrupted at the conduction block site, whereas the head and tail curves span the entire medium, from bottom to top, and circumvent the obstacle. 

\subsubsection{The M\"obius scroll \label{sec:mobius}}

The here-proposed formalism not only offers novel insight into certain phenomena but also predicts new states. The first example is a scroll wave around a conduction block line in the shape of a M\"obius band.

Its construction starts from the observations that head curves can turn around a phase defect surface, as in the twiston case, and that PDs in the linear-core regime are not necessarily oriented. Therefore, a head curve may wrap once around a phase defect surface, after which the surface is pasted together as a M\"obius strip. Our M\"obius scroll was created similarly as one would make such a strip with a sheet of paper: We sampled frames from half a rotation of a linear-core spiral wave in 2D BOCF model, isotropic medium, and pasted them along a circle in cylindrical coordinates in 3D while rotating the solution over 180$^\circ$ when the circle was traversed. Mathematically, we used toroidal coordinates $(r, \theta, \zeta)$:
\begin{align}
x&= (R + r\cos \theta) \cos \zeta \\
y&= (R + r \sin \theta) \sin \zeta \\
z&= r \sin \theta. 
\end{align}
If $\uu_0(\rho, \phi,t)$ is a 2D linear-core solution in polar coordinates $(\rho,\phi)$ with period $T$ and the middle of the core (PDL) at $x=0, y=0$, then we took as initial conditions in our simulation:
\begin{align*}
 \uu(r, \theta, \zeta) = \uu_0(\rho, \phi - \zeta/2, \pi t/T)
\end{align*}
As a result, we find a scroll ring that looks untwisted at a large distance, but whose head curve crosses the rim of the strip twice. Given previous paragraphs, one has two twistons here. The rim of the strip is one continuous pivot curve without changes in the direction of the pivot field. The rotation of the scroll wave around the M\"obius strip was stable for ten rotations. During this time, the core underwent precession as in 2D, causing the strip to rotate. We did not study its long-term behavior here. 

\begin{figure}
    \centering
    \includegraphics[width=0.3\textwidth]{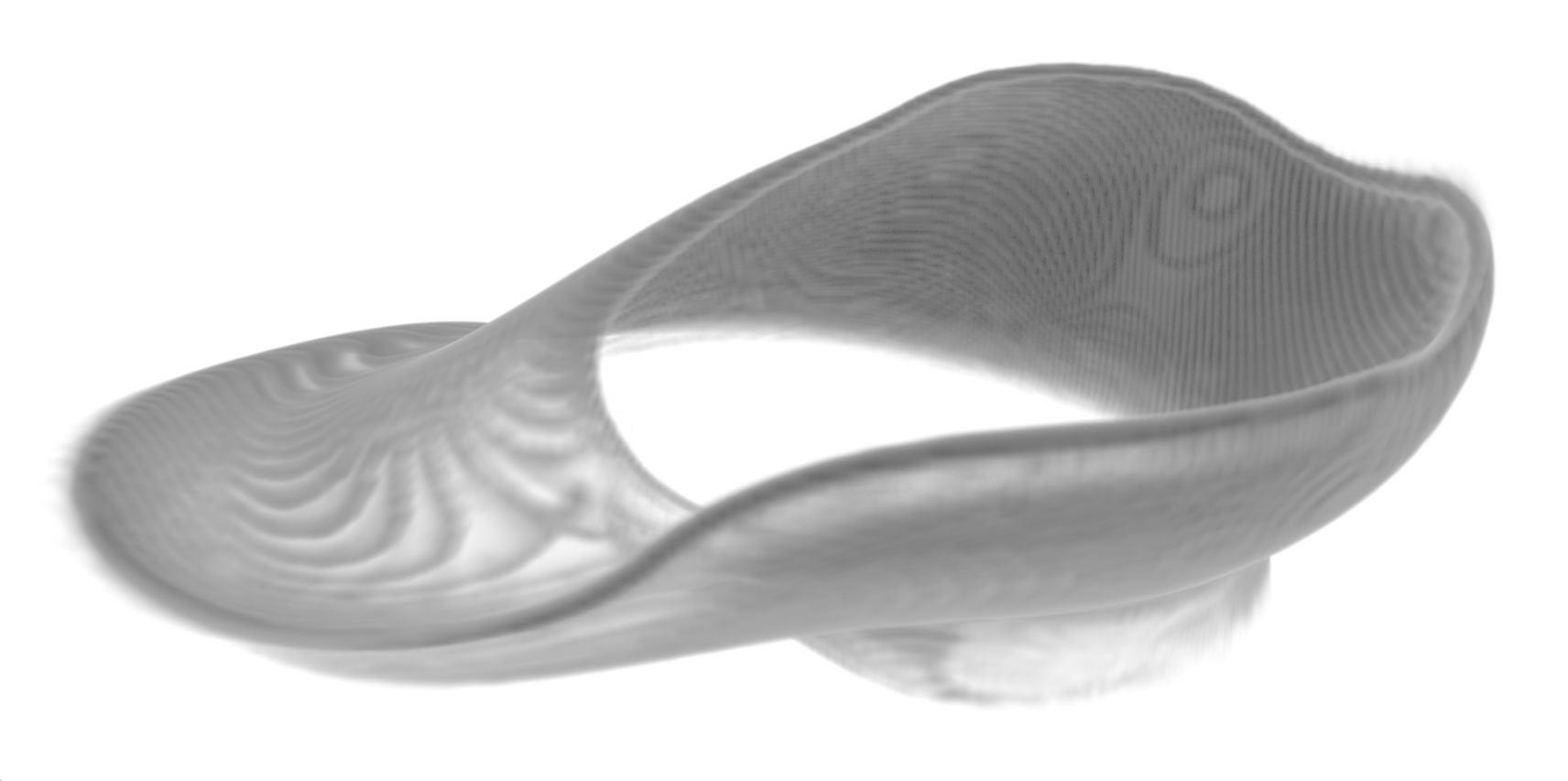}
    \caption{A M\"obius scroll, i.e. a scroll ring whose core is a M\"obius strip. For more details of the simulation, see Sec. \ref{sec:mobius} \label{fig:Mobius}}
\end{figure}

\subsubsection{Branching PDSs generate complex surface patterns \label{sec:branching}}

Based on our framework and initial numerical explorations, we conjecture that in inhomogeneous excitable media or 3D break-up regimes, there occur not only many wave blocks but that several of the wave blocks seen on the surface are connected by extended and branching PDs. This hypothesis could explain the nearly periodic patterns observed on the heart's surface during tachycardia. 

\begin{figure*}
       \centering
\begin{tabular}{c c c c}
\multicolumn{4}{c}{(a)}\\

\includegraphics[width=0.25\textwidth]{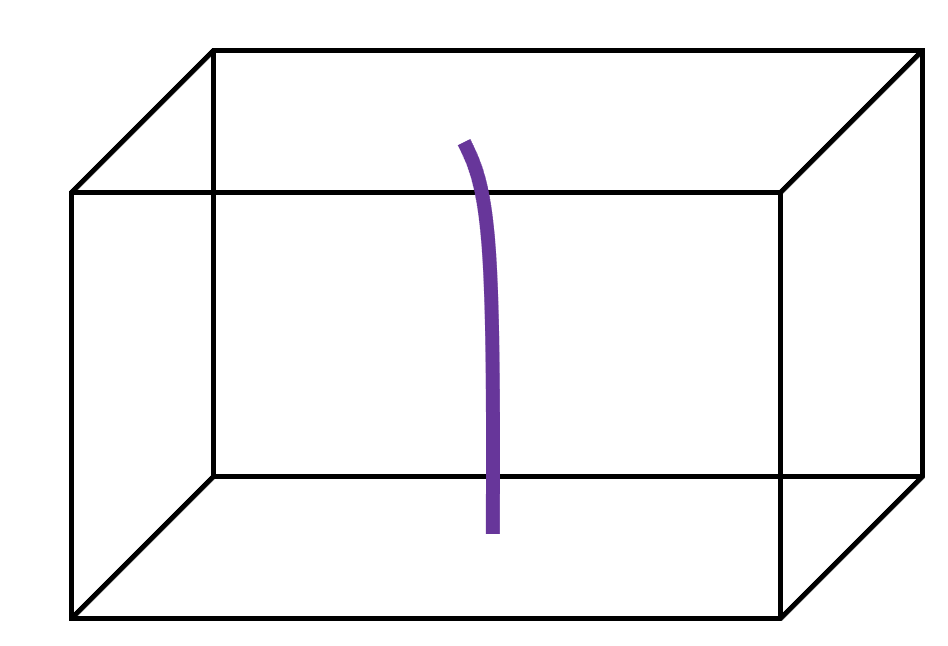}&
\includegraphics[width=0.25\textwidth]{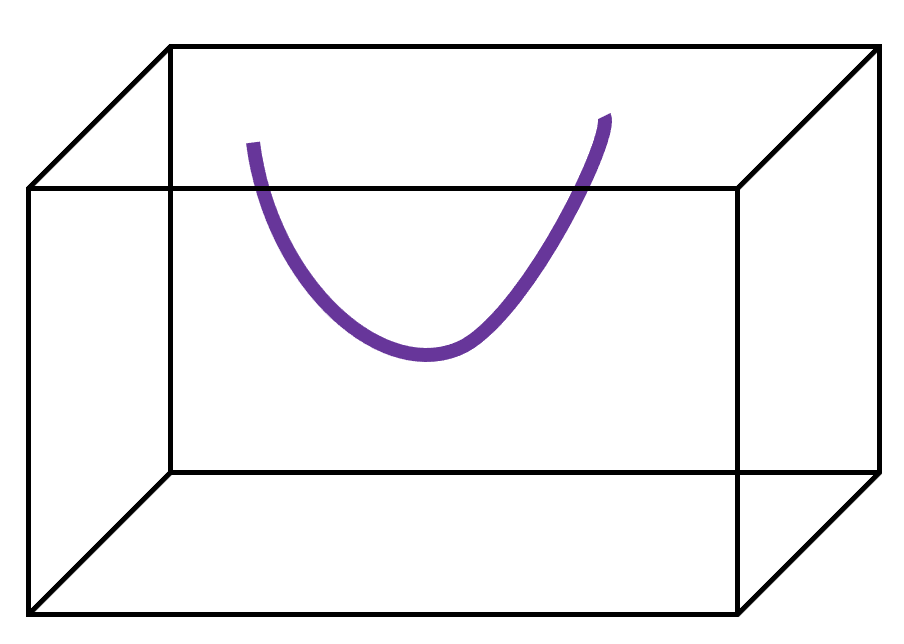}&
\includegraphics[width=0.25\textwidth]{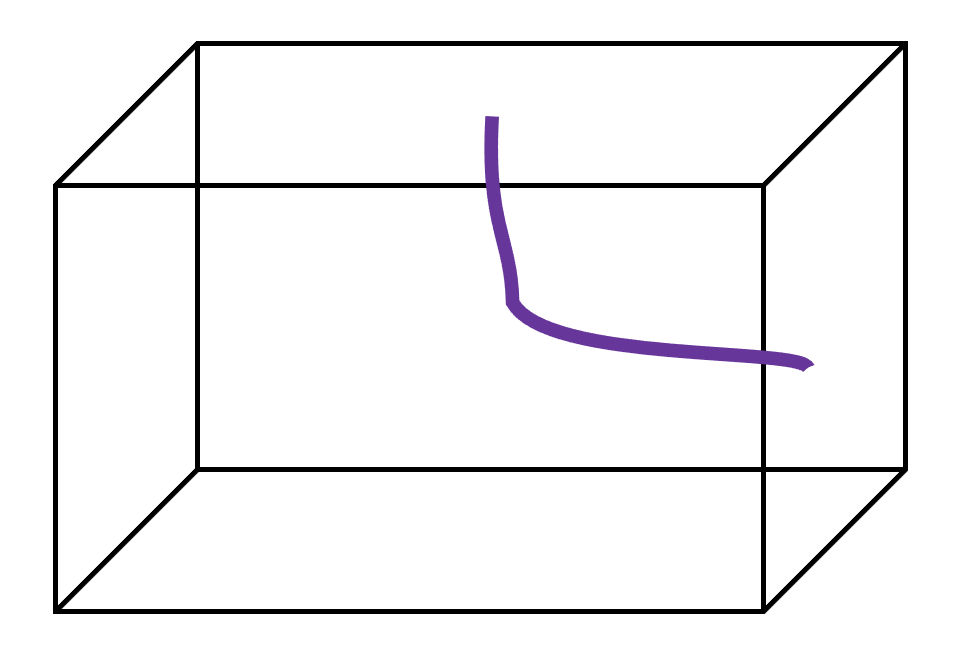}&
\includegraphics[width=0.25\textwidth]{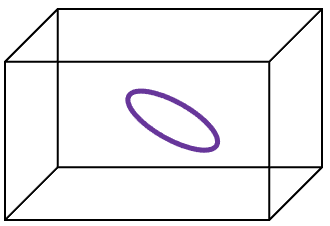}\\
\end{tabular}
\begin{tabular}{c c }
 (b) & (c) \\
 \includegraphics[width=0.3\textwidth]{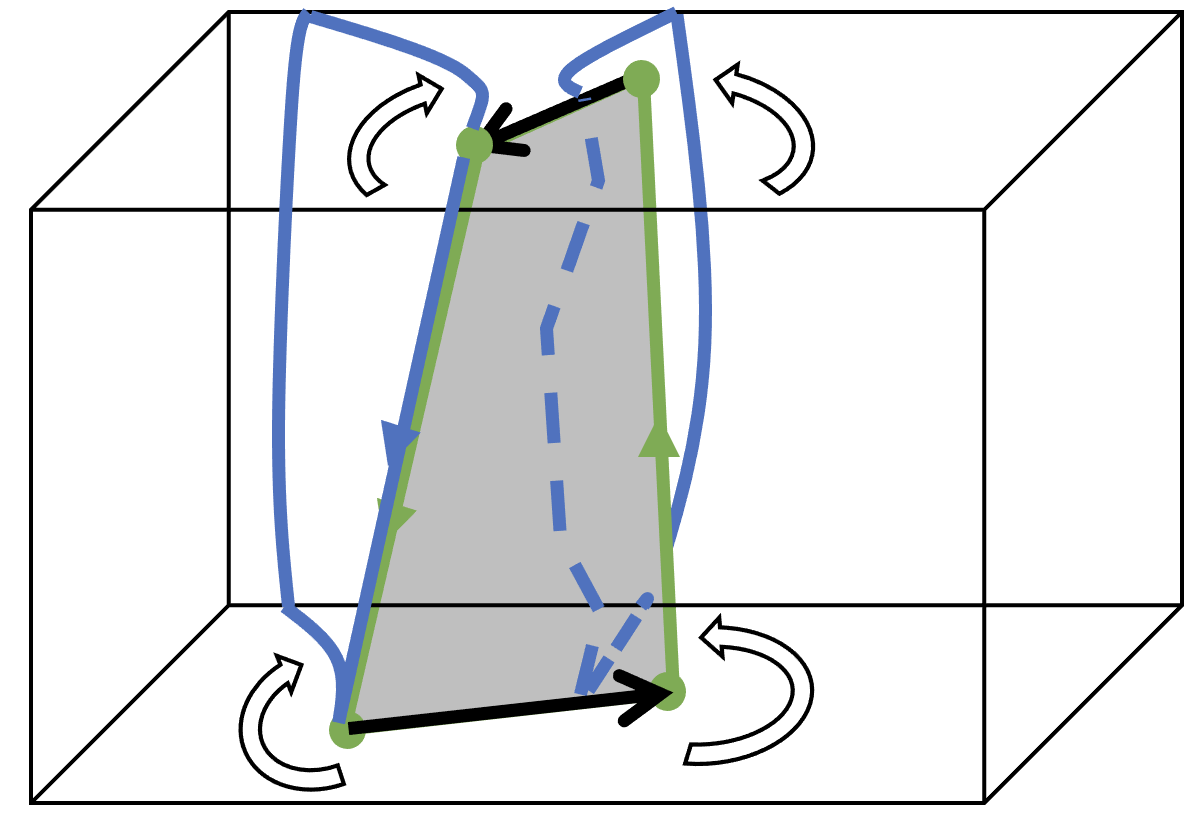} &
\includegraphics[width=0.7\textwidth]{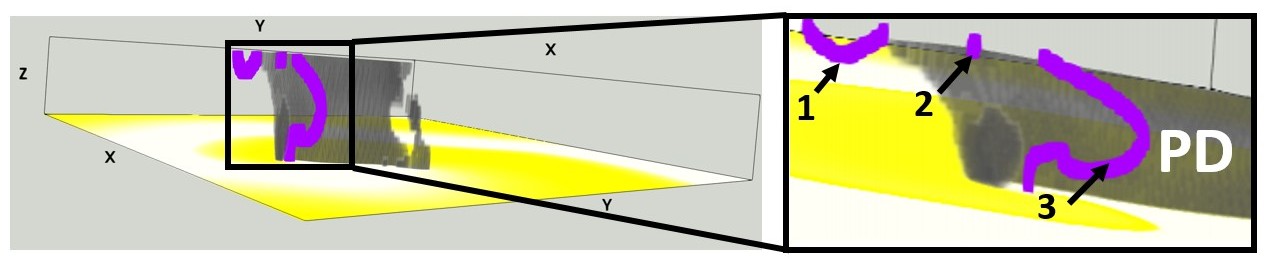}\\
\end{tabular}
\begin{tabular}{c c c}
(d) & (e) & (f) \\
\includegraphics[width=0.3 \textwidth]{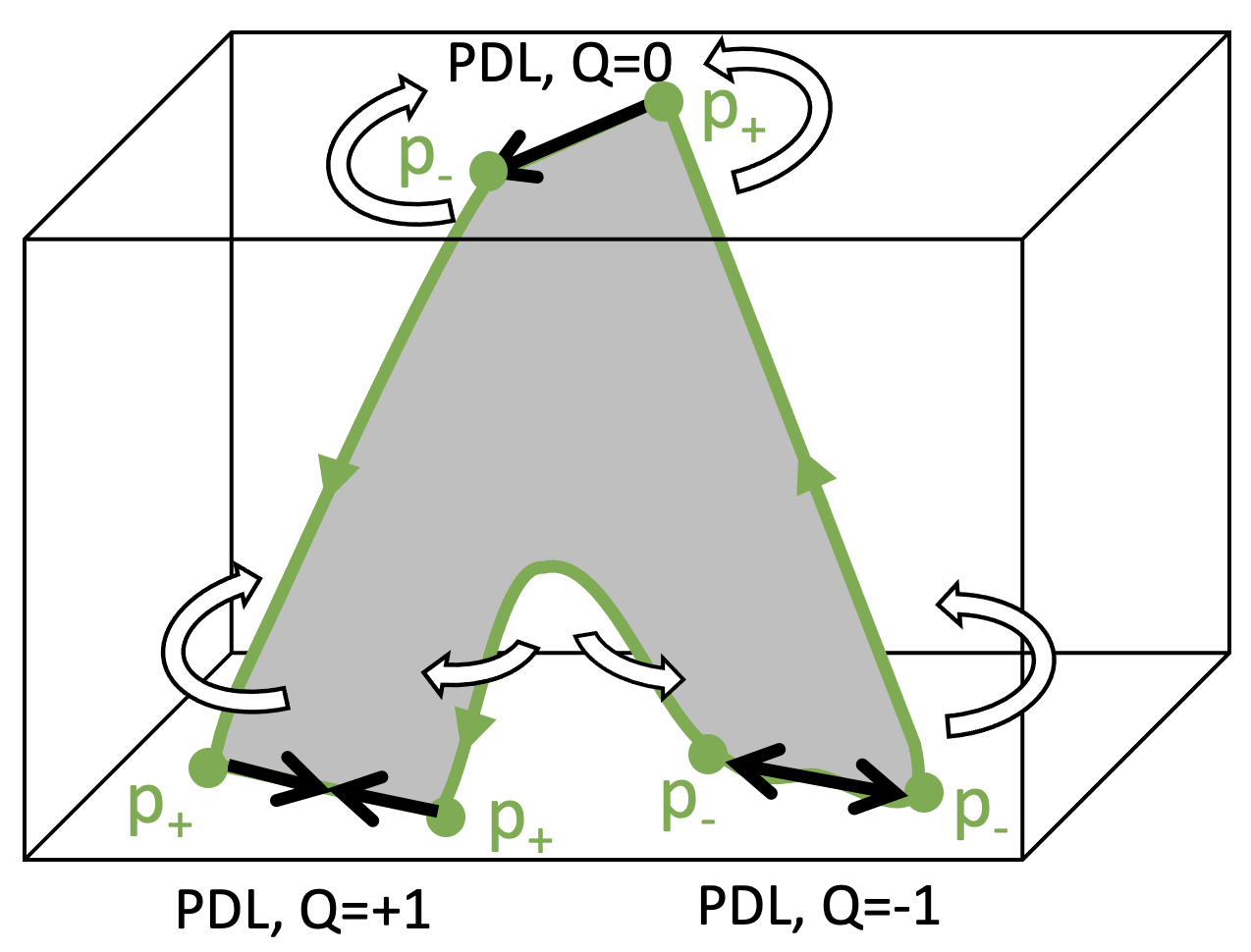} &
\includegraphics[width=0.3\textwidth]{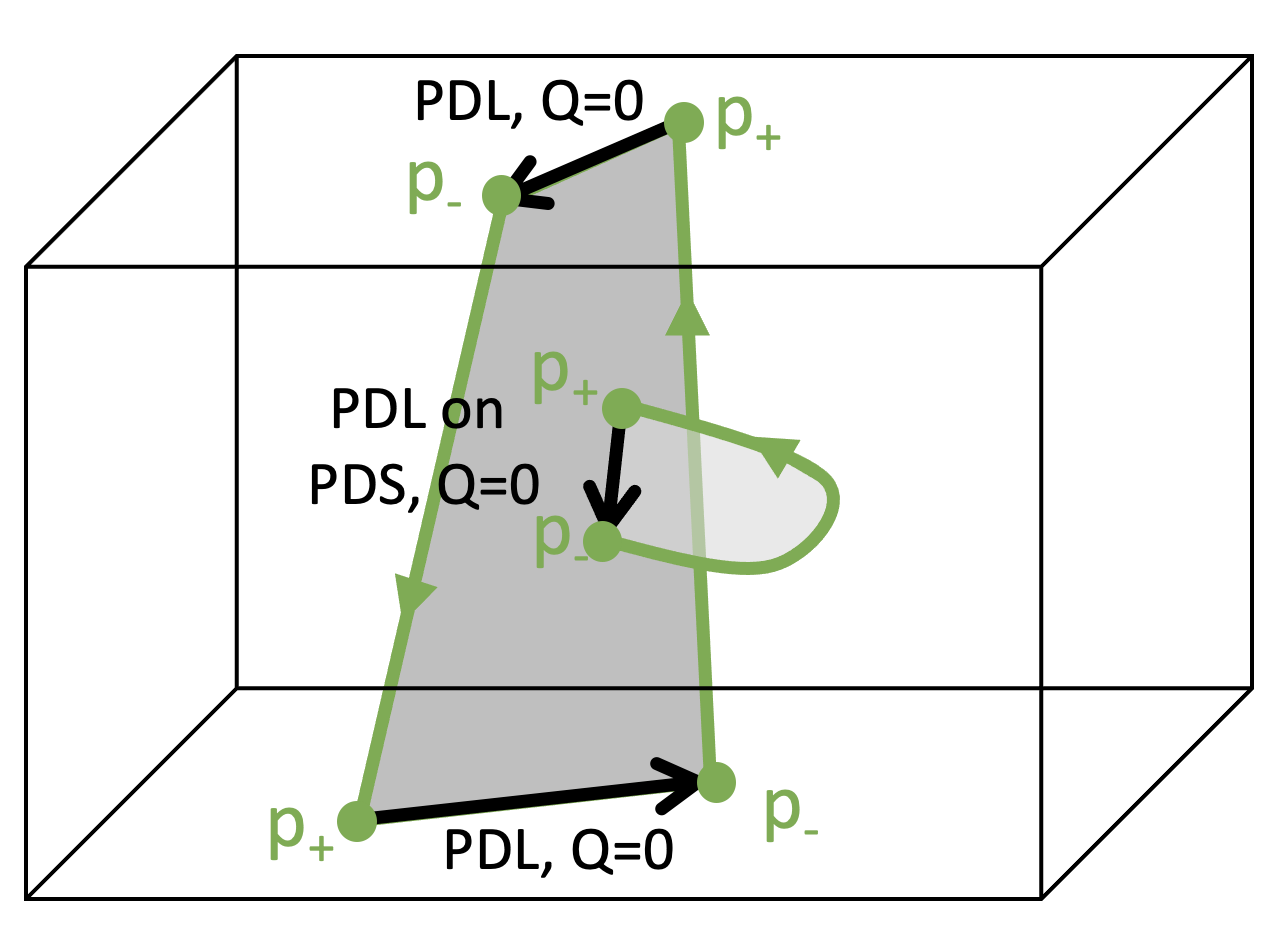} & 
  \includegraphics[width=0.33\textwidth]{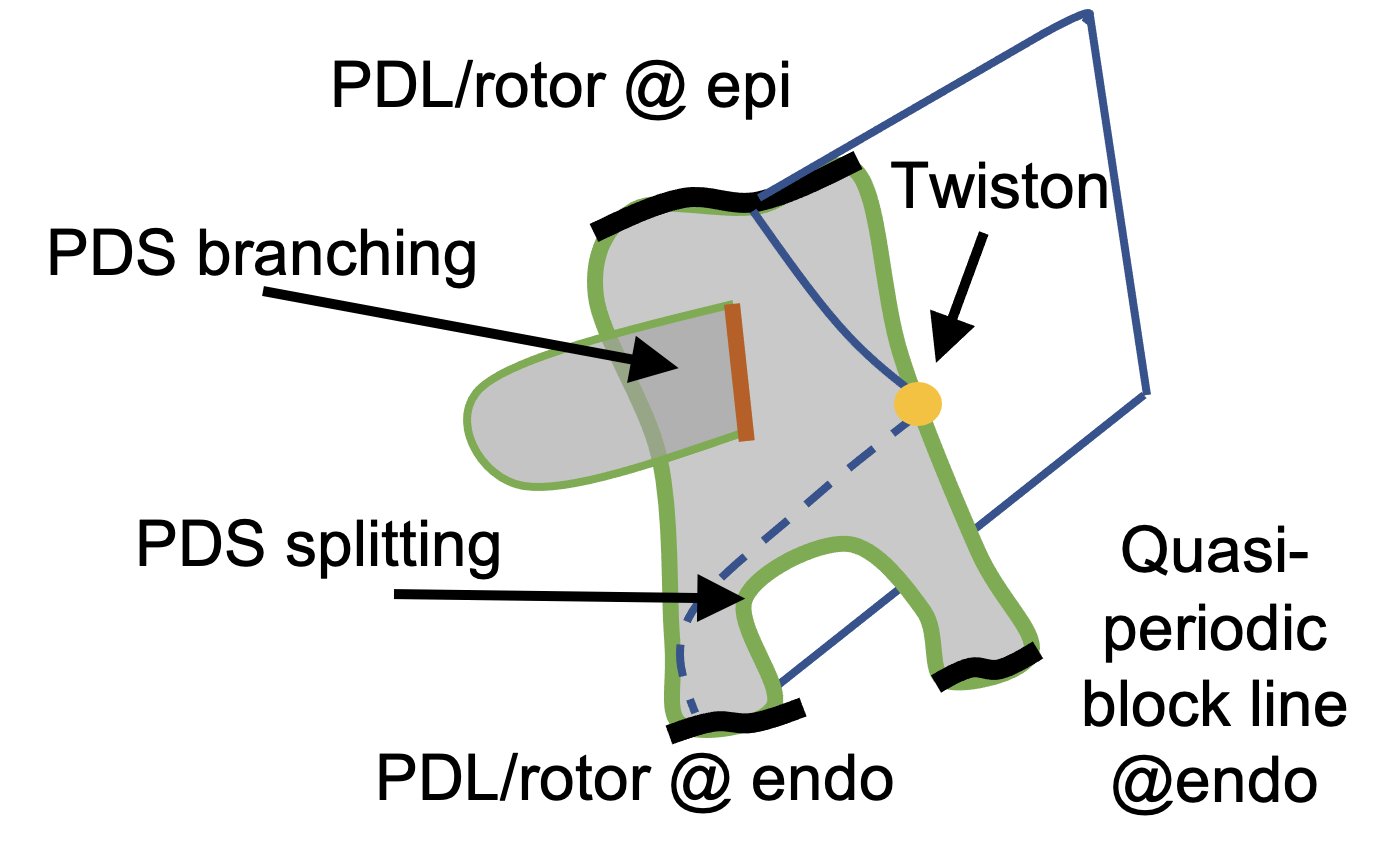} 
  \end{tabular}
    \caption{Complexity in the classical filament framework versus the phase defect framework. (a) In the classical theory, only I, U, L and O-shaped filaments can exist. However, when taking into account also the phase defect surfaces, more complex states are feasible. (b) A single PDS with multiple head curves on it, and thus carrying also multiple classical filaments. (c) Demonstration of the co-existence of head curves on the same PDS in a simulation with anisotropic geometry $D_1=1, D_2=0.25$ for which the fibre angle varies from $-60 ^{\circ}$ at the bottom to $60 ^{\circ}$ at the top with BOCF kinetics \cite{BuenoOrovio:2008}. (d) Phase defect surfaces can a split. (e) PDS can also have side branches, if conduction block arises during propagation around another conduction block. (f) An example of an hypothetical complex state that could in our framework occur during complex arrhythmia. \label{fig:topology}}
\end{figure*}

Fig. \ref{fig:topology} contrasts the few degrees of complexity in the classical viewpoint versus the many different topological structures possible in the here-proposed framework. Panel a. lists the topological states in the classical filament framework in 3D: Since filaments are curves that end on obstacles or medium boundaries, only I, U, L, or O-shaped filaments can occur. 

When taking into account conduction block surfaces, i.a. phase defects, too, we find that different head and tail curves can lie on the same phase defect surface. Fig. \ref{fig:topology}b shows an example of such a situation with a sketch. Here, a wave formed by a second stimulus is propagation along both sides around a conduction block. Since in 2D, two heads can lie on the same PDL, see top and bottom medium boundaries in Fig. \ref{fig:topology}b, two head curves can also lie on the same phase defect surface in 3D. We observed such states in a simulation of scroll wave dynamics in an inhomogeneous slab geometry, see panel c. The PDS connects one U-shaped filament with two I-shaped filaments. Thus, our framework can reveal connections between filaments that would be considered separate objects in the classical framework. It is also interesting to see how this looks to an observer, e.g. experimentalist or clinician, who only sees the top and bottom surfaces. At the bottom slice endocardium, one PDL and one PS are noted, while at the top slice epicardium, we see one PDL and four PSs. 

Panel \ref{fig:topology}d shows yet another configuration that is not captured by classical filament theory: Splitting PDS. This structure connects three PDLs on the surface, with total topological charges $Q= +1, Q=-1$, and $Q=0$. The mismatch in epi- and endocardial patterns is, in this case, not present in the phase defect view. Here, drawing the $\vec{P}$-field gives information on the topological charges seen on the medium surfaces. Note that there is no limit on the size or complexity of topologically different PDSs: They could in principle span the entire medium, connecting $N$ PDLs on the medium boundary.

Panel e. illustrates that during navigation of a wave front around a conduction block, a secondary conduction block may occur. As a result, a side branch to the initial PDS is formed. Here, we find a PDL on another PDS. From the viewpoint of treating PDs to some extent like inexcitable boundaries, this poses no problem to our framework. 

Finally, panel f. shows an example of a complex branching PDS combining many of the non-trivialities above. Even for a branching PDS, drawing its head shows that the PDLs visible on the medium's surface must have a total $Q$-charge equal to $0$. From this schematic drawing, we see that during each rotation of the front around the structure, the head will also sweep the rightmost `leg' which has $Q=0$. Hence, this phase defect will be visited by impeding fronts every rotation of the larger scroll wave, without the head ever making a full rotation around the PDL with $Q=0$. We conjecture here that this mechanism may be responsible for many of the `rotors' or 'wavelets' that make not a full turn on the cardiac surface that are observed during ventricular arrhythmias. 

\section{Discussion \label{sec:discussion}}

\subsection{Relation with other works and disciplines}

\subsubsection{Relation to topological models for cardiac excitation}

Our work is an extension of the classical phase analysis framework which resulted from a seminal paper by Gray et al. \cite{Gray:1995b}, and lead to classical filament theory in 3D \cite{Winfree:1984,Clayton:2005,Pertsov:1990b,Pertsov:1995,Pertsov:1999,Clayton:2005}.
Ribbon-like filaments have been suggested in experiments before \cite{Efimov:1999}, which comes close to our notion of phase defect surfaces. The ribbon model described in \cite{Echebarria:2006} resulted solely from describing twist of a circular-core scroll wave and was not related to linear rotor cores. 

A recent topological approach by Gurevitch and Grigoriev \cite{Gurevich:2019}, similar to previous work by Keener \cite{Keener:1998d}, highlights contour curves of two variables, thereby dividing the medium into excitable and recovering domains. However, the crossings of contour lines still correspond to classical tips, and the conduction blocks remain essentially untracked. 
Here, the partition of the medium in three regions, i.e. excited, unexcited, and a phase defect domain, includes the conduction blocks as PDs. 

We establish the topological model of head, tails, and pivots in three spatial dimensions, where two different viewpoints were used: (1) One could look at the medium boundary as a 'brane' for which the head, tail and pivot end on these medium boundaries, carrying a half-integer charge, or, (2) One could look at the medium boundary as a phase defect, for which the head, tail and pivot become continuous closed curves. Both viewpoints are useful, seen by, for example, the explanation of the isthmus-mediated ventricular tachycardia Sec.\ref{sec:loops} and the twiston analysis Sec. \ref{sec:twistons}. 

The 3D viewpoint itself is tightly connected to the 2D theory: The fact that heads and tails form closed curves implies that the total head charge and tail charge \cite{Arno:2023feynman} is zero over the surface. 

\subsubsection{Cross-fertilization between disciplines}

In this work, we have established a few remarkable links with other scientific domains, while unraveling the potential topological structures that underlie complex cardiac arrhythmia.

First, remark that when a phase defect surface, a structure of dimension $d=2$, ends on the medium boundary, that boundary must contain a phase defect line with co-dimension $d=1$ a topological charge $Q$ consistent with the PDS. Similarly, head, tail, and pivot curves with co-dimension $d=1$ end on the medium boundary in a point with co-dimension $d=0$ that is itself a head, tail, or pivot seen on the surface. This situation is reminiscent of string theory in theoretical physics, where branes can end on other branes \cite{Zwiebach:2004}, with a topological charge associated with the end points. Note further, that the vector charges $\vec{H}$, $\vec{T}$, $\vec{P}$ are also used in string theory \cite{Zwiebach:2004}.

Secondly, there is a link with Morse theory from mathematics \cite{matsumoto2002introduction}, which considers a lower-dimensional intersection of surfaces. In the present context, we see in the 3D space that taking cross-sections perpendicular to the Z-axis delivers 2D patterns and structures. Moving along Z can also be interpreted as forward evolution in time, such that the 3D theory may, as well, constrain or offer new insights into the evolution of 2D patterns. Similarly, one may wish to consider the time evolution of 3D excitable patterns in 4D space-time and look for further geometric insights, e.g. by identifying co-dimension four structures. 

\subsection{Limitations and outlook of the formalism}

\subsubsection{Structure of the state-space attractor}

An implicit assumption in our formalism is that the attractor for the local point-dynamics is simple, and has a ring-like or torus-like structure as shown in Fig. \ref{fig:torus}. This regime corresponds to an oscillatory or excitable system, where all elements follow approximately the same excitation loops. We are thus for now disregarding memory effects of the excitable elements. 
Additionally, in some real-world systems, such as repeated spiking in neural models \cite{Bressloff:2014} or early or delayed afterdepolarizations in the heart \cite{Luo:1994b,antzelevitch_overview_2011}, the dynamical attractor will be more complex. Therefore, an upcoming task is to accommodate those systems within the PD framework, e.g. by relieving that a single phase, i.e. element of $U(1)$, is sufficient to characterize the local state of the system.

\subsubsection{Numerical determination of structures}

Throughout this work, we have used suboptimal methods to track heads, tails, and pivots in 2D and 3D, see Sec. \ref{sec:methods}. More efforts will be needed to design robust algorithms to reconstruct these in 2D observations on the medium surface or 3D simulations or reconstructions \cite{Christoph:2018}, similar to our recent work on phase defects in 2D excitable media \cite{Kabus:2022}. To convey the concepts clearly, we drew the head curves, tail curves, and pivot curves separately in most figures, while in some cases the 3D equivalent of a compound state is present. E.g. Fig. \ref{fig:topology}b shows a `growth curve' at the left-hand side of the PDS. 
Fig. \ref{fig:Hren} presents a rendering of the excitation structures discussed here detected in a 3D heart simulation. We are convinced it would be useful to add such detection and presentation to simulation and visualization software in the near future. 

\subsubsection{Linear core regime and the zero-thickness limit}
The definition of heads and tails is independent of the shape of the phase defect, i.e. we can assign a head and tail for both circular and linear core. To define the pivot point or pivot lines, we assumed a regime where the PD contracts to a line in 2D or surface shape flattened region in 3D. However, due to electrotonic effects, biology or the diffusion in the models the phase defect can extend to a surface in 2D and a volume in 3D. Nevertheless, it is possible, to represent the two-dimensional phase defects as planar graphs, which enables a robust definition of the pivots \cite{Kabus:2022}. More efforts will be needed to extend this method to the three-dimensional case and devise robust algorithms. 

\subsubsection{Conservation of pivot charge}

The conservation of pivot charge is guaranteed in the linear-core regime, in cases where PDLs or PDSs move freely without branching or touching boundaries. Our formalism can be extended to those cases by considering the joints, as quasi-particles. This treatment falls outside our current scope.  

\subsubsection{Topology versus dynamics}

In the classical viewpoint, the motion of spirals, scrolls, and filaments is dominated by curvature effects, including filament tension \cite{Keener:1986, Wellner:2002, Biktashev:1994, Dierckx:2015PRL, Dierckx:2017PRL}. This theory is well-established for circular cores, and its extension to linear cores led to our investigation of phase defects \cite{Arno:2021, Kabus:2022}, which were independently discovered by Tomii et al. \cite{Tomii:2021}. Within our present scope, we only treated the topology of the patterns, and rewrote the basic terminology to classify excitation patterns for this purpose. Subsequent work will be devoted to seeing how the different structures of heads, tails, pivots, phase defects evolve in time.

\subsection{Outlook on cardiac applications \label{sec:outlookcardiac}}

The presented framework is of generic nature, applicable to all excitable media sustaining wave block or spiral wave formation. However, given our personal motivation and experience, we briefly outline some possible applications to cardiac arrhythmia mechanisms. 

\subsubsection{A method designed for complex cases}

The classical PS theory handles the cases of sustained rotors in 2D and 3D well. However, for the case of scroll wave initiation at the start of an arrhythmia, or short-lived pivots that are typically observed during fibrillation, an adequate description was missing. These regimes often occur in experimental and clinical data, where the tissue geometry and functional properties are markedly less homogeneous than in most numerical simulations. With these cases in mind, the phase defect formalism was developed: The examples above are intended to show the proposed method and its potential. Still, the new framework is an evolution rather than a revolution, since a classical filament, or set of PSs, can still be identified with a head curve. To this, we add the conduction block zone, as it is key to understanding the total pattern. 

\subsubsection{Fundamental mechanisms of arrhythmia initiation, perpetuation and management}

Upcoming work will be targeted to detect pivots, heads, and tails in experimental and clinical patterns, to see how often they occur and via which transitions they are formed or annihilated. Note that this step requires no advanced inversion, as optical voltage mapping or multi-catheter recordings of local activation times contain sufficient information. 

Of several severe arrhythmias such as ventricular tachycardia, atrial fibrillation, and ventricular fibrillation, the precise spatio-temporal organization is not entirely understood \cite{tabereaux_mechanisms_2009}. Given this work, the multi-wavelet hypothesis of cardiac fibrillation \cite{Jalife:1998b} can be cast into the language of pivots connected by PDLs to study the process quantitatively. The main question is to which extent the phase defects are connected by branching PD structures within the cardiac wall, as suggested in Fig. \ref{fig:topology}f. 

A forthcoming question is how current arrhythmia therapy can be viewed in the phase defect framework, to further optimize it. For example, pharmacological treatment or aging affects the dynamical attractor $\TT$ and therefore possibly the stability of the emergent structures. Another prevalent therapy is catheter ablation, in which part of the medium is scarred to prevent re-entrant patterns. Recent results of ablation of atrial fibrillation showed that certain pivots were responsible for sustaining the pattern \cite{seitz_af_2017}. It will be interesting to see if the phase defect framework can discriminate between the pivots and propose the correct ones to ablate. 

\subsubsection{From surface analysis to topology-based inversion}

Note that the analysis of PDs on surface patterns is only the first step on a roadmap exploiting the new topological insights. For, we here argue that if phase structures are observed on the epi- and endocardial surfaces of the heart, there are topological constraints on what lies in between: Heads and tails should properly connect to the other side, only touching each other (as a classical filaments) or phase defects. The pivot curves should also match on both sides, being connected by phase defects and their side branches. As such, there is a possibility to design topology-based models for the inversion of cardiac signals, which to a lesser or more intense degree exploit these constraints. 

\subsubsection{Role of heads and tails in electrogram generation}
Note further, that the concept of boundary heads and boundary tails is useful in the interpretation of cardiac electrograms. In the solid angle approximation to electrical signals from the heart, the measured unipolar potential is proportional to the solid angle spanned by the wave front \cite{Plonsey:1995}. Thereby, the electrical signal recorded by clinicians follows as a line integral from the head curves and tail curves that included also medium boundaries, see e.g. Figs. \ref{fig:headtail3D}, \ref{fig:hybrid}. Conversely, this implies that information on the location of heads and tails, and therefore also phase defects is directly encoded in the recorded electrical signals, which offers possibilities to include phase defects directly in inversion schemes. 

\section{Conclusion \label{sec:conclusions}}

A three-dimensional description of complex excitable patterns showing also conduction blocks is possible using three different types of closed string, i.e. head, tail, and pivot curves, which each carry a vector charge and live on phase defect surfaces. These surfaces of conduction block may split or branch to become arbitrarily complex, and hold the promise to be driving or sustaining several processes in excitable media. 

\section*{Acknowlegdements} The authors thank Piet Claus, Dani\"el A. Pijnappels, Tim De Coster, Olivier Bernus, Henri Verschelde, Alexander V. Panfilov and Joris Ector for helpful discussions and inspirational remarks. 

\bibliography{main.bbl}

\end{document}